\documentclass[a4paper,USenglish,cleveref,autoref,thm-restate]{scrartcl}
\date{}

\usepackage{mathtools} 
\DeclarePairedDelimiter{\abs}{\lvert}{\rvert}

\usepackage{calc}

\usepackage{onlyamsmath} 

\usepackage{xspace} 
\makeatletter
\def\?#1{}
\def\whp{w.h.p\@ifnextchar.{.\?}{\@ifnextchar,{.}{\@ifnextchar){.}{\@ifnextchar:{.:\?}{.\ }}}}}
\def\Whp{W.h.p\@ifnextchar.{.\?}{\@ifnextchar,{.}{.\ }}}
\makeatother

\def\lefttag#1{\tag*{\makebox[0pt][l]{\hspace*{-\textwidth}#1}}}
\def\numberthis{\addtocounter{equation}{1}\tag{\theequation}}

\let\originalleft\left
\let\originalright\right
\renewcommand{\left}{\mathopen{}\mathclose\bgroup\originalleft}
\renewcommand{\right}{\aftergroup\egroup\originalright}

\usepackage{braket} 

\makeatletter

\newcommand{\Prob}[1]{\Pr\left[#1\right]}
\newcommand{\Ex}[1]{\mathbb{E}\left[#1\right]}

\makeatother

\usepackage[textsize=tiny,colorinlistoftodos,obeyFinal,textwidth=1.2cm,disable]{todonotes} 

\usepackage{xcolor}
\definecolor{mygray}{HTML}{505050}
\definecolor{myred}{HTML}{d01a0b}
\definecolor{myorange}{HTML}{fc8720}
\definecolor{myyellow}{HTML}{f9f627}
\definecolor{myblue}{HTML}{405080}
\definecolor{mygreen}{HTML}{54e547}

\newcommand{\fullproofinappendix}[1]{The full proof can be found in #1.}

\usepackage{amssymb}
\usepackage{amsthm}
\usepackage[capitalize,nosort]{cleveref}
\usepackage{thm-restate}

\usepackage{booktabs}

\declaretheorem[style=plain, numbered=no, name=Corollary]{corollary*}
\declaretheorem[style=plain]{lemma}
\declaretheorem[style=plain, numbered=no, name=Lemma]{lemma*}

\declaretheorem[style=plain, numbered=no, name=Proposition]{proposition*}
\declaretheorem[style=plain]{theorem}
\declaretheorem[style=plain, numbered=no, name=Theorem]{theorem*}

\declaretheorem[style=definition, numbered=no, name=Definition]{definition*}

\AtBeginDocument{\newtheorem{observation}[theorem]{Observation}%
\newtheorem{claim}{Claim}[theorem]%
}
\crefname{observation}{Observation}{Observations} 
\crefname{claim}{Claim}{Claims} 

\def\N{\mathbb{N}}
\def\R{\mathbb{R}}

\def\x{\mathbf{x}}
\def\X{\mathbf{X}}

\DeclareMathOperator{\BinDistr}{Bin}

\def\dense{\medmuskip=2.0mu plus 2.0mu minus 2.0mu%
\thinmuskip=2.0mu%
\thickmuskip=2.0mu plus 5.0mu%
}

\usepackage[modulo,mathlines]{lineno}

\makeatletter
\newcommand*\AUXL@patchAmsMathEnvironmentForLineno[1]{
    \expandafter\let\csname old#1\expandafter\endcsname\csname #1\endcsname
    \expandafter\let\csname oldend#1\expandafter\endcsname\csname end#1\endcsname
    \renewenvironment{#1}
        {\linenomath\csname old#1\endcsname}
        {\csname oldend#1\endcsname\endlinenomath}
}
\newcommand*\AUXL@patchBothAmsMathEnvironmentsForLineno[1]{
    \AUXL@patchAmsMathEnvironmentForLineno{#1}
    \AUXL@patchAmsMathEnvironmentForLineno{#1*}
}
\AUXL@patchBothAmsMathEnvironmentsForLineno{equation}
\AUXL@patchBothAmsMathEnvironmentsForLineno{align}
\AUXL@patchBothAmsMathEnvironmentsForLineno{flalign}
\AUXL@patchBothAmsMathEnvironmentsForLineno{alignat}
\AUXL@patchBothAmsMathEnvironmentsForLineno{gather}
\AUXL@patchBothAmsMathEnvironmentsForLineno{multline}
\makeatother

\makeatletter
\def\Hy@Warning#1{} 
\makeatother

\let\epsilon\varepsilon

\usepackage[
		isbn=false,
		sortcites,
		maxbibnames=99,
		maxcitenames=2,
		maxalphanames=4,
		minalphanames=3,
		style=numeric,
		url=false,
		giveninits=true,
		backend=biber,
		doi=false,
	]{biblatex}
\DeclareCiteCommand{\nobracketcite}
  {\usebibmacro{prenote}}
  {\usebibmacro{citeindex}%
   \usebibmacro{cite}}
  {\textcolor{black}{\multicitedelim}}
  {\usebibmacro{postnote}}

\makeatletter
\newcommand\fullciteauthor[1]{\def\blx@maxcitenames{99}\citeauthor{#1}\def\blx@maxcitenames{2}}
\makeatother

\defbibenvironment{midbib}
  {\list
     {}
     {\setlength{\leftmargin}{\bibhang}%
      \setlength{\itemsep}{\bibitemsep}%
      \setlength{\parsep}{\bibparsep}}}
  {\endlist}
  {\item}

\usepackage{xpatch}
\DeclareFieldFormat*{citetitle}{\itshape #1}
\DeclareFieldFormat*{title}{\itshape #1\isdot}

\DeclareFieldFormat[inproceedings]{volume}{Volume #1}
\DeclareFieldFormat*{pages}{pages #1}
\AtEveryBibitem{
 \ifentrytype{inproceedings}{%
  \iffieldundef{series}{}{%
    \savefield{volume}{\theoriginalvolume}%
    \restorefield{number}{\theoriginalvolume}%
    \clearfield{volume}%
  }%
 }{}%
 \ifentrytype{book}{%
  \iffieldundef{series}{}{%
    \savefield{volume}{\theoriginalvolume}%
    \restorefield{number}{\theoriginalvolume}%
    \clearfield{volume}%
  }%
 }{}%
 \clearlist{address}%
 \clearlist{location}%
 \ifentrytype{book}{}{
  \clearname{editor}%
 }%
}
\AtEveryCitekey{\UseBibitemHook}

\renewcommand*{\multicitedelim}{\addcomma\space}

\renewbibmacro*{issue+date}{%
  \setunit{\addcomma\space}
    \iffieldundef{issue}
      {\usebibmacro{date}}
      {\printfield{issue}%
       \setunit*{\addspace}%
       \usebibmacro{date}}
  \newunit}

\renewbibmacro*{publisher+location+date}{%
  \printlist{location}%
  \iflistundef{publisher}
    {\setunit*{\addcomma\space}}
    {\setunit*{\addcolon\space}}%
  \printlist{publisher}%
  \setunit*{\addcomma\space}%
  \addcomma\space\usebibmacro{date}%
  \newunit}

\DeclareFieldFormat*{number}{\mkbibparens{#1}}
\DeclareFieldFormat[inproceedings]{number}{\addcomma\space volume #1\addcomma}
\DeclareFieldFormat[incollection]{volume}{\addcomma\space volume #1\addcomma}
\DeclareFieldFormat[article]{volume}{volume #1}

\renewbibmacro*{volume+number+eid}{%
  \setunit{\addcomma\space}
  \printfield{volume}%
  \setunit*{\addnbthinspace}
  \printfield{number}%
}

\xpatchbibdriver{incollection}{%
  \usebibmacro{maintitle+booktitle}%
  \newunit\newblock
}{%
  \usebibmacro{maintitle+booktitle}%
  \iffieldundef{volume}{
    \newunit\newblock
  }{
    \addcomma\space\newblock
  }
}{}{}

\xpatchbibdriver{book}{%
\newunit\newblock
\printfield{edition}%
}
{%
\setunit{\addcomma\space}\newblock
\printfield{edition}%
}{}{}

\DeclareCiteCommand{\citefauthor}
{\defcounter{maxnames}{99}%
\defcounter{minnames}{99}%
\defcounter{uniquename}{2}%
\boolfalse{citetracker}%
\boolfalse{pagetracker}%
\usebibmacro{prenote}}
{\ifciteindex{\indexnames{labelname}}{}%
\printnames{labelname}}
{\multicitedelim}
{\usebibmacro{postnote}}

\DeclareCiteCommand{\footnotecite}[\iffootnote\mkbibparens\mkbibfootnote]
  {}
  {\usedriver
     {\defcounter{minnames}{99}%
\renewbibmacro{in:}{. In }%
\defcounter{maxnames}{99}}%
{\thefield{entrytype}}}
  {\multicitedelim}
  {\usebibmacro{postnote}}

\DefineBibliographyStrings{english}{%
  mathesis = {Master's thesis},
}

\makeatletter
\let\@fnsymbol\@arabic
\makeatother

\crefformat{footnote}{#2\footnotemark[#1]#3} 
\author{%
Talley Amir\footnotemark[1],
James Aspnes\cref{author1},
Petra Berenbrink\footnotemark[2],\\
Felix Biermeier\cref{author2},
Christopher Hahn\footnotemark[3],
Dominik Kaaser\footnotemark[4],\\
and John Lazarsfeld\cref{author1}\\
}

\begin{document}

\allowdisplaybreaks

\title{Fast Convergence of $k$-Opinion Undecided State Dynamics
\mbox{in the} Population Protocol Model}

\let\subparagraph\paragraph
\def\paragraph#1{\subsubsection*{#1}}

\maketitle

\footnotetext[1]{\label{author1}Yale University, USA, firstname.lastname@yale.edu}
\footnotetext[2]{\label{author2}Universität Hamburg, Germany, firstname.lastname@uni-hamburg.de}
\footnotetext[3]{Universität Hamburg, Germany, firstname.lastname-1@uni-hamburg.de}
\footnotetext[4]{TU Hamburg, Germany, firstname.lastname@tuhh.de}
\footnotetext[5]{Petra Berenbrink: Supported by DFG Research Group ADYN under grant DFG 411362735}

\begin{abstract}
We analyze the convergence of the $k$-opinion Undecided State Dynamics (USD) in the \emph{population protocol} model.
For $k$=2 opinions it is well known that the USD reaches \emph{consensus} with high probability within $O(n \log n)$ interactions.
Proving that the process also quickly solves the consensus problem for $k>2$ opinions has remained open, despite analogous results for larger $k$ in the related parallel \emph{gossip} model.
In this paper we prove such convergence: under mild assumptions on $k$ and on the initial number of undecided agents we prove that the USD achieves plurality consensus within $O(k n \log n)$ interactions with high probability, regardless of the initial bias.
Moreover, if there is an initial \emph{additive} bias of at least $\Omega(\sqrt{n} \log n)$ we prove that the initial plurality opinion wins with high probability, and if there is a multiplicative bias the convergence time is further improved.
Note that this is the first result for $k > 2$ for the USD in the population protocol model.
Furthermore, it is the first result for the unsynchronized variant of the USD with $k>2$ which does not need any initial bias.
\end{abstract}
\section{Introduction}

The Undecided State Dynamics (USD) is a simple protocol designed for
distributed models of computation where $n$ indistinguishable
agents engage in pairwise interactions.
The protocol assumes that every agent initially supports one
of $k \ge 2$ \emph{opinions}, and the process evolves according
to the following transition rules: when an agent $x$ interacts
with an agent $y$ and the opinions of $x$ and $y$
differ, $x$ transitions to an \emph{undecided} state.
When $x$ interacts with $y$ and $x$ is undecided,
$x$ adopts the opinion of $y$. If $y$ is undecided, or
if its opinion is the same as that of $x$, no updates occur. 

Given its suitability as a primitive for other distributed
tasks, a substantial amount of recent work has analyzed this process
as a protocol for \emph{consensus}
under varying settings of two key problem parameters:
first, the exact distributed model of
pairwise interaction, and second, the number of opinions $k$.
The USD was originally introduced by 
Angluin et al.\ \cite{DBLP:journals/dc/AngluinAE08} for $k$=2 opinions in the \emph{population protocol} model,\footnote{%
    Independently, Perron et al.\ \cite{DBLP:conf/infocom/PerronVV09} 
    analyzed the two opinion USD in the \emph{asynchronous gossip}
    model of Boyd et al.\ \cite{DBLP:journals/tit/BoydGPS06}, which can be viewed as the 
    continuous time variant of the population protocol model. 
    For simplicity, our work focused on the latter model, although 
    our results extend easily to the former. 
}
where at every discrete time step a single pair of agents
is chosen uniformly at random to interact.
In this setting, Angluin et al.\ showed that the USD
reaches consensus (a configuration where all agents support the same opinion)
in $O(n \log n)$ interactions.\footnote{%
    Throughout, all stated results hold with high probability (w.h.p.),
    meaning with probability  $1 - n^{-c}$ for some $c > 0$. 
}
Moreover, those authors (and later Condon et al.\ \cite{DBLP:conf/dna/CondonHKM17}
via a simplified analysis) also showed that the 
process solves the \emph{approximate majority} problem, 
meaning the eventual consensus opinion is the one whose
initial support was larger, so long as
the initial \emph{bias} (the difference between the support of 
the two opinions) is sufficiently large (specifically, of order 
$\Omega(\sqrt{n \log n})$).

Separately, the USD has also been analyzed in the
parallel \emph{gossip} model of communication, where in
each synchronous round, \emph{every} agent selects an
interaction partner uniformly at random.
In this model, Clementi et al.\ \cite{DBLP:conf/mfcs/ClementiGGNPS18} 
showed convergence results for
the case of $k$=2 opinions that are analogous to those in the population
protocol model: the process reaches consensus in $O(\log n)$
synchronous rounds and additionally solves approximate majority
when the initial bias is at least $\Omega(\sqrt{n \log n})$. 
In this model, Becchetti et al.\ \cite{DBLP:conf/soda/BecchettiCNPS15} 
also analyzed the process in the higher-dimensional regime when $k > 2$. 
Assuming a large enough \emph{multiplicative} bias
in the initial supports of opinions, they show that the USD reaches \emph{plurality consensus} 
in $O(k \log n)$
parallel rounds, meaning that the eventual consensus opinion
is the one whose initial support was largest.\footnote{%
  Note that when $k > 2$, the initial support of largest
  may not be a majority,
  which is why the term \emph{plurality} is used.
}

Although the population protocol model can be viewed
as the asynchronous analog to the synchronous gossip model,
the differences in these interaction scheduling modes cause the USD
to exhibit significant qualitative differences when run in either
setting, even in the case when $k$=2. This can largely be attributed
to the observation that one round of parallel interactions in the gossip
model can lead to a constant fraction of agents changing
their opinion, whereas at most a single change of opinion can result from
each interaction in the population protocol model.
These differences, as Clementi et al.\ \cite{DBLP:conf/mfcs/ClementiGGNPS18} remark, 
have largely prevented
any general analysis techniques from transferring between the two
models. In particular, it has remained an open problem to analyze
the convergence rate of the USD in the population protocol model
when $k > 2$. 

\subsection{Our Contribution}

In this work, we close the aforementioned gap and analyze the 
USD in the population protocol model in the 
high dimensional $k > 2$ regime.%
\footnote{
Our analysis can also be applied when $k$=2 and recovers the existing 
convergence results \cite{DBLP:journals/dc/AngluinAE08, DBLP:conf/dna/CondonHKM17}
in this setting. 
}
In particular, under mild assumptions, 
we prove that the USD solves the plurality consensus problem 
in this model in $O(k \cdot n\log n)$ interactions.
Stated informally, we prove the following result:
\begin{theorem}[informal]
Consider the USD in the population protocol model, and
assume a sufficiently small number of initially undecided agents.
Then for any $ 2 \le k \le O(\sqrt{n}/\log^2 n)$
\begin{enumerate}
\item   
    If the initial support of the plurality opinion
    is at least $\Omega(\sqrt{n} \log n)$ larger (additively)
    than all other opinions, then the process reaches
    plurality consensus within $O(kn \log n)$ interactions,
\item
    If the initial support of the plurality opinion 
    is a constant multiplicative factor larger than all other opinions,
    then the process reaches plurality consensus within $O(kn + n \log n)$
    interactions,
\item
    The process reaches an arbitrary consensus configuration otherwise,
\end{enumerate}
where each statement holds with high probability. 
\end{theorem}
The exact statement of our main result is given in \cref{thm:main_theorem},
where the convergence rates have a more precise dependence on the magnitude of the opinion with largest initial support.
Roughly speaking, the convergence rate of our result is analogous to 
that of Becchetti et al.~\cite{DBLP:conf/soda/BecchettiCNPS15} for the gossip model: in that model, plurality consensus is reached within $O(k \cdot \log n)$ \emph{rounds}, 
while in the population protocol model, we show it takes $O(k \cdot n \log n)$ 
\emph{interactions}. However, unlike the result of Becchetti et al.,
our analysis only requires an \emph{additive} bias of 
$\Omega(\sqrt{n} \log n)$ to reach plurality consensus
(rather than a constant \emph{multiplicative} bias); 
it holds for larger $k = O(\sqrt{n} / \log^2 n)$ (compared to $k = O((n/\log n)^{1/3})$); 
and we show the process still reaches consensus when starting from a configuration with no initial bias 
(e.g., when the initial support of each opinion is $n/k$). 
On the other hand, when the initial configuration does contain a constant
multiplicative bias, our analysis gives a faster convergence rate than
in the additive bias regime. 
Moreover, our convergence rate under a multiplicative bias
is faster (when considering its corresponding \emph{parallel time})
than the rate given by Becchetti et al.\ when the support of the initially largest opinion is close to the average opinion support.\footnote{This is shown
explicitly in \cref{apx:becchetti-compare}.}
In this setting, our results for the population protocol model can be 
viewed as improvements to the analogous results of Becchetti et al.\ 
for the gossip model.
If there is a large multiplicative bias (larger than $\log n$) the results by Becchetti et al.\ give better bounds on the convergence time. 
Similar to previous analyses in both models \cite{DBLP:journals/dc/AngluinAE08, DBLP:conf/dna/CondonHKM17, DBLP:conf/soda/BecchettiCNPS15, DBLP:journals/dc/BecchettiCNPST17}
our analysis requires carefully defining a sequence of \emph{phases} 
throughout which the (qualitative and quantitative) behavior of the process varies.
The main challenge is to define appropriate potential functions that allow us to track the progress of the process.
In \cref{sec:main-result} we give an overview about the main ideas of our analysis.

\subsection{Related Works}

\paragraph{The Undecided State Dynamics}
The two-opinion USD was introduced independently by Angluin et al.~\cite{DBLP:journals/dc/AngluinAE08} for the population
protocol model and by Perron et al.~\cite{DBLP:conf/infocom/PerronVV09} for the closely related (continuous time) asynchronous gossip model.
Both works show that the process converges \whp in $O(n \log n)$ steps 
(respectively, $O(\log n)$ continuous time). 
Condon et al.\ \cite{DBLP:conf/dna/CondonHKM17} give an improved 
analysis for the two-opinion case in the population model
and show the process solves the approximate majority problem
assuming an initial additive bias of $\Omega(\sqrt{n \log n})$, 
which improves over the additive bias of $\omega(\sqrt{n} \log n)$
needed in the analysis of Angluin et al. Similar to our approach, 
both Angluin et al.\ and Condon et al.\ analyze the process in distinct
phases that depend on the number of undecided agents and the magnitude
of bias in the configuration. 
In particular, after introducing a suitable structure
of phases and sub-phases, the analysis of Condon et al.
reduces the convergence of the process to
analyzing a sequence of biased, one-dimensional
random walks. The boundaries imposed by the
phase structure are used to control the magnitudes
of the bias, and bounds on the number of interactions
needed to complete each phase are derived using 
standard concentration techniques.

In the parallel gossip model, the convergence of the USD for the
$k \ge 2$ opinion case was first studied by Becchetti et al.~\cite{DBLP:conf/soda/BecchettiCNPS15}. 
Central to their analysis is the introduction of the 
\emph{monochromatic distance}, which measures the uniformity 
(i.e., lack of bias) of a configuration.
Roughly speaking, this distance is the sum of squares 
of the support of each opinion, normalized by the square of the 
most popular opinion. They show convergence
within $O(\text{md}(\x) \cdot \log n)$ parallel rounds, where 
$\text{md}(\x)$ is the monochromatic distance of the initial
configuration, which is always bounded above by $k$.
This analysis only holds when the initial configuration 
has a multiplicative bias. In the two-color case, 
Clementi et al.\ \cite{DBLP:conf/mfcs/ClementiGGNPS18} later present a
tight analysis (giving convergence rates that hold for any 
initial configuration) without using the monochromatic
distance, but an analysis for $k > 2$ opinions, starting
from any initial configuration in the gossip model still remains open.

In a related strain of research, multiple works \cite{DBLP:conf/podc/GhaffariP16a,DBLP:conf/icalp/BerenbrinkFGK16,DBLP:conf/podc/BankhamerEKK20,DBLP:conf/soda/BankhamerBBEHKK22} have analyzed a \emph{synchronized} variant of the USD where the system alternates between two different phases in a synchronized fashion.
In the first phase, all agents perform one step of the USD.
In the second phase, all undecided agents adopt an opinion again.
The use of so-called \emph{phase clocks} that synchronize the agents allows for a polylogarithmic convergence time regardless of the initial opinion configuration.
This improved convergence time comes at the price of making the protocol ``less natural'': these 
protocols have a significant state overhead and are typically not \emph{uniform}, meaning that the transition function or state space depend on $n$.

\paragraph{Other Consensus Dynamics}

In the population protocol model, consensus for the case of $k$=2 opinions is commonly known as the \emph{majority problem}.
A large number of works 
\cite{DBLP:conf/podc/AlistarhGV15, DBLP:conf/soda/AlistarhAG18, DBLP:conf/wdag/BerenbrinkEFKKR18, DBLP:conf/podc/NunKKP20, DBLP:journals/dc/BerenbrinkEFKKR21,DBLP:conf/focs/DotyEGSUS21}
aim to identify the majority opinion even if the initial winning margin is as small as only $1$.
The best known result \cite{DBLP:conf/focs/DotyEGSUS21} solves this \emph{exact} majority problem in $O(n \log n)$ interactions using $O(\log n)$ states, both in expectation.
For more details on algorithmic advances in Population Protocols we refer the reader to the surveys by Els{\"{a}}sser and Radzik \cite{DBLP:journals/eatcs/ElsasserR18} and Alistarh and Gelashvili \cite{DBLP:journals/sigact/AlistarhG18}.

Less is known about exact plurality consensus protocols for $k > 2$ opinions.
One line of research focuses on the state space requirements to \emph{always} compute the exact plurality opinion.
In \cite{DBLP:conf/ciac/NataleR19} the authors show that always correct plurality consensus requires $\Omega(k^2)$ states.
The currently best known protocol requires $O(k^6)$ many states \cite{DBLP:conf/opodis/GasieniecHMSS16}.
In \cite{DBLP:conf/podc/BankhamerBBEHKK22} the authors relax the requirement to \emph{always} return the correct result. 
They present a protocol for $k>2$ opinions that may fail with small probability.
This negligible error probability allows them to break the lower bound and design a protocol that converges \whp in $O(n \cdot ({k \log n + \log^2 n}))$ interactions using $O(k + \log n)$ states.

\def\protocol{}

A related family of protocols are the $j$-Majority processes.
The idea is that every agent adopts the majority opinion among a random sample of $j$ other agents (breaking ties randomly).
The most simple variant (for $j$=1) is also known as the so-called \protocol{Voter} process \cite{DBLP:journals/iandc/HassinP01, DBLP:journals/networks/NakataIY00, DBLP:conf/podc/CooperEOR12, DBLP:conf/icalp/BerenbrinkGKM16, DBLP:conf/soda/KanadeMS19}.
Here, every agent adopts the opinion of a single, randomly chosen agent.
The protocols for $j$=2 and $j$=3 have been analyzed under the names of \protocol{TwoChoices} process \cite{DBLP:conf/icalp/CooperER14,DBLP:conf/wdag/CooperERRS15,DBLP:conf/wdag/CooperRRS17} and the \protocol{3-Majority} dynamics \cite{DBLP:journals/dc/BecchettiCNPST17, DBLP:conf/podc/GhaffariL18, DBLP:conf/podc/BerenbrinkCEKMN17}.
In the \protocol{TwoChoices} process, lazy tie-breaking towards an agent's original opinion is assumed.
Ghaffari and Lengler \cite{DBLP:conf/podc/GhaffariL18} show for the \protocol{TwoChoices} process with $k = O(\sqrt{n / \log n})$ and for \protocol{3-Majority} with $k = O({n^{1/3} / {\log n}})$ that consensus is reached in $O({k\cdot \log n})$ rounds \whp.
For arbitrary~$k$ they show that \protocol{3-Majority} reaches consensus in $O(n^{2/3} \log^{3/2} n)$ rounds \whp.
Schoenebeck and Yu \cite{DBLP:conf/soda/SchoenebeckY18} analyze the convergence time of a generalization of multi-sample consensus protocols for two opinions on complete graphs and Erdős-Rényi graphs.
In the \protocol{MedianRule} process \cite{DBLP:conf/spaa/DoerrGMSS11} the authors assume that opinions are ordered.
In every step every agent then adopts the median of its own opinion and two randomly sampled opinions.
This protocol reaches consensus in $O({\log k \log\log n + \log n})$ rounds \whp.
We remark that in contrast to the MedianRule the USD does not require a total order among the opinions.
For further references and additional protocols in similar models we refer the reader to the survey of consensus dynamics by Becchetti et al.\ \cite{DBLP:journals/sigact/BecchettiCN20}.

\section{Background and Overview of Results}

In this section, we first introduce some of the preliminaries 
and notation related to the population protocols model 
and the USD. We then provide a technical
overview of our main result.

\paragraph{Population Protocols}

We consider a population protocol for $n$ identical, anonymous agents,
where each agent is modeled as a finite state machine with state space $Q$. 
Agents interact in pairs drawn uniformly at random.
In an interaction $(u,v)$ agent $u$ is called responder and agent $v$ is called initiator.
We allow for agents to interact with themselves.
The population protocol is defined by its transition function $\delta: Q^2 \rightarrow Q^2$. 

The undecided state dynamics (USD) is defined as follows. 
Each agent has either one of $k$ opinions or it is undecided, i.e., $Q=\set{1,\ldots k,\bot}$ where $\bot$ stands for undecided. The \emph{undecided state} population protocol is given by the transition function
\begin{align*}
    (q,q') & \rightarrow 
    \begin{cases}
    (\bot,q') \text{ if } q,q' \neq \bot \land q \neq q'\\
    (q',q') \text{ if } q = \bot, q' \neq \bot\\
    (q,q') \text{ otherwise.}
    \end{cases}
\end{align*}
Observe that only the responder $q$ changes its state.

A configuration $\x(t)$ at time $t$ is a vector $(x_1(t), x_2(t),\ldots x_k(t), u(t))$ of length $k+1$.
For $1\le i\le k$, $x_i(t)$ is the number of agents of Opinion $i$ and $u(t) = n-\sum_{i=1}^{k} x_i(t)$ is the number of undecided agents.
In the beginning we assume $x_1(0)\geq x_2(0)\geq \dots \geq x_k(0)$. For $t>0$ we define $\max(t)$ as the index of the opinion with the largest support at step $t$ (if there are several opinions with the same maximum support we pick an arbitrary one). 
Furthermore we introduce the notation $x_{\max}(t) = x_{\max(t)}(t) =\max_{i \in [k]} \set{x_i(t)}$ for the support of the largest opinion at time $t$. 
Note that $x_{\max}(t)$ can refer to the support of different opinions over time. 

We call an Opinion $i$ \emph{significant} if $x_i(t) > x_{\max}(t) - \alpha \cdot \sqrt{n} \log n$ for some fixed constant $\alpha$.
An opinion that is not significant is called \emph{insignificant}.
A configuration $\mathbf{x}$ has an \emph{additive bias} $\beta$ if there exists an Opinion $m$ such that for all other opinions $i\neq m$ we have $x_m \geq x_i + \beta$.
We say that a configuration $\mathbf{x}$ has a \emph{multiplicative bias} $\alpha$ if there exists an Opinion $m$ such that for all other opinions $i\neq m$ we have $x_m \geq \alpha \cdot x_i$.
In the following we use upper case letters for random variables (for example $\mathbf{X}(t)$ and $U(t)$) and lower case letters ($\mathbf{x}(t)$ and $u(t)$) for fixed configurations or values.
\subsection{Main Result} \label{sec:main-result}
We now state our main theorem.
We remark that in our analysis we bound the convergence time in terms of $n/x_1(0)$, where $x_{\max}(0) = x_1(0)$ is the support of the initially largest opinion.
Under the assumptions of our theorem, however, we have $x_1(0) > n/(2k)$, which leads to the bounds in terms of $k$.

\smallskip

\noindent\colorbox{black!10}{\begin{minipage}{\textwidth-2\fboxsep}%
\begin{theorem}
    \label{thm:main_theorem}
    Let $c>0$ be an arbitrary constant and let $\mathbf{x}(0)$ be an initial configuration with 
    $k \leq c \cdot \sqrt{n}/\log^2(n)$ opinions with $u(0) \leq (n-x_1(0))/2$ and  $x_1(0) \geq x_i(0)$ for all $i \in [k]$.  
    Then \whp all agents agree on Opinion 1 within
    \begin{enumerate}
        \item $O\left(n \log n + n^2/x_1(0) \right) = O\left(n \log n + n \cdot k \right)$ interactions if $\mathbf{x}(0)$ has a multiplicative bias of at 
        least $1 + \epsilon$ for an arbitrary constant $\epsilon$.
        \item $O(n^2 \log n/x_1(0)) = O(k \cdot n \log n)$ interactions if $\mathbf{x}(0)$ has an additive bias of at least $\Omega\left( \sqrt{n} \log n \right)$.
    \end{enumerate}
    Without any bias all agents agree on a significant opinion within $O\left(n^2 \log n /x_1(0) \right) = O(k \cdot n \log n)$ interactions \whp.
\end{theorem}%
\end{minipage}%
}

\paragraph{Main Idea of the Analysis}
The straightforward approach in the analysis of consensus processes is to track the growth of the support of the plurality Opinion 1 via change of the ratio ${x_1(t)}/{x_i(t)}$ over time~$t$.
Unfortunately, the change of the support of a single opinion depends on the entire configuration, that is, the support of all other opinions and also the number of undecided agents.
Let us fix two opinions $i$ and $j$ with $x_i>x_j$.
Then it is possible for the support of Opinion $j$ to grow faster than the support of Opinion $i$ and vice versa, depending on the number of undecided nodes.
Hence, to track the progress of the plurality opinion one has to take a close look at the number of undecided nodes.
This, in turn, is heavily influenced by the support of all opinions.
To cope with this ``nonlinearity'' we use the potential function $Z_{\alpha}(t) = 
 n-  2u(t)  - \alpha\cdot x_{\max}(t)$, where we use different values of $\alpha$ for different phases.
We analyze the drift of $Z_{\alpha}(t)$ which allows us to show that the number of undecided agents quickly approaches an ``unstable equilibrium'' $u^*$.
Whenever the process is close to the equilibrium (which changes over time), we can perform a ``classical'' analysis and show, e.g., that bias between two agents doubles in a certain number of interactions.

Our analysis also handles the case when there is no bias at all.
For this we proceed in two steps.
First we show that the support difference between two arbitrary but fixed large opinions quickly reaches a value of $\sqrt{n}$ via an anti-concentration bound. From there we bound the probability that the opinions continue to drift apart. In our analysis  we rely heavily on existing concentration bounds for the
hitting times of one-dimensional random walks with drift, which we can use
after establishing the appropriate reductions and potential functions in each
phase of the process.
The analysis is divided into five parts that correspond to different \emph{phases} of the process.
The phases are listed in the following table: 

{\small
\begin{center}
\begin{tabular}{lllll}
\toprule
Phase & Section & End Condition & Running Time & Main Lemma\\
\midrule
1 & \cref{sec:phase1} & $u \geq (n-x_{\max})/2$ & $O(n\log n)$ & \cref{lem:phase1}\\
2 & \cref{sec:phase2}  & $\forall i: x_{\max} \geq x_{i} + \Omega(\sqrt{n}\log n)$ & $O(n^2\log n/x_{\max})$ & \cref{lem:phase2} \\
3 & \cref{sec:phase3}  & $\forall i: x_{\max} \geq 2 x_{i} $ & $O(n^2\log n/x_{\max})$ & \cref{lem:phase3} \\
4 & \cref{sec:phase4}  & $x_{\max} \geq 2n/3$ & $O(n^2/x_{\max} + n\log n)$ & \cref{lem:phase4}\\
5 & \cref{sec:phase5}  & $x_{\max} = n $ & $O(n\log n)$ & \cref{lem:phase5} \\ 
\bottomrule
\end{tabular}
\end{center}
}

Note that the process does not have to pass through all five phases. 
For example, the second phase is not needed if there is a large bias in the initial configuration. 
Our analysis shows that the identity of the majority opinion does not change after the end of the second phase (or not at all if a large enough additive bias is present from the beginning).

\section{Rise of the Undecided (Phase 1)}

\label{sec:phase1}
In this section we analyze the running time of Phase 1 which ends as soon as we have a sufficient number of undecided agents (\cref{lem:phase1}).  
Additionally we show that $x_1(0)$ decreases by at most a constant fraction \whp (\cref{lem:phase1-no-loss-of-bias}).
Furthermore, an additive and multiplicative bias is preserved as long as $\mathbf{x}(0)$ is an initial configuration with bias.
At the end of this section we show an upper bound on the number of undecided agents which holds during the whole running time of the process (\cref{lem:undecided_general_bounds_v1}).
This lemma will be used to estimate the running time of the remaining phases. 

In the analysis of \cref{lem:phase1} we use the 
potential function
\begin{equation*}
Z(t) = n - 2 u(t) - x_{\max}(t) \;.
\end{equation*}
Observe that 
Phase 1 ends as soon as $Z(t) \leq 0$, since in this case
$u(t') \ge n/2 - x_{\max}(t') / 2$.

\begin{lemma} 
\label{lem:phase1}
    Let $T_1 = \inf\{t \geq 0 ~|~ u(t) \geq n/2-x_{\max}(t)/2\}$.
    Then $\Prob{T_1 \leq \lceil 7 n\ln n \rceil} \geq 1-n^{-3}$.
\end{lemma}

\begin{proof}
To show the lemma we calculate the expected change in $Z(t)$ for $Z(t) \geq 0$ and apply a drift theorem from \cite{DBLP:series/ncs/Lengler20}.
There are three cases.
First we consider the case $U(t+1) = u(t)-1$.
In this case a decided agent interacts with an undecided agent, and the latter adopts the opinion of the decided agent.
Let $M(t) = \set{i \in [k] | x_i(t) = x_{\max}(t)}$ be the set of all opinions with maximum support at time $t$.
For each Opinion $i$, an undecided initiator interacts with a responder of Opinion $i$ with probability $x_i(t)\cdot u/n^2$.
If $i \in M(t)$, then $Z(t)$ increases by $1$.
Otherwise $Z(t)$ increases by $2$.

Next we consider the case $U(t+1) = u(t)+1$.
In this case a decided initiator interacts with a responder of a different opinion and becomes undecided.
For each Opinion $i$, this happens with probability $x_i(t) \cdot (n-u(t)-x_i(t))/n^2$.
If $i \in M(t)$, then $Z(t)$ decreases by $1$.
Otherwise $Z(t)$ decreases by $2$.

With the remaining probability a step is unproductive and $Z(t)$ does not change.
Using these cases, we bound the expected drift of $Z(t)$ as
\begin{align*}
\MoveEqLeft \Ex{ Z(t)-Z(t+1) | \mathbf{X}(t)=\mathbf{x} }\\
& = - \sum_{i \in M(t)} \frac{x_i \cdot u}{n^2} - 2\sum_{i \notin M(t)} \frac{x_i \cdot u}{n^2} + \sum_{i\in M(t)} \frac{x_i(n-u-x_i)}{n^2} + 2\sum_{i \notin M(t)} \frac{x_i(n-u-x_i)}{n^2}  \\
& \geq \sum_{i \in [k]} \frac{x_i(n-2u-x_{\max})}{n^2} + \sum_{i \notin M(t)} \frac{x_i(n-2u-x_{\max})}{n^2}  \\
&\geq \frac{(n-u)(n-2u-x_{\max})}{n^2}
\geq \frac{Z(t)}{2n},
\end{align*}
where we used that $x_i \leq x_{\max}$, $Z(t) = n-2u-x_{\max} \geq 0$, and $u < n/2$ by definition of Phase 1.
We now apply \cref{thm:mult_drift_tail_lengler_18} with $r = 3 \ln n$, $s_0 = n-2u(0)-x_{\max}(0) \leq n$, $s_{min} = 1$, $\delta = 1/(2n)$ and get
\begin{align*}
\Prob{T_1 > \lceil 7 n\ln n \rceil} 
& \leq \Prob{T_1 > \left \lceil{ \frac{6 \cdot \ln n + \ln(n-2u(0)-x_{\max}(0))}{1/(2n)} }\right \rceil }
\leq e^{-3 \cdot \ln(n)}
= n^{-3} \; . \qedhere
\end{align*}
\end{proof}

Given the bound on $T_1$, we proceed to show that both the support of the most popular opinion and the bias of the initial configuration do not decrease too much until time $T_1$. Recall that initially Opinion $1$ has the largest support.

\begin{lemma}[name=,restate=lemmaPhaseOneNoLossOfBias,label=lem:phase1-no-loss-of-bias] 
    \label{lem:phase1-keeping-bias}
   Let $\alpha, \varepsilon > 0$ be arbitrary constants. Then each of the following statements holds with probability at least $1-4n^{-3}$:
    \begin{enumerate}    
        \item If $x_1(0) - x_i(0) \geq \alpha \cdot \sqrt{n} \log n$, then  $X_1(T_1) - X_i(T_1) \geq \alpha/3 \cdot \sqrt{n} \log n$.
        \item If $x_1(0) \geq (1+\varepsilon) \cdot x_i(0)$, then $X_1(T_1) \geq (1+\varepsilon/(6+5\varepsilon)) \cdot X_i(T_1)$.
        \item For the largest opinion we have $X_1(T_1) \geq x_1(0)/3$.
      \end{enumerate}
\end{lemma}

\begin{proof}[Proof Sketch]
For the first statement we show that $\Ex{(X_1(t)-X_i(t))/(n-U(t))} \geq 0$ and apply a Hoeffding bound.
For the second statement we show that 
\[\Prob{X_1(t+1) = x_1 + 1 ~|~ \X(t) = \x} \leq \Prob{X_1(t+1) = x_1 - 1 ~|~ \X(t) = \x}\]
such that we can bound the development of $X_1$ by a fair random walk.
This enables us to relate the multiplicative bias to the additive bias.
The third statement is derived from the first statement by choosing an Opinion $i$ with $x_i(0) = 0$.
The full proof can be found in \cref{apx:omitted-proofs-phase1}.
\end{proof}

Next we prove the upper bound on the number of undecided agents.
The lemma shows that the number of undecided agents stays close to a  threshold value $u^* = n \cdot (k-1)/(2k-1)\approx n/2$. 
Intuitively, this threshold $u^*$ can be regarded as an (unstable) equilibrium for the number of undecided agents: in configurations with more than $u^*$ undecided agents it is more likely that an undecided agent becomes decided than vice versa, whereas in configurations with less than $u^*$ undecided agents it is more likely that a decided agent becomes undecided than vice versa.

\begin{lemma}[name=,restate=lemmaPhaseOneUndecidedGeneralBounds] 
    \label{lem:undecided_general_bounds_v1}
    Assume $u(0) \leq (n-x_{\max}(0))/2$.
    Then
    \begin{align*}
        \Prob{\text{for all } t \in [n^3] \colon u(t) \leq \frac{n}{2} - \frac{1}{5c} \cdot \sqrt{n} \log(n)} > 1-n^{-3}.
    \end{align*}
\end{lemma}
\begin{proof}[Proof Sketch]
We first prove the claim for $u(t) > u^* + 3 \cdot \sqrt{n \log n}$.
At the end of the full proof we show how the lemma statement follows out of this.
We model the number of undecided agents over time $t$ as a non-lazy random walk $Z(t)$ with state space $\set{0, \dots, n-1}$.
Then we couple $Z(r)$ with a random walk $W(r)$ on the integers with a reflecting barrier at $0$ and otherwise fixed transition probabilities.
For $W(r)$ we can derive a bound on the probability
$\Prob{\exists t \in [n^3] : W(t)\geq 3\cdot \sqrt{n\log n}}$.
The bound follows since in this case $Z(r) \leq W(r) + Z(0)$.
To conclude the proof we show that the lemma statement follows from our bound stated in terms of $u^*$.
The full proof can be found in \cref{apx:omitted-proofs-phase1}.
\end{proof}

\section{Generation  of an Additive Bias (Phase 2)}
\label{sec:phase2}
Recall that $T_1$ is defined as the end of Phase 1.
In this section we consider configurations at time $T_1$ without any additive bias.
These configurations will have several significant opinions.
We define $T_2$ as the first time $t \geq T_1$ where $\mathbf{x}(t)$ has only one opinion left which is significant. 

Note that $x_{\max}(t))\ge x_{\max}(0)/2 = \Omega(\sqrt{n}\cdot \log^2(n))$ for each interaction $t$ in this phase.
This follows from \cref{lem:undecided_general_bounds_v1} together with the pigeonhole principle.
In \cref{lem:phase2} we show that \whp the running time of this phase is $O(n^2 \cdot \log n / x_{\max}(T_1))$.
To show that result we first need a lower bound (as opposed to the upper bound of \cref{lem:undecided_general_bounds_v1}) on the number of undecided agents.
Again, this bound holds until the end of the process.

\begin{lemma}[name=,restate=lemmaPhaseTwoUndecidedLowerBound] 
\label{lem:undecided>n/2-x1/2-o(n)}
  $ \Prob{\text{for all } t \in [T_1,n^3]\colon u(t) \geq n/2 - x_{\max}(t)/2 - 8\sqrt{n \cdot \ln n}} \geq 1- n^{-5} \;.$
\end{lemma}   

\begin{proof}[Proof Sketch]
Recall that for the proof of \cref{lem:phase1} we defined $Z(t) = n-2u(t)-x_{\max}(t)$.
We then showed that we have a drift towards zero.
We use this for a drift analysis following Theorem 6 in \cite{lengler2018drift}.
\fullproofinappendix{\cref{apx:omitted-proofs-phase2}}
\end{proof}

In the following lemma we show that the support of the largest opinion does not shrink by more than a factor of two during Phase 2. 

\begin{lemma}[name=,restate=lemmaPhaseTwoMaxNoShrinking] 
\label{lem:phase2-max-no-shrinking}
    Let $c> 0$ be an arbitrary constant and define $T = c\cdot n^2 \cdot \log n / x_{\max}(T_1)$.
    Then
    \begin{align*}
        \Prob{\text{for all }t \in [T_1,T_1+T] \colon x_{\max}(t) \geq x_{\max}(T_1)/2} \geq 1-n^{-5} .
    \end{align*}
\end{lemma}

In \cref{lem:small-opinions-do-not-grow} we first show that ``small opinions'' remain small (they only double their support).
With small opinion we mean opinions having a support which have support at most $20\sqrt{n\log n}$ and are thus at least a polylogarithmic factor smaller compared to $x_{\max}(t)$.
Then in the second part we show that insignificant opinions remain insignificant.
Recall that an Opinion $i$ is insignificant if $x_{\max}(t)-x_i(t) = \Omega(\sqrt{n}\log n)$.
\begin{lemma}[name=,restate=LemmaSmallOpinionsDoNotGrow] 
\label{lem:small-opinions-do-not-grow}
    Let $c,c'> 0 $ be  arbitrary constants and define $T= c\cdot n^2 \cdot \log n / x_{\max}(T_1)$.  Assume for Opinion $j$ there exists a time $t_0 \in [T_1,T_1+T]$ with 

    \begin{enumerate}
    \item $x_j(t_0) \leq 20\sqrt{n\log n}$. 
    Then
    \begin{align*}
        \Prob{\text{for all }t \in [t_0,T_1+T] \colon x_j(t) \leq 40\sqrt{n\log n}} \geq 1-2n^{-3} .
    \end{align*}
    \item $x_{\max}(t_0)-x_j(t_0) \geq c'\cdot \sqrt{n}\log n$.
    Then 
    \begin{align*}
        \Prob{\text{for all }t \in [t_0,T_1 +T] \colon x_{\max}(t)-x_i(t) \geq c'/2 \cdot\sqrt{n}\log n} \leq 1-2n^{-3} .
    \end{align*}
    \end{enumerate}
\end{lemma}

\begin{proof}[Proof Sketch]
    In the first part we bound the probability for a small Opinion $j$ to grow using \cref{lem:undecided>n/2-x1/2-o(n)}.
    This probability is sufficiently small for Opinion $j$ not to double.
    In the second part, we make a case distinction based on the size of $x_j(t_0)$.
    If $x_j(t_0)$ is small, then the support of $x_j$ does not double (see Part 1) while $x_1$ keeps at least half of its support (\cref{lem:phase2-max-no-shrinking}).
    Otherwise, we use \cref{lem:undecided>n/2-x1/2-o(n)} to show that the bias is likely to increase.
    Then the second part follows from the gambler's ruin problem.
    \fullproofinappendix{\cref{apx:omitted-proofs-phase2}}
\end{proof}

The following lemma constitutes the foundation of the application of the drift result from \cite{DBLP:conf/spaa/DoerrGMSS11} which will be used in the proof of \cref{lem:phase2}.
In the first part of \cref{lem:significant_opionion_create_small_bias} we consider two important opinions with (almost) the same support. 
We use an anti-concentration result 
to show that their support difference quickly reaches $\Omega{(\sqrt{n})}$.
In the second part we again consider two important opinions and give precise bounds on the probability that their support difference increases by a constant factor. Our proof is based on gambler's ruin problem.
The proof of this result can be found in \cref{apx:omitted-proofs-phase2}.

\begin{lemma}[name=,restate=lemmaPhaseTwoDoublingImportant] 
\label{lem:significant_opionion_create_small_bias}
    Fix two opinions $i$ and $j$ and assume there exists $t_0 \geq T_1$ with $x_i(t_0) \geq x_j(t_0) \geq x_{\max}(t_0)-4\alpha\sqrt{n}\log n$ .
    Let   $T = 40\cdot n^2/x_{\max}(T_1)$.
    Then 
    \begin{enumerate}
    \item If $x_i(t_0)-x_j(t_0) < 4\alpha \cdot \sqrt{n}$ then
    \begin{align*}
        \Prob{X_i(t_0+T)-X_j(t_0+T) \geq 4\alpha \cdot \sqrt{n}} \geq e^{-\frac{\alpha^2}{16}} .
    \end{align*}
    \item If $x_i(t_0)-x_j(t_0)  \geq 4\alpha\cdot \sqrt{n}$ then 
    {\small\dense
    \begin{align*}
    \MoveEqLeft\Prob{X_i(t_0+T)-X_j(t_0+T) \geq \min\left\lbrace \frac{3(x_i(t_0) -x_j(t_0))}{2}, 4\alpha \sqrt{n}\log n\right\rbrace} \geq 1-e^{-\frac{x_i(t_0)-x_j(t_0)}{\sqrt{n}}}
    \end{align*}
    }
    \end{enumerate}
\end{lemma}

Now we are ready to analyze the running time of Phase 2.

\begin{lemma} 
\label{lem:phase2}
Let \, $T_2 = \inf\set{ t \geq T_1  | \exists i \in [k] : \forall j\neq i: x_{i}(t)-x_j(t) \geq \alpha\sqrt{n}\log n}$. 
Then
\begin{align*}
    \Prob{T_2-T_1 \leq 40\cdot c\cdot n^2 \cdot \log n/x_{\max}(T_1)} \geq 1-2n^{-2} .
\end{align*}
\end{lemma}
\begin{proof}
We define 
\begin{align*}
    \hat{T} = \inf\set{t\geq T_1 | u(t) \notin \left[\frac{n-x_{\max}(t)}{2}-8\cdot \sqrt{n\ln n}, \frac {n}{2}\right] \mbox{ or } x_{\max}(t)<x_{\max}(T_1)/3}
\end{align*}
as a stopping time and $(\hat{\mathbf{X}})_{t\ge T_1}$ as the process with $\hat{\mathbf{X}}(t) = \mathbf{X}(t)$ for all $t\leq \hat{T}$ and $\hat{\mathbf{X}}(t) = \mathbf{X}(\hat{T})$ for $t>\hat{T}$.
From  \cref{lem:undecided>n/2-x1/2-o(n)} it follows that $u(t) \geq (n-x_{\max}(t))/2-8\cdot \sqrt{n\ln n} $ for all $t \in [T_1,n^3]$, \whp.
From \cref{lem:undecided_general_bounds_v1} it follows that $ u(t) \leq n/2$  for all $t \in [T_1,n^3]$, \whp.
Finally, \cref{lem:phase2-max-no-shrinking} gives us that $ x_{\max}(t) \geq x_{\max}(T_1)/3$ for all $t \in [T_1,T_1+ cn^2 \log n / x_{\max}(T_1)]$, \whp.
Thus, $\hat{T}-T_1 =\Omega(n^2\cdot \log n/x_{\max}(T_1) )$ \whp and we can assume that $(\mathbf{X})_{t\ge T_1}$ and $(\hat{\mathbf{X}})_{t\ge T_1}$ are identical for $t \in [T_1,T_1 + O(n^2\cdot \log n/x_{\max}(T_1))]$.

Recall that an Opinion $i$ is \emph{significant} at time $t$ if $x_i(t) > x_{\max}(t) - \alpha\sqrt{n}\log n$. 
In the following we call an Opinion $i$ \emph{important}  at time t if $x_i(t) > x_{\max}(t) - 4\cdot \alpha\sqrt{n}\log n$.
In the following we will show that for each pair of important opinions $i$ and $j$ at time $T_1$ at least one of them becomes unimportant.
Furthermore, we show that no unimportant opinion ever becomes significant.
From this follows that after $O(n^2/\hat{x}_{max}(T_1) \cdot \log n)$ only one significant opinion remains.

First we consider a pair of opinions $i$ and $j$ which are important at time $T_1$ and show that \whp at least one of them becomes unimportant within the next $\tau=40\cdot c n^2 \cdot \log n/\hat{x}_{max}(T_1) $ interactions.

We divide the interactions from $[T_1, T_1+\tau] $ into $c_1\log n$ subphases of length  $ 40\cdot n^2/\hat{x}(T_1)$ each.
For $1\le i \le c\log n$ we define $\ell_1= 1$ and $\ell_i=1+(i-1)\cdot n^2/\hat{x}(T_1)$.
Then the $i$th subphase contains interactions $\ell_i$ to $(\ell_{i+1}-1)$. 
Furthermore, we define $t_i$ is the first  interaction in subphase $i$.

Now we fix an arbitrary subphase $i$ and we consider two cases. 
If $\hat{x}_i(t_i)-\hat{x}_j(t_i) < 4\alpha\sqrt{n}$ then
 it follows from \cref{lem:significant_opionion_create_small_bias} 
\begin{align}
    \MoveEqLeft\Prob{\hat{X}_i(t_{i+1})-\hat{X}_j(t_{i+1}) \geq 4\alpha\sqrt{n}} \geq e^{-\frac{\alpha^2}{16}}  \label{eq:nobias-create-small-bias_cond_1}
\end{align}
 Otherwise, if  $\hat{x}_i(t_i)-\hat{x}_j(t_i) \geq 4\alpha\sqrt{n}$ then
{\small 
\begin{align}
    \MoveEqLeft\Prob{\hat{X}_i(t_{i+1})-\hat{X}_j(t_{i+1}) \geq \min\set{(3/2) \cdot (\hat{x}_i(t_i)-\hat{x}_j(t_i)),4\alpha\sqrt{n}\log n}} \geq 1-e^{-\frac{\hat{x}_i(t_i)-\hat{x}_j(t_i)}{\sqrt{n}}} \label{eq:nobias-create-small-bias_cond_2}
\end{align}
}
In either case we call such subphase successful.

In the following we  show that in the interval $[T_1, T_1+\tau]$ there is a sufficient amount of consecutive successful subphases such that at least one of the two opinions becomes unimportant. 
To do so, we define a function  $f: [1,c_1\log n] \rightarrow [0,\log \log n]$ which counts the consecutive number of successful subphases.
\begin{align*}
    f(i) = 
    \begin{cases}
        0\qquad \mbox{ if } \ \abs{\hat{x}_i(t_i)-\hat{x}_j(t_i)} < 4\alpha\sqrt{n} \\
        j\qquad\mbox{ if } \ (3/2)^{j-1} \cdot 4\alpha\sqrt{n} \leq \abs{\hat{x}_i(t_i)-\hat{x}_j(t_i)} < (3/2)^j \cdot 4\alpha\sqrt{n}
    \end{cases}
\end{align*}
Note that either Opinion $i$ or Opinion $j$ is unimportant at the beginning of subphase $i$  if $f(t_i)=\log\log n$.

We define a random walk $W$ over the state space  $[0,\log\log n]$ as follows. $W$ has a reflective state $0$ and an absorbing state $\log\log n$. 
Initially, $W(1) = 0$.
For  $w \in [0,\log\log n -1]$ the transition probabilities are defined  as follows
\begin{align*}
        &\Prob{W(t+1) = 1 ~|~ W(t)=0} = e^{-\frac{\alpha^2}{16}} \\
        &\Prob{W(t+1) = w +1 ~|~ W(t) = w} = 1-e^{-2^w} \\
        &\Prob{W(t+1) = 0 ~|~ W(t) = w  } = e^{-2^w} . 
\end{align*}
To show that  either Opinion $i$ or Opinion $j$ becomes unimportant, which is equivalent to our function $f$ taking on the value $\log\log n$,  we define  coupling between $f(i)$ and $W(i)$ such that $f(i) \geq W(i)$ for all $i \in [1,c_1 \log n]$.

For $i = 1$ the claim holds trivially since we have $W(1) = 0$ and $f(1)\geq 0$.
Now assume for $i \geq 1 $ that $f(i) \geq W(i)$.
Now we consider two cases. 
In the first case  assume $\abs{\hat{x}_i(t_i)-\hat{x}_j(t_i)} < 4\alpha\sqrt{n}$.
Then we know  $f(i) = 0$ and hence, $W(i)=0$.
It follows from \cref{eq:nobias-create-small-bias_cond_1} and $\abs{\hat{x}_i(t_i)-\hat{x}_j(t_i)} \geq 0$
\begin{align*}
    \Prob{f(i+1) \geq f(i) + 1  ~|~ f(i) = 0 } \geq  e^{-\frac{\alpha^2}{16}} \text{ and }\\
    \Prob{f(i+1) \geq 0  ~|~ f(i) = 0 } < 1- e^{-\frac{\alpha^2}{16}} 
\end{align*}
Likewise, from the definition of $W$ it follows
\begin{align*}
    \Prob{W(i+1) = W(i) + 1 ~|~ W(i) = 0} = e^{-\frac{\alpha^2}{16}} \text{ and } \\
    \Prob{W(i+1) = 0 ~|~ W(i) = 0} = 1- e^{-\frac{\alpha^2}{16}}
\end{align*}
Hence, we can couple the to processes such that the following holds:  whenever $W(i)$ is increased by one then $f(i)$ is increased, too. 
Whenever $f(i)$ is decreased $W(i)$ jumps back to zero. 

In the second case we assume 
\begin{equation*}
    4\alpha\sqrt{n} \leq \abs{\hat{x}_i(t_i)-\hat{x}_j(t_i)} < \min\set{2(\hat{x}_i(t_i)-\hat{x}_j(t_i)),4\alpha\sqrt{n}\log n} \;.
\end{equation*} 
Then it follows from \cref{eq:nobias-create-small-bias_cond_2} and $\abs{\hat{x}_i(t_i)-\hat{x}_j(t_i)} \geq 0$
\begin{align*}
    \Prob{f(i+1) \geq f(i) + 1  ~|~ f(i) = 0 } \geq  1-e^{-(\hat{x}_i(t_i)-\hat{x}_j(t_i))/\sqrt{n}} \text{ and }\\
    \Prob{f(i+1) \geq 0  ~|~ f(i) = 0 } < e^{-(\hat{x}_i(t_i)-\hat{x}_j(t_i))/\sqrt{n}}
\end{align*}
Likewise, from the definition of $W$ it follows
\begin{align*}
    \Prob{W(i+1) = W(i) + 1 ~|~ W(i) = m} =  1-e^{-2^m}\text{ and } \\
    \Prob{W(i+1) = 0 ~|~ W(\ell) = m} =  e^{-2^m}
\end{align*}
Observe that 
\begin{align*}
    1-e^{-(\hat{x}_i(t_i)-\hat{x}_j(t_i))/\sqrt{n}}
    \geq 1-e^{-2^{f(i)}}
    \geq 1-e^{-2^m} .
\end{align*}
Again, we can couple the to processes such that $f(i) \geq W(i)$.

Finally an application of \cref{lem:drift_nobias} that  \whp  there exists $i \in [1,c_1 \log n]$ such that $W(i) = \log\log n$. From this follows that there exists a  time $t' \leq [T_1,T_1+\tau]$ such that $\hat{x}_i(t')-\hat{x}_j(t') \geq 4\alpha\sqrt{n}\log n$.
This implies, in turn,  that at least  Opinion  $j$ is unimportant.
From Statement 2 in \cref{lem:small-opinions-do-not-grow} it follows that $\hat{x}_{max}(t)-\hat{x}_j(t) \geq 2\alpha\sqrt{n}\log n$ for all $t\in [t',T_1+\tau]$ \whp.
Hence, the Opinion $j$ does not become significant during the time interval.
Finally a union bound over all pairs of initial important opinions at time $T_1$ yields  that all but a single opinion of those important opinions becomes insignificant in the time interval \whp.

\medskip

Now we show that none of the  unimportant opinions at time $T_1$ ever becomes significant during $[T_1,T_1+\tau]$. 
First we fix an Opinion $j$ which is unimportant at time $T_1$.
Again from Statement 2 in \cref{lem:small-opinions-do-not-grow} it follows that $\hat{x}_{max}(t)-\hat{x}_j(t) \geq 2\alpha\sqrt{n}\log n$ for all $t\in [T_1,T_1+\tau]$ \whp. Hence, all unimportant opinions at time $T_1$ does not become significant during the time interval by a union bound.
At last the statement follows because all but a single opinion becomes insignificant and hence, $T_2-T_1 \leq \tau$.  
\end{proof}

\section{From Additive to Multiplicative Bias (Phase 3)}
\label{sec:phase3}
Recall that $T_2$ is defined as the end of Phase 2, and $\mathbf{x}(T_2)$ is a configuration with an additive bias of $\Omega(\sqrt{n}\log n)$.
In the following we assume w.l.o.g.\ that $x_1(T_2)\ge x_2(T_2)\ldots \ge x_k(T_2)$.

We start our analysis of Phase 3 with \cref{lem:phase3-max-no-shrinking} where we show that the support of the largest opinion does not shrink by more than a factor of two. The lemma is the equivalent to Statement 2 of \cref{lem:small-opinions-do-not-grow} from Phase 2. The proof can be found in \cref{apx:omitted-proofs-phase3}.
\begin{lemma}[name=,restate=LemmaPhaseThreeMaxNoShrinking] 
\label{lem:phase3-max-no-shrinking}
    Let $c> 0$ be an arbitrary constant and define $T = c \cdot n^2 \cdot \log n / x_1(T_2)$.
    Then
    \begin{align*}
        \Prob{ \text{for all }t \in [T_2,T_2+T] \colon x_1(t) \geq x_1(T_1)/2} \geq 1-n^{-5} .
    \end{align*}
\end{lemma}

We proceed to show 
that the support difference between Opinion $1$ and each other opinion doubles every $O(n^2/x_{1}(T_2))$ interactions until the ratio between the support of both opinions is sufficiently large.
This will be used in \cref{lem:phase-additive-multi-bias} to show that after $O(\log n \cdot n^2/x_{1}(T_2))$ interactions we reach \whp a configuration with a constant factor multiplicative bias.

\begin{lemma}[name=,restate=lemmaPhaseThreeDoubleBiasSingleSubphase] 
    \label{lem:additive_bias_growth_v2}
    Fix an Opinion $i \neq 1$ and assume there exists $t_0 \geq T_2$ with $x_i(t_0) \geq 20\sqrt{n\log n}$ and 
    $x_1(t_0) -x_i(t_0) \geq \alpha \sqrt{n}\log n$.
    Let $ T = 420\cdot n^2 / x_1(T_2)$ and let $\Delta_0 = x_1(t_0)-x_i(t_0)$. Then 
  {\dense
    \begin{align*}
        \Prob{ \exists t \in [t_0,t_0+T] \colon \!
        x_1(t) - x_i(t) \geq \min\{2\cdot \Delta_0,\  3\cdot x_i(t)\}  \text{ or }  
        x_i(t) < 20\sqrt{n\log n} } \geq 1-2n^{-3}\!\!.
    \end{align*}
    }
\end{lemma}

\begin{proof}[Proof Sketch]
The proof follows the analysis of the classical Gambler's ruin problem.
That is, starting with $\Delta = x_1(t)-x_i(t)$ we track the evolution of this quantity throughout a sequence of $O(n^2/x_1(t))$ interactions and show that it  reaches  $2\Delta$  before $\Delta/2$.
Here we rely on the bounds on the number of undecided agents (\cref{lem:undecided_general_bounds_v1} and \cref{lem:undecided>n/2-x1/2-o(n)}) and on the lower bound  Opinion $1$ which holds \whp during time $[T_2,T_2+ 420\cdot n^2 \cdot \log n / x_1(T_2)]$ (\cref{lem:phase2-max-no-shrinking}).
\fullproofinappendix{\cref{apx:omitted-proofs-phase3}}
\end{proof}

Now we are ready to analyze the running time of Phase 3.
\begin{lemma} 
\label{lem:phase-additive-multi-bias}
\label{lem:phase3}
    Assume that $\mathbf{x}(T_2)$ is a configuration with $x_1(T_2)-x_i(T_2) \geq \alpha\sqrt{n}\log n$ for all $i \neq 1$. 
    Let $T_3 = \inf\set{t \geq T_2 | \forall i \neq 1: x_1(t) \geq 2x_i(t)}$.
    Then 
    \begin{align*}
        \Prob{T_3-T_2 \leq 420\cdot n^2 \cdot \log n / x_1(T_2)} \geq 1-2n^{-2}.
    \end{align*}
\end{lemma}

\begin{proof}
The main idea of this proof is to repeatedly apply \cref{lem:additive_bias_growth_v2} to each Opinion $i \neq 1$ 
until either the support of Opinion $1$ becomes larger than $2n/3$ or the support of Opinion $i$ becomes less than $20\cdot \sqrt{n\log n}$.
In both cases it then follows that the ratio between the support of Opinion $1$ and Opinion $i$ is larger than two, and there is a time where there is a multiplicative bias between the first opinion and each other opinion.

    Let
    \[\hat{T} = \inf\left\{t\geq T_2 ~\left|~ u(t) \notin \left[({n-x_{\max}(t)})/2-8\cdot \sqrt{n\ln n},  {n}/{2}\right]\ \text{ or } \ x_{1}(t)<x_{1}(T_2)/2\right.\right\}
    \]
    be a stopping time. We define $(\hat{\mathbf{X}})_{t\ge T_2}$ as the process with $\hat{\mathbf{X}}(t) = \mathbf{X}(t)$ for all $t\leq \hat{T}$ and $\hat{\mathbf{X}}(t) = \mathbf{X}(\hat{T})$ for $t>\hat{T}$.
    From  \cref{lem:undecided>n/2-x1/2-o(n)} it follows that $u(t) \geq (n-x_{\max}(t))/2-8\cdot \sqrt{n\ln n} $ for all $t \in [T_2,n^3]$, \whp.
    From \cref{lem:undecided_general_bounds_v1} it follows that $ u(t) \leq n/2$  for all $t \in [T_2,n^3]$, \whp.
    Finally, \cref{lem:phase3-max-no-shrinking} gives us that $ x_{\max}(t) \geq x_{\max}(T_2)/2$ for all $t \in [T_2,T_2+ cn^2 \log n / x_{\max}(T_2)]$, \whp.
    Thus, $\hat{T}-T_2 =\Omega(n^2\cdot \log n/x_{1}(T_2) )$ \whp and we can assume that $(\mathbf{X})_{t\ge T_2}$ and $(\hat{\mathbf{X}})_{t\ge T_2}$ are identical for $t \in [T_2,T_2 + O(n^2\cdot \log n/x_{\max}(T_2))]$.
  
    Let $\tau = 420\cdot n^2 \cdot \log n / x_{\max}(T_2)$ and fix an Opinion $i \neq 1$ at time $T_2$ with $x_i(T_2) \geq 20\sqrt{n\log n}$.
    We divide the interactions from $[T_2,T_2+\tau]$ into $\log n$ subphases of length $420\cdot n^2/\hat{x}_1(T_2)$ each.
    For $1\leq j \leq \log n$ we define $\ell_1 = 1 $ and $\ell_j = 1+(j-1) \cdot 420 \cdot n^2/\hat{x}_1(T_2)$.
    Then the $j$th subphase contains interactions $\ell_j $ to $(\ell_{j+1}-1)$.
    Furthermore, we define $t_j$ is the first interaction in subphase $j$.
    Now fix an arbitrary subphase $j$.
It follows from \cref{lem:additive_bias_growth_v2} 
   that there exists a time $t' \in [t_{j},t_{j+1}]$ such that \whp either
   $\hat{x}_1(t) - \hat{x}_i(t) \geq \min\{2\cdot (\hat{x}_1(t_j)-\hat{x}_i(t_j)),\  3\cdot \hat{x}_i(t)\}$ or $\hat{x}_i(t) < 20\sqrt{n\log n}$.

   \medskip

We apply  \cref{lem:additive_bias_growth_v2} to each subphase. From the union bound over all subphases and all opinions it follows that after at most $\log n$ subphases \whp there exists for each Opinion $i$ a time $t_i' \in [T_2,T_2+\tau]$ with 
    either (a) $\hat{x}_1(t_i')-\hat{x}_i(t_i') \geq 2n/3$ or (b) $ \hat{x}_i(t_i') < 20\sqrt{n\log n} $ or (c) $\hat{x}_1(t_i') \geq 4\cdot \hat{x}_i(t_i')$.
   In the following we consider three cases.

    \paragraph{Case  (a)}
    There exists an Opinion $i\neq 1$ such that 
    $\hat{x}_1(t_i') - \hat{x}_i(t_i') \geq 2n/3$.
    Hence, we have at $t'_i$ a constant multiplicative bias between Opinion $1$ and all other opinions $i \neq 1$.
    From this the statement follows immediately with $T_3=t'_i$.
    
    \paragraph{Case (b)}
    For  Opinion $i$ there exists a $t'_i$ such that  $\hat{x}_i(t_i') < 20\sqrt{n\log n}$.
    From \cref{lem:small-opinions-do-not-grow}(1) it follows that $\hat{x}_i(t) \leq 40\sqrt{n\log n}$ for all $t\in [t_i',T_2+\tau]$ \whp.
    Additionally we know $\hat{x_1}(t) \geq \hat{x}_1(T_2)/2 \geq c'\sqrt{n}\log^2 n$  for all $t\in [t_i',T_2+\tau]$. 
    Hence, $\hat{x}_1(t)/\hat{x}_i(t) \gg 2$ for all $t\in [t_i',T_2+\tau]$ and, from the viewpoint of Opinion $i$ we have that $T_3$ can take on an arbitrary value in $[t_i',T_2+\tau]$.

    \paragraph{Case (c)}
    For Opinion $i$ there exists a $t_i'$ such that $\hat{x}_1(t_i') \geq 4 \cdot \hat{x}_i(t_i')$.
    From the claim below it follows that \whp $\hat{x}_1(t) \geq 2\hat{x}_i(t)$ for all $t \in [t_i',T_2+\tau]$ and from the viewpoint of Opinion $i$ we have that $T_3$ can take on an arbitrary value in $[t_i',T_2+\tau]$.
    
\medskip

Now \cref{lem:phase3} follows either immediately from Case (a).
Or we can apply Case (b) or Case (c) for each Opinion $i\neq 1$ and then we can choose $T_3=T_2+\tau$.
It remains to show the following claim.
The proof can be found in \cref{apx:omitted-proofs-phase3}.

\global\def\dcmqedhack{\qedhere\global\def\dcmqedhack{}}

\begin{claim}[name=,restate=claimPhaseThreeLossOfMultiBias]
        \label{claim:phase3-loss-of-bias}
        Let $j$ be an arbitrary subphase and let $t_0 \in [t_j,t_{j+1}]$.
        Fix an Opinion $i$ and assume $\hat{x}_i(t_0) \in [20\cdot \sqrt{n \log n}, \hat{x}_1(t_0)/4]$. 
        Then $\hat{x}_1(t) \geq 2 \cdot \hat{x}_i(t)$ for all $t \in [t_0,T_2+\tau]$.\dcmqedhack{}
    \end{claim}
\end{proof}

\section{From Multiplicative Bias To Absolute Majority (Phase 4)}
\label{sec:phase4}
Recall that $T_3$ is the end of Phase 3 and $\mathbf{X}(T_3)$ is a configuration with multiplicative bias.
In this version of the paper we assume that the bias is at least two, the proof of the case of a $(1+\epsilon)$-bias for any constant $\epsilon$ is deferred to the full version of this paper. It follows from a slightly more involved calculation.
In the following we assume w.l.o.g.\ that $x_1(T_3) > x_2(T_3)\ldots \ge x_k(T_3)$.
The main result for this phase is \cref{lem:phase4}, where we show that the multiplicative bias is grown into a unique majority opinion with support at least $2n/3$ within $O(n \log n + n^2/x_1(T_3))$ interactions, \whp.
To do so we first need an improved  bound on the number of undecided agents which we reach at time 
$T_3+O(n \log n)$. Additionally we have to show that in the meantime that both $x_1$ and the multiplicative bias decrease only by a small constant fraction (\cref{lem:phase4_x1_does_not_shrink} and \cref{lem:phase4_multiplicative_bias}). The proofs of both lemmas are similar to the proofs of \cref{lem:phase2-max-no-shrinking} and \cref{claim:phase3-loss-of-bias}, respectively, and can be found in \cref{apx:omitted-proofs-phase4}.

\begin{lemma}[name=,restate=lemmaPhaseFourXmaxNotShrinking] 
    \label{lem:phase4_x1_does_not_shrink}
    Let $c> 0$ be an arbitrary constant and define $T = c\cdot n^2 \log n / x_1(T_3)$.
    Then
    \begin{align*}
        \Prob{\text{for all }t \in [T_3,T_3+T] \colon x_1(t) \geq x_1(T_3)/2} \geq 1-n^{-5} .
    \end{align*}
\end{lemma}
    
\begin{lemma}[name=,restate=lemmaPhaseFourUndecidedImprovedBoundMultiplicativeBias] 
\label{lem:phase4_multiplicative_bias}
    Assume that $\mathbf{x}(T_3)$ is a configuration with $x_1(T_3) \geq 2 \cdot x_i(T_3)$ for all $i \neq 1$.
    Then \[
   \text{ for all } ~ i\neq 1 \quad \Prob{\text{for all } t \in [T_3,111\cdot n^2/x_1(T_3)]  \colon x_1(t) \geq 7/4 \cdot x_i(t)} \geq 1-2n^{-3}  .\]
\end{lemma}

Next we improve the lower bound on the number of undecided agents from  \cref{lem:undecided_general_bounds_v1}.
Recall that $T_4$ is the end of Phase 4, defined as $T_4 = \inf\set{t \geq T_3 |  x_1(t) \geq 2n/3 }$.

\begin{lemma}[name=,restate=lemmaPhaseUndecidedGrowImprovedBound] 
\label{lem:undecided_grow_improved_bound}
    Let $T_u = \inf\set{t \geq T_3 |  u(t) \geq n/2 - 7/8 \cdot x_{1}(t)}$.
    Then
    \[ \Prob{\min(T_4,T_u)-T_3 \leq \lceil 7 n\ln n \rceil } \geq 1-4n^{-3}.\]
\end{lemma}

\begin{proof}[Proof Sketch]
    The proof is similar to the proof of \cref{lem:phase1}.
    The main difference is that we use a modified potential function $Z(t) = n - 2u(t)- 7/8 \cdot x_1(t)$  instead of $Z(t) = n - 2u(t)-x_{1}(t)$.
    The expression for the expected drift of this modified potential function becomes slightly more complicated, and to bound it we require the multiplicative bias from \cref{lem:phase4_multiplicative_bias}.
    The full proof can be found in \cref{apx:omitted-proofs-phase4}.
\end{proof}

Now we are ready to analyze the running time of Phase $4$.
\begin{lemma} 
    \label{lem:phase4}
    Assume that $\mathbf{x}(T_3)$ is a configuration with $x_1(T_3) \geq 2\cdot x_i(T_3)$ for all $i \neq 1$.
    Then there exists a constant $c$ such that
    Then
    \begin{align*}
        \Prob{T_4-T_3 \leq 7 n\ln n + 444 \cdot n^2/x_1(T_3)} \geq 1-2n^{-2} .
    \end{align*}
\end{lemma}
\begin{proof}
To show the statement we require the following two auxiliary results.
First we establish in \cref{claim:phase4_undecided_improved_bound_holds} that the improved bound on the undecided agents from \cref{lem:undecided_grow_improved_bound}  holds throughout the remainder of the phase.
As before, we define $T_u = \inf\set{t \geq T_3 |  u(t) \geq n/2 - 7/8 \cdot x_{1}(t)/2}$ and recall that $T_4$ denotes the end of the phase.
The proof follows along the lines of the proof of \cref{lem:undecided>n/2-x1/2-o(n)} with the new $Z(t)$, and can be found in \cref{apx:omitted-proofs-phase4}.

\begin{claim}[name=,restate=claimPhaseUndecidedImprovedBoundHoldsUntilNextPhase]
\label{claim:phase4_undecided_improved_bound_holds}
   $ \displaystyle
       \Prob{\text{for all } t \in [T_u,\min\{n^3, T_4\}] \colon u(t) \geq n/2 - 7/16 \cdot  x_1(t) - 8 \cdot \sqrt{n \ln n}} \geq 1-4n^{-3}. $
\end{claim}

Next, in \cref{claim:mult-bias-x1-grows-fast} we bound the number of interactions until the support of Opinion $1$ has doubled.
Similarly to \cref{lem:additive_bias_growth_v2}, the proof uses the classical gambler's ruin problem
to show that in a sequence of $c\cdot n^2/x_1(t)$ interactions the support of Opinion 1 doubles \whp before it halves.
\fullproofinappendix{\cref{apx:omitted-proofs-phase4}}

\begin{claim}[name=,restate=claimPhaseFourDoubleSupportMaximumSingleSubphase] 
\label{claim:mult-bias-x1-grows-fast}
    Let $\x(t)$ be a configuration with $u(t) \geq n/2 - 7/16 \cdot x_{1}(t)- 8 \cdot \sqrt{n \ln n}$ and $x_1(t)< 2n/3$.
   We define $t' = c\cdot n^2/x_1(t)$ for a suitable chosen constant $c$. Then  
    \begin{align*}
        \Prob{\exists t' \in [t,t+t'] \colon x_1(t') \geq 2\cdot x_1(t)~ \text{ or } ~ x_1(t) \geq 2n/3} \geq 1-n^{-3}.
    \end{align*}
\end{claim}

With these two auxiliary claims we are now ready to show the lemma.
We start with a brief overview of the proof.
The proof is similar to the proof of \cref{lem:phase-additive-multi-bias} but we only have to consider the analog to Case (a). We repeatedly apply \cref{claim:mult-bias-x1-grows-fast} to Opinion $1$.
Then  the support of the largest opinion, $x_1(t)$ doubles every $O(n^2/x_1(t))$ interactions
until its support becomes larger than $2n/3$.
After doubling at most $\log n$ times, we reach a configuration where $x_1(t) \geq 2n/3$. This will be our time $T_4$.

    \bigskip
    
    To show that there exists a $t$ with  $x_1(t) \geq 2n/3$ we define
    \begin{align*}
        \hat{T} = \inf\set{t\geq T_3 +t_0 | u(t) \notin [(n- 7/16 \cdot x_{1}(t))-8\cdot \sqrt{n\ln n}, n/2] \text{ or } x_1(t) < x_1(T_3)/2}
    \end{align*}
    as a stopping time.
    Here $t_0$ is defined as $\inf\set{t \colon u(t_0) \geq (n- 7/16 \cdot x_1(t_0))} $.
    From  \cref{lem:undecided_grow_improved_bound} it follows \whp that $t_0 \leq T_3 + 7n\ln n$. 
    
    Let $(\hat{\mathbf{X}}(t))_{t\geq T_3+t_0}$ denote the process with $\hat{\mathbf{X}}(t) = \mathbf{X}(t)$ for all $t\leq \hat{T}$ and $\hat{\mathbf{X}}(t) = \mathbf{X}(\hat{T})$ for $t>\hat{T}$. From \cref{claim:phase4_undecided_improved_bound_holds} it follows that $u(t) \geq (n-x_{1}(T_3+t_0))/2-8\cdot \sqrt{n\ln n} $ for all $t \in [T_3+t_0,n^3]$, \whp.
    From \cref{lem:undecided_general_bounds_v1} it follows that $ u(t) \leq n/2$  for all $t \in [T_1,n^3]$, \whp.
    Finally, \cref{lem:phase2-max-no-shrinking} gives us that $ x_{1}(t) \geq x_{1}(T_3)/2$ for all $t \in [T_3,T_3+ cn^2 \log n / x_{1}(T_3)]$, \whp.
    Thus, $\hat{T}-(T_3+t_0) =\Omega(n^2/x_{1}(T_3) )$ \whp and we can assume that $(\mathbf{X})_{t\ge T_3+t_0}$ and $(\hat{\mathbf{X}})_{t\ge T_3+t_0}$ are identical for $t \in [T_3+t_0,T_3+t_0 + O(n^2\cdot \log n/x_{1}(T_3))]$.

    To track the progress of Opinion 1 we divide the interactions from $[T_3+t_0,T_3+t_0 + c\cdot n^2/x_{1}(T_3)]$ into subphases of varying length. 
    Let $T_{(0)} = T_3+t_0$ and define for $1\le \ell \le \log n $
    \begin{align*}
        T_{(\ell)} = \inf\set{t\geq T_{(0)} |  \hat{x}_1(t) \geq 2^{\ell} \cdot \hat{x}_1(T_{(0)}) \mbox{ or } \hat{x}_1(t) \geq 2n/3 } .
    \end{align*}
    We call the interactions in the interval $\big[T_{(\ell-1)}, T_{(\ell)}\big)$ \emph{subphase} $\ell$.
    Note that by definition of $T_{(\ell)}$, the support of $x_1$ doubles in every subphase (or $x_1 \geq 2/3 n$ and Phase 4 ends). In more detail, 
     for a fixed but arbitrary subphase $\ell$ it follows 
    from \cref{claim:mult-bias-x1-grows-fast} that the length of subphase $\ell$ is at most $ c\cdot n^2/\hat{x}(T_{(\ell-1)}) \leq c\cdot n^2/(2^{\ell-1} \cdot \hat{x}_1(T_{(0)})$, \whp.
    Hence, it follows that there exists a time $t' \in [T_{(\ell-1)},T_{(\ell-1)} + c\cdot n^2/\hat{x}(T_{(\ell-1)})]$ such that $\hat{x}_1(t') \geq 2^\ell \cdot \hat{x}_1(T_{(0)})$ or $\hat{x}_1(t') \geq 2/3 \cdot n$, \whp. From the union bound over all subphases we get that after at most $\log n$ subphases there exists \whp a time $t' \in [T_{(0)},T_{(\log n)}]$ such that $\hat{x}_1(t') \geq 2n/3$.
    This holds since otherwise
    $
        \hat{x}_1(t') \geq 2^{\log n} \cdot \hat{x}_1(T_{(0)}) \geq n \cdot \sqrt{n}\log^2 n > n .
    $, a contradiction.
    
     Summing up the length of all subphases for $c=111$ gives us
    \begin{align*}
        \sum_{i=1}^{\log n} \frac{c \cdot n^2}{  2^{i-1} \cdot \hat{x}_1(T_{(0)}) } 
        = \frac{c \cdot n^2}{\hat{x}_1(T_{(0)})} \cdot \sum_{i=1}^{\log n} \frac{1}{2^{i-1}} 
        \leq 2\cdot c \cdot \frac{n^2}{\hat{x}_1(T_{(0)})} 
    \end{align*}
    and hence, $T_4-T_3 \leq 7n\ln n + 4\cdot c \cdot \frac{n^2}{\hat{x}_1(T_3)}$ as claimed.
\end{proof}

\section{From Absolute Majority to Consensus (Phase 5)}
\label{sec:phase5}
Recall that $T_4$ is the end of Phase 4 and $\mathbf{X}(T_4)$ is a configuration where the support of the largest opinion, $x_{\max}(T_4)$ is at least $2n/3$.
The fifth phase ends when all agents agree on the Opinion $\max(T_4)$.
In the following we assume w.l.o.g.\ that $x_1(T_4) \geq x_2(T_4) \geq \ldots \geq x_k(T_4)$.
In \cref{lem:convergence_phase5} we show that the running time of this phase is $O( n\log n)$. 
This result is shown via a coupling where we couple the USD on configuration $\mathbf{X}(T_4)$ with the USD on a configuration $\tilde{\mathbf{X}}$ with two opinions only.  We show that the time until all agents agree starting from configuration $\mathbf{X}(T_4)$ is majorized by the time starting from configuration $\tilde{\mathbf{X}}$ (see \cref{lem:k-undecided-part5}).

\begin{lemma} 
    \label{lem:convergence_phase5}
    \label{lem:phase5}
    Assume that $\mathbf{x}(T_4)$ is a configuration with $x_1(T_4)\ge 2n/3$.
    Let {\dense$T_5 = \inf\set{t\geq T_4 | x_1(t) = n}$}. 
   \begin{align*}
   \lefttag{\textit{Then}}
       \Prob{T_5-T_4 \leq c\cdot n \log n} \geq 1-n^{-3} .
   \end{align*}
\end{lemma}

W.l.o.g.\, we assume that $T_4 = 0$.
To show this lemma we couple our process $(\mathbf{X}(t))_{t\in \mathbb{N}}$ with $k$ opinions with a process $(\tilde{\mathbf{X}}(t))_{t\in \mathbb{N}}$ with $2$-opinions. 
$\tilde{\mathbf{x}}$ is defined as follows.  $\tilde{x}_1(0)=x_1(0), \tilde{x}_2(0) = \sum_{i=2}^k x_i(0)$ and $\tilde{u}(0) = u(0)$.
We will show in \cref{lem:k-undecided-part5} that there exists a coupling such that for all $t\geq 0$ we have $x_1(t) \geq \tilde{x}_1(t)$.
From this follows that for all $t\ge 1$ we have $\Prob{x_1(t)=n} \ge \Prob{\tilde{x}_1=n}$.

Since $x_1(0) \geq 2n/3$, we have $\tilde{x}_1(0) - \tilde{x}_2(0) \geq n/3$. For $k=2$ it follows from \cite{DBLP:journals/dc/AngluinAE08} that $(\tilde{\mathbf{X}}(t))_{t\in \mathbb{N}}$ converges \whp to $\tilde{x}_1(t) = n$ in $O(n \log n)$ interaction.

For technical reasons, we show the stronger invariant   $x_1(t) \geq \tilde{x}_1(t)$ and $x_1(t) + u(t) \geq \tilde{x}_1(t) + \tilde{u}(t)$ using a step-by-step coupling.

\begin{lemma} 
    \label{lem:k-undecided-part5}
    Consider the two processes $(\mathbf{X}(t))_{t\in \mathbb{N}}$ with $k$ opinions and $(\tilde{\mathbf{X}}(t))_{t\in \mathbb{N}}$ with $2$-opinions where $\tilde{x_1}(0)=x_1(0), \tilde{x}(2) = \sum_{i=2}^k x_i(0)$ and $\tilde{u}(0) = u(0)$.
    Then there exists a coupling between the two processes such that $\forall t\geq 0$:
    \begin{align}
        \label{eq:k-undecided-part5-hypothesis}
        x_1(t) \geq \tilde{x}_1(t) \text{ ~ and ~ } x_1(t) + u(t) \geq \tilde{x}_1(t) + \tilde{u}(t)
    \end{align}
\end{lemma}

\begin{proof}
    We prove the lemma by induction over $t$.
    Obviously, the claim holds for $t=0$.
    Fix a time step $t$ where \eqref{eq:k-undecided-part5-hypothesis} holds. 
    We show that \eqref{eq:k-undecided-part5-hypothesis} holds for time $t+1$.
    In the following, we omit $t$ if clear from the context.
We represent the $n$ agents of the configurations $\mathbf{x}(t)$ and $\tilde{\mathbf{x}(t)}$ by $n$-dimensional vectors $\mathbf{v}(t)$ and $\tilde{\mathbf{v}}(t)$ which are sorted as follows. 
    {\small
    \begin{align*}
        \tilde{v}_i(t) & = \begin{cases}
            1 & \text{ if } i \in [\tilde{x}_1]\\
            \bot & \text{ if } i-\tilde{x}_1 \in [u]\\
            2 & \text{ if } i-\tilde{x}_1-u \in \left[\sum_{j=2}^k x_j\right]\\
            \bot & \text{ if } i-\tilde{x}_1-\tilde{x}_2 \in \left[u+1, \tilde{u}\right]\\
            2 & \text{ otherwise}
        \end{cases} & &
        v_i(t) & = \begin{cases}
            1 & \text{ if } i \in [\tilde{x}_1]\\
            \bot & \text{ if } i-\tilde{x}_1 \in [u]\\
            2, \dots, k & \text{ if } i-\tilde{x}_1-u \in \left[\sum_{j=2}^k x_j\right]\\
            1 & \text{ if } i-\tilde{x}_1-u-\tilde{x}_2 \in \left[x_1-\tilde{x}_1\right]\\
            \bot & \text{ otherwise }
        \end{cases}
    \end{align*}
    }
    The definition results in the following two cases.
\paragraph{Case 1: $\tilde{u}(t)\ge u(t)$}
In this case the vectors are sorted as follows.
\begin{center}
    \begin{tabular}{cc|c|cccc||c|c}
        $\tilde{\textbf{v}}$ = & 1 \ldots 1 & $\bot$ \ldots $\bot$ & 2 \ldots 2 & 2 \ldots 2 & \ldots & 2 \ldots 2 & $\bot$ \ldots $\bot$ & 2 \ldots 2 \\
        \hline
        $\textbf{v}$ = & 1 \ldots 1 & $\bot$ \ldots $\bot$ & 2 \ldots 2 & 3 \ldots 3 & \ldots & k \ldots k & 1 \ldots 1 & 1 \ldots 1\\
        & \multicolumn{3}{l}{$| \leftarrow$} & a & \multicolumn{2}{r}{$\rightarrow |$} & \multicolumn{2}{l}{ }
    \end{tabular}
\end{center}

\paragraph{Case 2: $\tilde{u}(t)< u(t)$}
In this case the vectors are sorted as follows.

    \begin{center}
    \begin{tabular}{cc|c|cccc||c|c}
        $\tilde{\textbf{v}}$ = & 1 \ldots 1 & $\bot$ \ldots $\bot$ & 2 \ldots 2 & 2 \ldots 2 & \ldots & 2 \ldots 2 & 2 \ldots 2 & 2 \ldots 2 \\
        \hline
        $\textbf{v}$ = & 1 \ldots 1 & $\bot$ \ldots $\bot$ & 2 \ldots 2 & 3 \ldots 3 & \ldots & k \ldots k & 1 \ldots 1 & $\bot$ \ldots $\bot$\\
        & \multicolumn{3}{l}{$| \leftarrow$} & a & \multicolumn{2}{r}{$\rightarrow |$} & \multicolumn{2}{l}{ }
    \end{tabular}
    \end{center}
    
    We will use the identity coupling, both processes choose the same pair  $(i, j) \in [n]^2$ uniformly at random.
    Hence, the next interaction is $(v_i(t), v_j(t))$ in the $k$-opinion process and $(\tilde{v}_i(t),\tilde{v}_j(t))$ in the two-opinion process. 
    Let $a=\tilde{x}_1(t)+\min(u(t),\tilde{u}(t))+\sum_{j=2}^k x_j(t)$.
    We split the analysis into four cases and exemplify the proof on the first case $i,j \leq a$.

\paragraph{ Case 1: $i,j \leq a$.}
If $\tilde{v}_i(t)=v_i(t)$ and $\tilde{v}_j(t)=v_j(t)$ both processes perform the same transition and the inductive step is trivially fulfilled. If $\tilde{v}_i(t)\neq v_i(t)$ and $\tilde{v}_j(t)\neq v_j(t)$ we have that $\tilde{v}_i(t),\tilde{v}_j(t)=2$
and $v_i(t),v_j(t) > 2$. Then $v_i(t+1)=\bot$ and $v_i(t+1)=2$ resulting in $u(t+1)\ge u(t)$.
We have two cases.

\medskip

\noindent\begin{minipage}{0.5\textwidth-1em}
\tabcolsep=2.5pt
If $\tilde{v}_i(t)=v_i$ but $\tilde{v}_j(t)\neq v_j(t)$ the following transitions are possible. 
\begin{center}
\begin{tabular}{llll|ll}
\toprule
$v_i(t)$  & $\tilde{v}_i(t)$  & $v_j(t)$ & $\tilde{v}_j(t)$ & $v_i(t+1)$ & $\tilde{v}_i(t+1)$\\
\midrule
1       & 1                & $>2$     & 2              & $\bot$ & $\bot$            \\
$\bot$    & $\bot$             & $>2$     & 2              &  $>2$  & 2        \\
2       & 2                & $>2$     & 2              &  $\bot$ & 2    
\\
\bottomrule
\end{tabular}
\end{center}
\end{minipage}\hfill\begin{minipage}{0.5\textwidth-1em}
If $\tilde{v}_i(t)\neq v_i(t)$ and $\tilde{v}_j(t)= v_j(t)$ the following transitions are possible.
\begin{center}
\tabcolsep=2.5pt
\begin{tabular}{llll|ll}
\toprule
$v_i(t)$  & $\tilde{v}_i(t)$  & $v_j(t)$ & $\tilde{v}_j(t)$ & $v_i(t+1)$ & $\tilde{v}_i(t+1)$\\
\midrule
$>2$       & 2   & 1      & 1       & $\bot$ & $\bot$            \\
$>2$       & 2   & $\bot$ & $\bot$  &  $>2$  & 2        \\
$>2$      & 2   & 2     & 2       &  $\bot$ & 2         \\
\bottomrule
\end{tabular}
\end{center}
\end{minipage}

\medskip

\noindent The inductive step now follows in both cases since in all cases $u(t)\ge \tilde{u}(t)$.
The remaining cases follow analogously and can be found in \cref{apx:omitted-proofs-phase5}.
\end{proof}

\section{Conclusions} \label{sec:conclusions}

We show fast convergence
of the USD in the population model, where the exact convergence
rates depend on the magnitude of support of the initial plurality 
opinion and the type of bias (if any) in the initial configuration.
Although our result can be viewed as an improvement over 
the existing, analogous convergence rates for the process
in the gossip model \cite{DBLP:conf/soda/BecchettiCNPS15},
our analysis does not readily transfer to that model. 
Thus it remains open to prove convergence of the $k > 2$ opinion
USD with \emph{no initial bias} in the gossip model, and moreover
to understand whether there exists a unified analysis for
analyzing the process in both models simultaneously. 
Additionally, analyzing the process
for $k = \omega(\sqrt{n} / \log^2 n)$ opinions is left for future work.

\printbibliography
\clearpage
\appendix
\section*{Appendix}
\section{Auxiliary Results}
\label{apx:auxiliary-results}
In this appendix we state a number of auxiliary results that we use throughout our analysis.
\subsection{Random Walks}
\begin{lemma}
    \label{lem:biased_reflective_random_walk}
    Let $W(t)$ be the random variable at time $t$ of a random walk on the positive integers with a reflective border at $0$ and $W(0) = 0$.
    Let $p$ be the probability of a $+1$-step.
    Let $q>p$ be the probability of a $-1$-step everywhere except for the origin.
    Let $r=1-p-q$ be the probability of remaining in place ($1-p$ for the origin).
    Let $T_m = \inf\set{t \geq 0 ~|~ W(t) \geq m}$.
    Then $\Pr[T_m \leq n^c] \leq n^c \cdot (p/q)^m$.
\end{lemma}

\begin{proof}
    The stationary distribution $W$ is given by
    $\Prob{W=n} = (p/q)^n \cdot (1 - p/q)$ and therefore
    $\Prob{W\geq n} = (p/q)^n$.
    Since $W(0) = 0$, it holds for all finite $t$ that
    $\Prob{W(t)\geq m} \leq Pr[W\geq m] = (p/q)^m$.
    The result follows from the union bound over $n^c$ steps.
\end{proof}

\begin{lemma}[\cite{feller68}]
\label{lem:random_walk_no_ruin_probability}
If we run an arbitrarily long sequence of independent trials, each with success probability at least $p$, then the probability that the number of failures ever exceeds the number of successes by $b$ is at most $((1-p)/p)^b$.
\end{lemma}

\begin{lemma}[Gambler's Ruin \cite{feller68}]
\label{lem:gambler's_ruin}
Let $Z(t)$ be the  random variable at time $t$ of a random walk on positive integers $[0,a]$  with absorbing states at $0$ and $b$ and $Z(0) = a$ for $a,b \in \N$ with $0<a<b$.
Let $p$ be the probability of a $+1$-step and  $q = 1-p$ be the probability of a $-1$-step with $p\neq q$ everywhere except for the absorbing states.
Let $T_{ruin} = \inf\{t\geq 0 ~|~Z(t) = 0\}$ and $T_{win} = \inf\{t\geq 0 ~|~Z(t) = b\}$.
Then
\[ 
 \Pr[T_{ruin}] = \frac{(q/p)^b-(q/p)^a}{(q/p)^b -1} \quad\mbox{ and }\quad \Pr[T_{win}] = 1-\Pr[T_{ruin}].
\]
\end{lemma}

\subsection{Drift Theorems}

\begin{theorem}[Theorem 18 of \cite{DBLP:series/ncs/Lengler20}]
    \label{thm:mult_drift_tail_lengler_18}
    Let $(X_t)_{t\geq0}$ be a sequence of non-negative random variables with a finite state space $\mathcal{S} \subseteq \mathbb{R}_0^+$ such that $0 \in \mathcal{S}$. Let $s_{min} \coloneqq \min(\mathcal{S}\setminus\{0\})$, and let $T \coloneqq \inf\{t\geq 0 ~|~ X_t = 0\}$. Suppose that $X_0 = s_0$, and that there exists $\delta > 0$ such that for all $s \in \mathcal{S}\setminus\{0\}$ and all $t\geq 0$, 
    \begin{align*}
        \Ex{X_t-X_{t+1} ~|~ X_t = s} \geq \delta s.
    \end{align*}
    Then, for all $r \geq 0$,
    \begin{align*}
        \Prob{ T > \left \lceil{\frac{r+\ln(s_0/s_{min})}{\delta}} \right \rceil} \leq e^{-r}.
    \end{align*}
\end{theorem}

The following lemma summarizes the drift analysis introduced in \cite{DBLP:conf/spaa/DoerrGMSS11}.
It functions as the basis of our analysis in \cref{sec:phase2}, which shows that in the case without bias the support of two fixed, large opinions drifts apart.
The original statement and proof are due to \cite{DBLP:conf/spaa/DoerrGMSS11}.
For convenience, we give in the following a slightly adapted and condensed version of the proof.

\begin{lemma}[Modified version of \cite{DBLP:conf/spaa/DoerrGMSS11}]
    \label{lem:drift_nobias}
    Let $W(t)$ be the random variable at time $t$ of a random walk on the state space $[0,\log\log n]$ with a reflective state $0$ and absorbing state $\log\log n$ and initially $W(0)=0$
    The transition probabilities are defined for every $t \in \N$ and $\ell \in [1,\log\log n-1]$ as follows
    \begin{align*}
        &\Pr[W(t+1) = 1 ~|~ W(t)=0] = p \\
        &\Pr[W(t+1) = \ell +1 ~|~ W(t) = \ell] = 1-e^{-2^\ell} \\
        &\Pr[W(t+1) = 0 ~|~ W(t) = \ell  ] = e^{-2^\ell}
    \end{align*}
    where $p\leq 1$ is an arbitrary constant.
    Let $T$ be the first time that $W(T) = \log\log n$, i.e., $W$ reaches the absorbing state.
    Then $T = O(\log n)$ \whp.
\end{lemma}
\begin{proof}
    We consider a sequence of attempts $Z_1,Z_2,\ldots$ such that $W$ reaches the absorbing state $\log\log n$. 
    
    The attempts are identical distributed and each (unsuccessful) attempt can be described by a random variable $B$ that denotes the number of consecutive successes (right steps of $W$) starting with $W(t) = 0 $ before its first fail ( falling back to state $0$).
    Note that a successful attempt ends up in the absorbing state $\log\log n$.
    We show that each attempt is successful with at least constant probability and then apply Chernoff bounds to conclude that $O(\log n)$ attempts are sufficient to provide at least one successful attempt.
    Additionally, we show that $t = O(\log n)$, i.e., the total number of trials sum up over all attempts is $O(\log n)$.
    We start with the first statement.
    For any $\ell \in [1,\log\log n -1]$ we have
    \begin{align*}
        \Prob{B=\ell} 
        = p \cdot \prod_{j=1}^{\ell-1} (1-e^{-2^j}) \cdot e^{-2^\ell}
        \leq p\cdot e^{-2^\ell}
        \leq e^{-2^\ell}
    \end{align*}
    and hence,
    \begin{align*}
        \Prob{B < \log\log n}
        &= \sum_{\ell = 0}^{\mathclap{\log\log n -1}} \Prob{B = \ell} \\
        &= \Prob{B= 0} + \sum_{\ell = 1}^{\mathclap{\log\log n -1}} \Prob{B = \ell} \\
        &\leq (1-p) + \sum_{\ell = 1}^{\mathclap{\log\log n -1}} p \cdot e^{-2^\ell} \\
        &\leq (1-p) + p \cdot \sum_{\ell = 1}^{\infty} e^{-2^\ell} \\
        &\leq (1-p) + 0.2 \cdot p \\
        &= 1- 0.8 \cdot p
    \end{align*}
    Therefore each attempt $Z_i$ is successful with  probability at least $1-(1-0.8p) = 0.8p$.
    Now consider $r = c\log n$ random variables $S_1,\ldots,S_r$ each indicates whether the attempt $Z_i$ is successful. 
    We know $\Prob{S_i = 1} \geq 0.8p$ for every $i\leq r$.
    An application of Chernoff bounds (\cref{lemma:chernoff_poisson_trials}) yields at least one successful attempt \whp.

    Now we continue with the second part of the statement.
    From the first part we already know that $r = c\log n$ attempts are sufficient.
    We upper bound the total number of steps of the random walk until it reaches the absorbing state by upper bound the total number of steps of $r = c\log n$ unsuccessful attempts.
    In order to do that we define new independent random variables  $Z_i' = Z_i +1$ for each $i \leq r$ and $Z' = \sum_{i=1}^r Z_i'$.
    Observe that $\Prob{Z_i' = \ell} = \Prob{Z_i = \ell-1}$ for every $\ell \in [1,\log\log n -1]$.
    Using our results from the first part we know that
    \begin{align*}
        \Prob{Z_i' = \ell} 
        \leq\begin{cases}
            p \cdot e^{-2^{\ell-1}} &, \ell \in [2,\log\log n -1] \\
            p &, \ell = 1 .
        \end{cases}
    \end{align*}
    By simple calculation it is easy to see that $e^{-2^{\ell-1}} \leq e^{-2(\ell-1)}$ for $\ell \geq 2$ and hence,  $\Prob{Z_i' = \ell} \leq p\cdot e^{-2(\ell-1)} $ for all $ \ell \in [1,\log\log n -1] $.
    Therefore 
    \begin{align*}
        \Ex{Z'}
        = \sum_{i=1}^r \Ex{Z_i'}
        = c\log n \cdot  \sum_{\ell = 1}^{\mathclap{\log\log n -1}} \ell \cdot \Prob{Z_i' = \ell}
        \leq c\log n \cdot p \cdot \sum_{\ell = 1}^{\mathclap{\log\log n -1}} \ell \cdot e^{-2^{\ell-1}}
        \leq p \cdot c \cdot \log n .
    \end{align*}
    This allows us to apply the following Chernoff bound (\cref{lem:chernoff-bound-exponential-tail}) that yields for $Z' = \sum_{i=1}^{r} Z_i'$, $\mu = \Ex{Z'}$, $\varepsilon = 2$ and $c = 6$
    \begin{align*}
        \Prob{Z' \geq c'\log n}
        \leq \Prob{Z' \geq (1+\varepsilon) p\cdot c \cdot \log n + O(r)} 
        \leq e^{-\frac{\varepsilon^2 \cdot c \cdot \log n}{2(1+\varepsilon)}}
        \leq n^{-4}
    \end{align*}
    and hence, the second part of the statement holds.
\end{proof}
\subsection{Anti-Concentration Bounds}
\begin{lemma}[Lemma 4 of \cite{DBLP:journals/siamcomp/KleinY15}]
\label{lemma:reverse_chernoff_v2}
    Let $X \sim \BinDistr(n,p)$ with $\mu = np$.
    For any $\delta \in (0,1/2]$ and $p \in (0,1/2]$, assuming $\delta^2 \mu \geq 3$, it holds that
    \[
        \Pr[X \geq (1+\delta)\mu] \geq e^{-9\delta^2 \mu}
    \]
    \[
        \Pr[X \leq (1-\delta)\mu] \geq e^{-9\delta^2 \mu}
    \]
\end{lemma}

\subsection{Concentration Bound}

\begin{theorem}[\cite{DBLP:books/daglib/0012859}, Theorem $4.4$, $4.5$]
\label{lemma:chernoff_poisson_trials}
    Let $X_1, \dots, X_n$ be independent Poisson trials with $\Pr[X_i = 1] = p_i$ and let $X = \sum X_i$ with $\Ex{X} = \mu$. Then the following Chernoff bounds hold: \\
    For $0 < \delta' \leq 1$:
    \[
        \Pr[X > (1+\delta') \mu] \leq e^{-\mu {\delta'}^2 / 3}.
    \]
    For $0 <\delta' < 1$:
    \[
        \Pr[X < (1-\delta') \mu] \leq e^{-\mu {\delta'}^2 / 2},
    \]
\end{theorem}

\begin{lemma}[Full version of \cite{DBLP:conf/spaa/DoerrGMSS11}]
    \label{lem:chernoff-bound-exponential-tail}
    Suppose that $X_1,\ldots,X_n$ are independent random variables on $\N$, such that there is a constant $\gamma >0$ with $\Prob{X_i=k}\leq \gamma(1-\delta)^{k-1}$ for every $k \in \N$.
    Let $X=\sum_{i=1}^n X_i$, $\mu = \Ex{X}$.
    Then it holds for all $\varepsilon>0$ that 
    \begin{align*}
        \Prob{X\geq (1+\varepsilon)\mu + O(n)} \leq e^{-\frac{\varepsilon^2 n}{2(1+\varepsilon)}} .
    \end{align*}
\end{lemma}

\begin{theorem}[rephrased, based on \cite{hoeffding63}]
\label{thm:chernoff-hoeffding-bound}
    Let $X_1,X_2,\ldots,X_n$ be independent random variables  and $a_i\leq X_i \leq b_i $ $(i=1,2,\ldots,n)$, then for $\lambda>0$
    \begin{align*}
        \Prob{\sum_{i=1}^n X_i - \Ex{\sum_{i=1}^n X_i} \geq \lambda}
        \leq e^{-\frac{2\lambda^2}{\sum_{i=1}^n (b_i-a_i)^2}} .
    \end{align*}
\end{theorem}

\begin{lemma}
    \label{lem:hoeffding-dependent-tailbound_v1}
    Consider a sequence of $\tau$ random variables $Z_1,\ldots,Z_\tau$ w.r.t.\ a sequence of random vectors $\mathbf{X}(0),\ldots, \mathbf{X}(\tau-1)$.
    Let $ \mathbf{X}_{<i} = \{\mathbf{X}(0),\ldots,\mathbf{X}(i-1)\}$ for all $i \leq \tau$.
    Let $Z = \sum_{i=1}^\tau Z_i$ and $\mu= \sum_{i=1}^\tau \mu_i$ with $\mu_i = E[Z_i ~|~ X_{<i}]$ for $i \leq \tau$.
    Assume $a\leq Z_i \leq b$ for all $i \leq \tau$.
    Then for all $\lambda > 0$
    \begin{align*}
        \Pr[Z -\mu <  - \lambda] \leq e^{-\frac{2\lambda^2}{\tau(b-a)^2}} 
    \end{align*}
\end{lemma}
\begin{proof}
    We follow the standard proof technique for Hoeffding bounds.
    For any $t<0$ we have 
    \begin{align*}
        \Pr[Z -\mu<  - \lambda]
        = \Pr\left[ e^{t(Z-\mu)} > e^{-t\lambda} \right]
        \leq \frac{E\left[ e^{t(Z-\mu)} \right]}{e^{-t\lambda}}
    \end{align*}
    where in the last inequality we apply  Markov's inequality.
    First we consider the term $E\left[ e^{t(Z-\mu)} \right]$.
    Since we do not assume any independence among the $Z_i$'s we utilize the concept of conditional independence via the law of total expectation. 
    That is,
    \begin{align*}
        E\left[ e^{t(Z-\mu)} \right]
        &= E\left[ e^{\sum_{i=1}^\tau t(Z_i-\mu_i)} \right]
    \\  &= E\left[ E\left[ e^{\sum_{i=1}^\tau t(Z_i-\mu_i)} ~|~ \mathbf{X}_{<\tau} \right] \right]
    \\  &= E\left[ e^{\sum_{i=1}^{\tau-1} t(Z_i-\mu_i)} \cdot E\left[ e^{t(Z_\tau-\mu_\tau)} ~|~ \mathbf{X}_{<\tau} \right] \right]
    \\  &= E\left[ e^{\sum_{i=1}^{\tau-1} t(Z_i-\mu_i)}  \right] \cdot  E\left[ e^{t(Z_\tau-\mu_\tau)} ~|~ \mathbf{X}_{<\tau} \right]
    \\  &= E\left[ e^{\sum_{i=1}^{\tau-2} t(Z_i-\mu_i)} \cdot E\left[ e^{t(Z_{\tau-1}-\mu_{\tau-1})} ~|~ \mathbf{X}_{<\tau-1} \right] \right] \cdot  E\left[ e^{t(Z_\tau-\mu_\tau)} ~|~ \mathbf{X}_{<\tau} \right]
    \\  &= E\left[ e^{\sum_{i=1}^{\tau-2} t(Z_i-\mu_i)}  \right] \cdot E\left[ e^{t(Z_{\tau-1}-\mu_{\tau-1})} ~|~ \mathbf{X}_{<\tau-1} \right] \cdot  E\left[ e^{t(Z_\tau-\mu_\tau)} ~|~ \mathbf{X}_{<\tau} \right]
    \\  &= \prod_{i=1}^\tau E\left[ e^{t(Z_i-\mu_i)} ~|~ \mathbf{X}_{<i} \right]
    \end{align*}
    Due to the conditional expected value  we cannot directly apply Hoeffding's lemma to yield an upper bound on this expression.
    Recall that this result states for any real valued random variable $W$ such that $a\leq W\leq b$ almost surely that for all $\lambda\in \R$ 
    \begin{align*}
        E\left[ e^{\lambda(W)} \right] \leq e^{\lambda\cdot E[W] \frac{\lambda^2(b-a)^2}{8} } .
    \end{align*}
    Fortunately we can derive a  conditional version as well.
    The key is to define  new random variables $W_i = Z_i-\mu_i $ for all $i \leq \tau$ and observe that $E[W_i ~|~ X_{<i}] = 0 $.
    Using the convexity of $e^{\lambda x}$ we get
    \begin{align*}
        E\left[ e^{tW_i} ~|~ X_{<i} \right]
        &\leq E\left[ \frac{b-W_i}{b-a} \cdot e^{ta} + \frac{W_i-a}{b-a} \cdot e^{tb} ~|~ X_{<i} \right]
        \\&= \frac{b-E[W_i ~|~ X_{<i}]}{b-a} \cdot e^{ta} + \frac{E[W_i ~|~ X_{<i}]-a}{b-a} \cdot e^{tb}
        \\&= \frac{b}{b-a} \cdot e^{ta} + \frac{-a}{b-a} \cdot e^{tb}
    \end{align*}
    The remaining steps to proof the conditional version of Hoeffding's lemma are identical to the original proof.
    Thus,
    \begin{align*}
        E\left[ e^{\lambda(Z_i-\mu_i)} ~|~ X_{<i} \right] \leq e^{\frac{\lambda^2(b-a)^2}{8} }.
    \end{align*}
    At last we combine this  with the calculation  from the beginning and obtain 
    \begin{align*}
        \Pr[Z -\mu<  - \varepsilon]
        \leq \frac{E\left[ e^{t(Z-m)} \right]}{e^{-t\varepsilon}}
        \leq e^{\frac{\tau \cdot t^2(b-a)^2}{8} + t\varepsilon}  
    \end{align*}
    By optimizing the choice of $t<0$ we set $t = -4\lambda/(\tau(b-a)^2)$ and the desired result follows.
\end{proof}

\section{General Observations about the USD}
\label{apx:transition-probabilities}
In this appendix we give bounds for the transition probabilities of the USD.
In the following we abbreviate $r^2(t)=\sum_{i=1}^k x_i^2(t)$. 

In our analysis we (mostly) track the number of undecided agents $U(t)$ over time $t$.
We denote the transition probabilities for the number of undecided agents as follows:

\begin{align*}
        p_-(t) & = \Prob{U(t+1) = u(t)-1 ~|~ \mathbf{X}(t)=\mathbf{x}(t) },\\
        p_+(t) & = \Prob{U(t+1) = u(t)+1~|~\mathbf{X}(t)=\mathbf{x}(t) } \text{ and}\\
        \tilde{p}_+(t) & = \Prob{U(t+1) = u(t)+1 ~|~ \mathbf{X}(t)=\mathbf{x}(t), \mathbf{X}(t+1) \neq x(t)}.\\
    \end{align*}

\begin{observation}
    \label{obs:probabilitiesxxx}
    Consider the process in an arbitrary configuration $\mathbf{x}(t)$.
    Then the following holds.
    \begin{enumerate}
        \item $p_-(t) = u(t) \cdot \left( \sum_{i=1}^k x_i(t) \right) \cdot n^{-2} = u(t) \cdot (n-u(t)) \cdot n^{-2}$.
        \item $p_+(t) = \left( \sum_{i=1}^k x_i(t)  \cdot (n-u(t)-x_i(t)) \right) \cdot n^{-2} = ((n-u(t))^2-r^2(t)) \cdot n^{-2}$.
    \end{enumerate}
\end{observation}
The observation follows immediately from the definition of the process.

\begin{observation}
    \label{obs:probabilitiesyyy}
    Consider the process in an arbitrary configuration $\mathbf{x}(t)$ and let $\epsilon \geq 0$ be an arbitrary constant.
    Let $u^* = n \cdot (k-1)/(2k-1) $.
    If $u(t) \geq u^*+\epsilon \cdot n$, then $\tilde{p}_+(t)  \leq \frac{1}{2} - \frac{\epsilon}{2}$.
\end{observation}

\begin{proof}
    We first give an expression for $\tilde{p}_+$.
    Note that $p_- + p_+$ is the probability of a so-called productive step and $\tilde{p}$ is the probability with which $u$ increases conditioned on a productive step.
    Thus,
    \begin{align*}
        \tilde{p}_+ 
        & = \frac{p_+}{p_- + p_+}
        = \frac{1}{2} - \frac{ p_- - p_+ }{ 2(p_- + p_+) }.
    \end{align*} 
    Next, we give an expression for $(p_- - p_+) / (p_- + p_+)$ that only depends on $u$, $r^2$ and $n$ (which is constant over time).
    \begin{align*}
        \frac{ p_- - p_+ }{ p_- + p_+ }
        &= \frac{u\cdot (n-u) \cdot n^{-2} - ((n-u)^2-r^2) \cdot n^{-2}}{u \cdot (n-u) \cdot n^{-2} + ((n-u)^2-r^2) \cdot n^{-2}}
        = \frac{u\cdot (n-u) - (n-u)^2 + r^2}{u\cdot (n-u) + (n-u)^2 - r^2}\\
        & = \frac{(2u-n)(n-u)+r^2}{n(n-u) - r^2}
        = \frac{(2u-n)(n-u) - (2u-n)r^2/n + 2u\cdot r^2/n}{n(n-u)-r^2}\\
        & = \frac{2u-n}{n} \cdot \frac{(n-u)-r^2/n}{(n-u)-r^2/n} + \frac{2u\cdot r^2/n}{n(n-u)-r^2}
        = \left(\frac{2u}{n}-1\right) + \frac{2u\cdot r^2}{n^2(n-u)-n\cdot r^2}
    \end{align*}
    Fix some $u \in [0,n]$.
    Note that $r^2$ is maximal if all $n-u$ decided agents agree on one opinion, i.e., $r_{max}^2 = (n-u)^2$ and that $r^2$ is minimal if the support is equally distributed over all opinions, i.e, $r_{min}^2 = \sum_{i=1}^k (n-u)^2/k^2 = (n-u)^2/k$.
    Therefore, we can bound the expression $n^2 (n-u)-n \cdot r^2$ as follows.
    \begin{align*}
        n^2 (n-u)- n\cdot r^2 \geq n^2 \cdot (n-u) - n (n-u)^2 = n \cdot u (n-u) \geq 0.
    \end{align*}
    Now, it is obvious that 
    $( p_-(t) - p_+(t))/(p_-(t) + p_+(t)) $ is increasing for increasing $u$ and for increasing $r^2$.
   Therefore, $\tilde{p}_+$ is maximal for minimal $u$ and for minimal $r^2$, i.e., $u = u^*+\epsilon \cdot n$ and $r^2 = (n-(u^*+\epsilon \cdot n))^2/k$.
    Thus, with $u^* = n \cdot (k-1)/(2k-1)$, we have 
    \begin{align*}
        p_{max} 
        & \leq \frac{1}{2} - \frac{ (u^* + \epsilon \cdot n) (n - (u^* + \epsilon \cdot n)) - (n - (u^* + \epsilon \cdot n))^2 + (n-(u^*+\epsilon \cdot n))^2/k } { 2 \left( (u^* + \epsilon \cdot n) (n - (u^* + \epsilon \cdot n)) + (n - (u^* + \epsilon \cdot n))^2 - (n-(u^*+\epsilon \cdot n))^2/k \right)}\\
        & = \frac{1}{2} - \frac{(n-(u^* + \epsilon \cdot n))\left( (u^* + \epsilon \cdot n) - (n - (u^* + \epsilon \cdot n))(1-1/k) \right)}{2 (n-(u^* + \epsilon \cdot n)) \left( (u^* + \epsilon \cdot n) + (n - (u^* + \epsilon \cdot n))(1-1/k) \right)}\\
        & = \frac{1}{2} - \frac{(u^* + \epsilon \cdot n) - (n - (u^* + \epsilon \cdot n))(1-1/k)}{2 \left( (u^* + \epsilon \cdot n) + (n - (u^* + \epsilon \cdot n))(1-1/k) \right)}\\
        & = \frac{1}{2} - \frac{\epsilon (2k-1)^2}{2(\epsilon(2k-1)+2k(k-1))}
        \leq \frac{1}{2} - \frac{\epsilon}{2}
    \end{align*}
\end{proof}

In the next lemma we bound the number of undecided agents.
We show that \whp the number of undecided agents is less than $n/2$ for $k = O(\sqrt{n}/\log(n))$ and at most $n/2 + o(n)$ for arbitrary $k$ during the whole process.
    
Similar to the previous observation  we give probabilities for the support of the opinions in the USD.
\begin{observation}
\label{obs:prob_opinion_single_interaction}
Fix an Opinion $i$. Then
\begin{enumerate}
    \item $p_+^{(i)}(t) = \Prob{X_i(t+1) = x_i+1 ~|~ \mathbf{X}(t) = \mathbf{x}} = u \cdot x_i \cdot n^{-2}$,
    \item $p_-^{(i)} = \Prob{X_i(t+1) = x_i-1 ~|~ \mathbf{X}(t)= \mathbf{x}} = x_i \cdot (n-u-x_i) \cdot n^{-2}$ and
    \item $\tilde{p}_+^{(i)} = \Prob{X_i(t+1)  = x_i +1 ~|~ \mathbf{X}(t) = \mathbf{x}, X_i(t+1) \neq x_i}
        = \frac{p_+^{(i)}(t)}{p_+^{(i)}(t) + p_-^{(i)}(t)}\\
        \phantom{\tilde{p}_+^{(i)} } = \frac{1}{2} + \frac{p_+^{(i)}(t)-p_-^{(i)}(t)}{2\cdot (p_+^{(i)}(t) + p_-^{(i)}(t))}$. 
\end{enumerate}
\end{observation}

\begin{observation}
    \label{obs:conditional_probabilities}
    Fix two opinions $i$ and $j$. Let $\Delta(t) = X_i(t)-X_j(t)$. Then
    {\small
    \begin{enumerate}
        \item $p_+^{(ij)}(t) = \Prob{\Delta(t+1) = x_i-x_j+1 ~|~ \mathbf{X}(t) = \mathbf{x}}
        = p_+^{(i)}(t) + p_-^{(j)}(t)
        = \frac{u\cdot x_i + x_j\cdot(n-u-x_j)}{n^2}$,
        \item $p_-^{(ij)}(t) = \Prob{\Delta(t+1) = x_i-x_j-1 ~|~ \mathbf{X}(t) = \mathbf{x}}
         = p_-^{(i)}(t) + p_+^{(j)}(t)
         = \frac{u\cdot x_j + x_i\cdot(n-u-x_i)}{n^2}$ and
         \item $\tilde{p}_+^{(ij)}(t) = \Prob{\Delta(t+1) = x_i-x_j +1~|~\mathbf{X}(t) = \mathbf{x}, \Delta(t+1) \neq \Delta(t)} 
         = \frac{p_+^{(ij)}(t)}{p_+^{(ij)}(t) + p_-^{(ij)}(t)}\\
         \phantom{\tilde{p}_+^{(ij)}(t) } = \frac{1}{2} + \frac{p_+^{(ij)}(t)-p_-^{(ij)}(t)}{2\cdot (p_+^{(ij)}(t) +p_-^{(ij)}(t))}$.
    \end{enumerate}
    }
\end{observation}

\section{Omitted Proofs}

In this appendix we give the full formal proofs of our analysis. For convenience we restate the lemmas before the corresponding proofs.

\subsection{Omitted Proofs of Section 3 (Phase 1)}
\label{apx:omitted-proofs-phase1}

\lemmaPhaseOneNoLossOfBias*
\begin{proof}
    We start with the proof of the first statement.
    Fix an Opinion $i \neq 1$.
    We consider \begin{align*}
        \Psi_t = \frac{x_1(t)-x_i(t)}{n-u(t)}
    \end{align*} and show via a version of the Hoeffding bound (see \cref{lem:hoeffding-dependent-tailbound_v1}) that this quantity  does not decrease significantly throughout the first phase. Recall that $T_1$ is defined as the first time $t$ where $u(t)\geq n/2-x_{\max}(t)/2$ and that by definition $x_{\max}(0) = x_1(0)$.
        
    Let $\hat{T} = \inf\set{t\geq 0~|~u(t) \geq n/2}$ be a stopping time and
    let $(\mathbf{\hat{X}}(t))_{t\in \N}$ denote the  process with 
    $\hat{X}(t)=X(t)$ for all $t\le \hat{T}$ and $\hat{X}(t)=X(\hat{T})$ for
    $t>\hat{T}$.
    From \cref{lem:undecided_general_bounds_v1} it follows $\hat{T} \geq n^3$ with probability at least $1-n^{-3}$. Also, from \cref{lem:phase1} it follows that $T_1 \leq 7\cdot n\ln n $ with probability $1-n^{-3}$.
    Thus, \whp  $(\mathbf{X}(t))_{t\in \N}$ and $(\mathbf{\hat{X}}(t))_{t\in \N}$ behave the same during the first phase.
    
    In the following we define $Z_{t+1} = \Psi_{t+1}-\Psi_t$ and $\mu_{t+1} = \Ex{Z_{t+1}~|~\mathbf{X_{<t+1}}}$.
    Similarly we define $\hat{\Psi}(t)$,$\hat{Z}(t)$ and $\hat{\mu}_{t+1}$ for $(\mathbf{\hat{X}}(t))_{t\in \N}$.
    Our goal is to apply \cref{lem:hoeffding-dependent-tailbound_v1} to  $\hat{Z}_1,\hat{Z}_2,\ldots $.
    First we calculate $\hat{\mu}_{t+1} = \Ex{\hat{Z}_{t+1}~|~\mathbf{\hat{X}}_{<t+1}}$.
    Similar to the proof of \cref{lem:phase1} we distinguish between $3$ cases:
    $\hat{u}(t+1) = \hat{u}(t) +1$, $\hat{u}(t+1) = \hat{u}(t)-1$ and $\hat{u}(t+1) = \hat{u}(t)$. 
    For readability, we use in the following $\hat{u}$ instead of $\hat{u}(t)$ and $\hat{x}_i$ instead of $\hat{x}_i(t)$ for all $1\le i\le k+1$. We get
    \begin{align*}
        \MoveEqLeft \Ex{\hat{Z}_{t+1}~|~\mathbf{\hat{X}}_{<t+1} \land \hat{u}(t+1) = \hat{u}-1}\\
        & = \frac{\hat{x}_1}{n-\hat{u}} \cdot \left( \frac{\hat{x}_1-\hat{x}_i+1}{n-\hat{u}+1} \right) + \frac{\hat{x}_i}{n-\hat{u}} \cdot \left(\frac{\hat{x}_1-\hat{x}_i-1}{n-\hat{u}+1} \right) \\
        & \phantom{={}} + \frac{n-\hat{u}-\hat{x}_1-\hat{x}_i}{n-\hat{u}} \cdot \left( \frac{\hat{x}_1-\hat{x}_i}{n-\hat{u}+1} \right) - \frac{\hat{x}_1-\hat{x}_i}{n-\hat{u}}\\
        & = \frac{\hat{x}_1-\hat{x}_i}{n-\hat{u}+1} \cdot \left( \frac{\hat{x}_1+\hat{x}_i+(n-\hat{u}-\hat{x}_1-\hat{x}_i)}{n-\hat{u}} \right) + \frac{\hat{x}_1-\hat{x}_i}{(n-\hat{u})(n-\hat{u}+1)} - \frac{\hat{x}_1-\hat{x}_i}{n-\hat{u}}\\
        & = \frac{(n-\hat{u})(\hat{x}_1-\hat{x}_i) + \hat{x}_1-\hat{x}_i - (n-\hat{u}+1)(\hat{x}_1-\hat{x}_i)}{(n-\hat{u})(n-\hat{u}+1)}\\
        & = \frac{(\hat{x}_1-\hat{x}_i)-(\hat{x}_1-\hat{x}_i)}{(n-\hat{u})(n-\hat{u}+1)}
        = 0
    \end{align*}
    
    {\dense \begin{align*}
        \MoveEqLeft \Ex{\hat{Z}_{t+1}~|~\mathbf{\hat{X}}_{<t+1} \land \hat{u}(t+1) = \hat{u}-1} \\
        &=\Ex{\hat{\Psi}_{t+1}-\hat{\Psi}_t~|~\mathbf{\hat{X}}_{<t+1} \land \hat{u}(t+1) = \hat{u}+1} \\ &
        = \frac{\hat{x}_1(n-\hat{u}-\hat{x}_1)}{(n-\hat{u})^2-\hat{r}^2} \cdot \left( \frac{\hat{x}_1-\hat{x}_i-1}{n-\hat{u}-1} \right) + \frac{\hat{x}_i(n-\hat{u}-\hat{x}_i)}{(n-\hat{u})^2-\hat{r}^2} \cdot \left( \frac{\hat{x}_1-\hat{x}_i+1}{n-\hat{u}-1} \right) 
        \\&\qquad+ \frac{((n-\hat{u})^2-\hat{r}^2)-\hat{x}_1(n-\hat{u}-\hat{x}_1)-\hat{x}_i(n-\hat{u}-\hat{x}_i)}{(n-\hat{u})^2-\hat{r}^2} \cdot \left( \frac{\hat{x}_1-\hat{x}_i}{n-\hat{u}-1} \right) - \frac{\hat{x}_1-\hat{x}_i}{n-\hat{u}}
        \\&= \frac{\hat{x}_1-\hat{x}_i}{n-\hat{u}-1} \cdot \left( \frac{(n-\hat{u})^2-\hat{r}^2}{(n-\hat{u})^2-\hat{r}^2} \right) - \frac{\hat{x}_1(n-\hat{u}-\hat{x}_1)-\hat{x}_i(n-\hat{u})(n-\hat{u}-\hat{x}_i)}{((n-\hat{u})^2-\hat{r}^2)(n-\hat{u}-1)} - \frac{\hat{x}_1-\hat{x}_i}{n-\hat{u}}
        \\&= \frac{((n-\hat{u})^2-\hat{r}^2)(\hat{x}_1-\hat{x}_i)(n-\hat{u}) - \hat{x}_1(n-\hat{u})(n-\hat{u}-\hat{x}_1)+\hat{x}_i(n-\hat{u})(n-\hat{u}-\hat{x}_i)}{(n-\hat{u})(n-\hat{u}-1)((n-\hat{u})^2-\hat{r}^2)} 
        \\ &\phantom{={}}-\frac{((n-\hat{u})^2-\hat{r}^2)(n-\hat{u}-1)(\hat{x}_1-\hat{x}_i)}{(n-\hat{u})(n-\hat{u}-1)((n-\hat{u})^2-\hat{r}^2)}
        \\&= \frac{(n-\hat{u})(\hat{x}_1^2-\hat{x}_i^2)-\hat{r}^2(\hat{x}_1-\hat{x}_i))}{(n-\hat{u})(n-\hat{u}-1)((n-\hat{u})^2-\hat{r}^2)}
        \\&= \frac{\hat{x}_1-\hat{x}_i}{n-\hat{u}} \cdot \left( \frac{(\hat{x}_1+\hat{x}_i)(n-\hat{u}) - \hat{r}^2}{(n-\hat{u}-1)((n-\hat{u})^2-\hat{r}^2)} \right)
    \end{align*} }
    Note that if the number of undecided agents does not change, there is no change at all and hence $\Ex{\hat{Z}_{t+1}~|~\mathbf{\hat{X}}_{<t+1}\land \hat{u}(t+1)=\hat{u}} = 0$.
    By combining these results we obtain
    { \dense \begin{align*}
        \MoveEqLeft \Ex{\hat{Z}_{t+1}~|~\mathbf{\hat{X}}_{<t+1}}\\
        & = 0 \cdot \Prob{\hat{u}(t+1)=\hat{u}-1} + \left(\frac{\hat{x}_1-\hat{x}_i}{n-\hat{u}} \cdot  \frac{(\hat{x}_1+\hat{x}_i)(n-\hat{u}) - \hat{r}^2}{(n-\hat{u}-1)((n-\hat{u})^2-\hat{r}^2)} \right) \cdot \Prob{\hat{u}(t+1)=\hat{u}+1}\\
        & \phantom{={}} + 0 \cdot \Prob{\hat{u}(t+1)=u}\\
        & = \left(\frac{\hat{x}_1-\hat{x}_i}{n-\hat{u}} \cdot  \frac{(\hat{x}_1+\hat{x}_i)(n-\hat{u}) - \hat{r}^2}{(n-\hat{u}-1)((n-\hat{u})^2-\hat{r}^2)} \right) \cdot \frac{(n-\hat{u})^2-\hat{r}^2}{n^2}\\
        & = \frac{\hat{x}_1-\hat{x}_i}{n-\hat{u}} \cdot \left( \frac{(\hat{x}_1+\hat{x}_i)(n-\hat{u}) -\hat{r}^2}{(n-\hat{u}-1)n^2} \right)\\
        & = \frac{\hat{x}_1-\hat{x}_i}{n-\hat{u}} \cdot \left( \frac{\sum_i \hat{x}_i(\hat{x}_1-\hat{x}_i) +\hat{x}_i(n-\hat{u})}{(n-\hat{u}-1)n^2} \right)\\
        &\geq 0
    \end{align*} }
    

    Now we show that $\abs{\hat{Z}_{t+1}} \leq a$ with $a=4/(n-2)$ for all $t < \tau$.
To do so we simply consider every possible outcome of
\begin{align*}
    \frac{\hat{x}_1(t+1)-\hat{x}_i(t+1)}{n-\hat{u}(t+1)} -\frac{\hat{x}_1(t)-\hat{x}_i(t)}{n-\hat{u}(t)},
\end{align*}
 i.e., either  the number of undecided agents $\hat{u}$ is increased and decreased by one, or $\hat{x}_i$ or $\hat{x}_1$ are increased by one, or In the following we consider all six cases, 

\begin{align*}
    &\abs*{\frac{\hat{x}_1-\hat{x}_i}{n-\hat{u}-1} - \frac{\hat{x}_1-\hat{x}_i}{n-\hat{u}} }
    \;=\; \abs*{\frac{\hat{x}_1-\hat{x}_i}{(n-\hat{u})(n-\hat{u}-1)} }
    \;=\; \frac{\hat{x}_1-\hat{x}_i}{(n-\hat{u})(n-\hat{u}+1)} 
\\  &\abs*{\frac{\hat{x}_1-\hat{x}_i}{n-\hat{u}+1} - \frac{\hat{x}_1-\hat{x}_i}{n-\hat{u}} }
    \;=\; \abs*{\frac{-(\hat{x}_1-\hat{x}_i)}{(n-\hat{u})(n-\hat{u}+1)}}
    \;=\; \frac{\hat{x}_1-\hat{x}_i}{(n-\hat{u})(n-\hat{u}+1)}
\\  &\abs*{\frac{\hat{x}_1-\hat{x}_i+1}{n-\hat{u}+1} - \frac{\hat{x}_1-\hat{x}_i}{n-\hat{u}}}
    \;=\; \abs*{\frac{n-\hat{u}-(\hat{x}_1-\hat{x}_i)}{(n-\hat{u})(n-\hat{u}+1)}}
    \;=\; \frac{n-\hat{u}-(\hat{x}_1-\hat{x}_i)}{(n-\hat{u})(n-\hat{u}+1)} 
\\  &\abs*{\frac{\hat{x}_1-\hat{x}_i-1}{n-\hat{x}+1} - \frac{\hat{x}_1-\hat{x}_i}{n-\hat{u}}}
    \;=\; \abs*{-\frac{n-\hat{u}+(\hat{x}_1-\hat{x}_i)}{(n-\hat{u})(n-\hat{u}+1)}}
    \;=\; \frac{n-\hat{u}+(\hat{x}_1-\hat{x}_i)}{(n-\hat{u})(n-\hat{u}+1)}  
\\  &\abs*{\frac{\hat{x}_1-\hat{x}_i-1}{n-\hat{u}-1} - \frac{\hat{x}_1-\hat{x}_i}{n-\hat{u}}}
    \;=\; \abs*{\frac{-(n-\hat{u})+(\hat{x}_1-\hat{x}_i)}{(n-\hat{u})(n-\hat{u}-1)}}
    \;=\; \frac{n-\hat{u}-(\hat{x}_1-\hat{x}_i)}{(n-\hat{u})(n-\hat{u}-1)}
\\  &\abs*{\frac{\hat{x}_1-\hat{x}_i+1}{n-\hat{u}-1} - \frac{\hat{x}_1-\hat{x}_i}{n-\hat{u}}} \numberthis\label{eq:bounded_Z_i}
    \;=\; \abs*{\frac{n-\hat{u}+(\hat{x}_1-\hat{x}_i)}{(n-\hat{u})(n-\hat{u}-1)}}
    \;=\; \frac{n-\hat{u}+(\hat{x}_1-\hat{x}_i)}{(n-\hat{u})(n-\hat{u}-1)} 
\end{align*}

It is easy to see that \cref{eq:bounded_Z_i} results in the largest change in $\hat{Z}_{t+1}$ (by comparing the equations pairwise) and hence,
    \begin{align*}
        |\hat{Z}_{t+1}| \leq
        \frac{\hat{x}_1-\hat{x}_i}{(n-\hat{u})(n-\hat{u}-1)} 
        \leq \frac{n-\hat{u}+(\hat{x}_1-\hat{x}_i)}{(n-\hat{u})(n-\hat{u}-1)}
        \leq \frac{(n-\hat{u})+(n-\hat{u})}{(n-\hat{u})(n-\hat{u}-1)} = \frac{2}{n-\hat{u}-1}
    \end{align*}
    By definition of $\hat{T}$ we know $\hat{u}(t) < n/2$   and thus $\abs{\hat{Z}_{t+1}} < 4/(n-2)$. 
    Observe that for every $t$ we have
    \begin{align*}
        \sum_{i=1}^t \hat{Z}_i = \sum_{i=1}^t (\hat{\Psi}_i-\hat{\Psi}_{i-1}) = \hat{\Psi}_t - \hat{\Psi}_0 .
    \end{align*}
    Finally, the application of \cref{lem:hoeffding-dependent-tailbound_v1} with $\abs{\hat{Z}_{t+1}} \leq 4/(n-2)$ , $\tau = 7n\ln n$, $\lambda = \sum_{i=1}^\tau \hat{\mu}_i + c'\log n / \sqrt{n}$  and $c' = 15\cdot \sqrt{3}\cdot \ln 2$ yields
    \begin{align*}
        \MoveEqLeft \Prob{\hat{\Psi}_\tau-\hat{\Psi}_0 < -c'\cdot \frac{\log n}{\sqrt{n}}}
        = \Prob{\sum_{i=1}^\tau \hat{Z}_i - \sum_{i=1}^\tau \hat{\mu}_i < -\lambda}
        \leq \exp\left(-\frac{2\lambda^2}{\tau \cdot (\frac{8}{n-2})^2}\right)\\
        & = \exp\left(-\frac{2(\sum_{i=1}^\tau \mu_i + c'\log n / \sqrt{n})^2}{\tau \cdot (\frac{8}{n-2})^2}\right)
        \leq \exp\left(-\frac{2(c'\log n / \sqrt{n})^2}{\tau \cdot (\frac{8}{n-2})^2}\right)\\
        & = \exp\left(-\frac{c'^2 \cdot \log^2 n}{ \frac{224\cdot n^2 \cdot \ln n}{(n-2)^2}}\right)
        = \exp\left(-\frac{c'^2 \cdot \ln n }{225\cdot \ln^2 2}\right) 
        \leq n^{-3}.
    \end{align*}

     Recall that from \cref{lem:phase1} it follows that $T_1 \leq 7\cdot n\ln n $ with probability $1-n^{-3}$. Thus $\Psi_\tau > \Psi_0 - c' \cdot \log n/\sqrt{n}$ with probability at least $1-n^{-3}$.
    Again, since $u(t) < n/2$, we have
    \begin{align*}
        && \frac{x_1(T_1)-x_i(T_1)}{n-u(T_1)} = \Psi_{T_1} & > \Psi_0 - c' \cdot \log n/\sqrt{n} 
        = \frac{x_1(0)-x_i(0)}{n-u(0)} - c' \cdot \log n/\sqrt{n}\\
        &\Leftrightarrow & x_1(T_1)-x_i(T_1) & > (x_1(0)-x_i(0)) \frac{n-u(T_1)}{n-u(0)} - (n-u)(T_1) c' \cdot \log n/\sqrt{n}\\
        &\implies & x_1(T_1)-x_i(T_1) & > \frac{x_1(0)-x_i(0)}{2} - c' \sqrt{n} \log n
    \end{align*}
    Due to the assumption $x_1(0)-x_i(0) \geq c\sqrt{n}\log n $ with $c=6c'$ we have $x_1(T_1)-x_i(T_1) > (x_1(0)-x_i(0))/3$.
    By application of the union bound over all opinions $i \neq 1$  the first statement holds with probability at least $1-n^{-2}$.

\medskip

    For the second statement we will show that \whp $x_1(T_1) \leq 2 x_1(0)$. Then the third statement can be shown as follows. Fix an Opinion $i$ with $x_1(0)/x_i(0) = 1 + \varepsilon$ for some $\varepsilon = \Omega(1)$.
    Since $x_1 \geq (n-u)/k$ and $k\le c \cdot \sqrt{n}/\log^2(n)$ we have $x_1 = \Omega(\sqrt{n} \log^2 n)$ and therefore $x_1(0)-x_i(0) = \Omega(\sqrt{n} \log n)$.
    From the second statement, we have that $x_i(T_1) \leq x_1(T_1)-(x_1(0)-x_i(0))/3$.
    Thus, 
    \begin{align*}
        \frac{x_1(T_1)}{x_i(T_1)} 
        & \geq \frac{x_1(T_1)}{x_1(T_1)-(x_1(0)-x_i(0))/3}
        = \frac{1}{1-\frac{x_1(0)-x_i(0)}{3x_1(T_1)}}
        = \frac{1}{1- \frac{x_1(0)/x_i(0) - 1}{3x_1(T_1)/x_i(0)}}\\
        & = \frac{1}{1 - \frac{\varepsilon x_i(0)}{3x_1(T_1)}}
        \stackrel{*}{\geq} \frac{1}{1-\frac{\varepsilon}{6(1+\varepsilon)}}
        = 1 + \frac{\varepsilon}{6+5\cdot \varepsilon}
    \end{align*}
    where for * we used that $x_1(T_1) \leq 2 x_1(0)$.
    The third statement then follows from the union bound over all opinions $i \neq 1$.

    It remains to show that \whp $x_1(T_1) \leq 2 x_1(0)$.
    We first show that it is more likely for $x_1$ to decrease than to increase.
    Recall the probabilities of Opinion $1$ to increase or decrease from 
    \cref{obs:prob_opinion_single_interaction}.
    Then,
    \begin{align*}
        \MoveEqLeft \Prob{X_1(t+1) = x_1+1~|~\mathbf{X}(t) = \mathbf{x}, \mathbf{X}(t+1) \neq \mathbf{x}}
        = \frac{u \cdot x_1 \cdot n^{-2}}{u \cdot x_1 \cdot n^{-2} + x_1(n-u-x_1)\cdot n^{-2}}\\
        & = \frac{1}{2} - \frac{u \cdot x_1 - x_1(n-u-x_1) }{2\left(u \cdot x_1 + x_1(n-u-x_1)\right)}
        = \frac{1}{2} - \frac{n-2u-x_1}{2(n-x_1)}
        \leq \frac{1}{2} - \frac{n-2u-x_{\max}}{2(n-x_1)}.
    \end{align*}
    Before the end of the first phase we have $2u(t) < n - x_{\max}(t)$, so
    we can bound the probability to be less than $1/2$ for all $t < T_1$.

    We relate $x_1(t)$ to the fair non-lazy random walk $W(0), W(1), \ldots, W(T_1)$ on $\mathbb{N}$ starting at $W(0) = 0$.
    It follows 
    that \whp $W(t) < x_1(0)$ for all $t \leq 7 n \ln n$.
    It follows from \cref{lem:phase1} that \whp $T_1 \leq 7 n \ln n$.
    By the union bound, \whp there exists a $t < 7 n \ln n$ such that $t = T_1$ and $x_1(t) \leq W(t) + x_1(0) < 2 \cdot x_1(0)$ which completes the proof.

    The third statement is a direct consequence of the first statement.
    To see this, consider $n$ agents with $k$ opinions and let $\mathbf{x}$ be a vector of size $k+1$ that denotes the configuration ($x_{k+1}$ is the number of undecided agents).
    Let $\mathbf{y}$ be a vector of size $k+2$ where $y_{k+2} = x_{k+1}$, $y_{k+1} = 0$ and otherwise $y_i = x_i$.
    The process does not depend on whether we use $\mathbf{x}$ or $\mathbf{y}$ to describe the configuration and thus $y_{k+1}(t) = 0$ for all $t \geq 0$.
    Therefore, any statement that holds for $\mathbf{x}$ also holds for $\mathbf{y}$ (using $k+1$ instead of $k$).
    The second statement assumes that $y_1(0) = x_1(0) = \Omega(\sqrt{n}\log^2 n)$.
    Hence, we can apply the first statement to $\mathbf{y}$ and get $x_1(T_1) = y_1(T_1)-0 = y_1(T_1) - y_{k+1}(T_1) \geq (y_1(0)-y_{k+1}(0))/3 = y_1(0)/3 = x_1(0)/3$.
\end{proof}

\lemmaPhaseOneUndecidedGeneralBounds*
\begin{proof}
In order to show the lemma we define the threshold $u^* = n \cdot (k-1)/(2k-1)$ and prove that $u(t) \leq u^* + 6 \cdot \sqrt{n \log n}$.
Then at the end of the proof we show how the lemma statement follows from this.
We model the number of undecided agents over time $t$ as a random walk $U(t)$ with state space $\set{0, \dots, n-1}$ and denote the corresponding non-lazy random walk $Z(r)$.
The transition probabilities of $Z(r)$ are denoted as
    \begin{align*}
    \Prob{Z(r+1)=z(r)+1~|~Z(r)=z(r)} &=\tilde{p}^+(r) \\
    \Prob{Z(r+1)=z(r)-1~|~Z(r)=z(r)} &=1 - \tilde{p}^+(r).\lefttag{\text{and}}
    \end{align*}

    Unfortunately the transition probabilities of $Z(r)$ depend on the configuration at time $r$ and thus the random walk is not time-homogeneous.
    However, we can bound the transition probabilities as follows.
    Let $\epsilon = 3 \cdot \sqrt{\log n/n}$.
    From a lengthy but straightforward calculation (see \cref{apx:transition-probabilities}) it follows that $\tilde{p}(r)^+ \leq 1/2-\epsilon/2$ if $Z(r) \geq u^* + \epsilon \cdot n$.

    In order to bound the probability that at some time $r \in [n^3]$ we have $U(r) \geq u^*+2\epsilon \cdot n$ it is sufficient to bound the probability that at some time $r \in [n^3]$ we have $Z(r) \geq u^* + 2\epsilon \cdot n$.
    For $Z(r)$ we know that we have a drift ``in the right direction'' between $u^* + \epsilon \cdot n$ and $u^* + 2 \epsilon \cdot n$.
    It therefore suffices to bound the probability that a random walk
    traverses from $u^* + \epsilon \cdot n$ to $u^* + 2 \epsilon \cdot n$.
    To do so we define a random walk $W(r)$ on the non-negative integers with a reflecting barrier at $0$ and otherwise transition probabilities
    \begin{align*}
        \Prob{W(r+1)=w(r)+1~|~W(r)=w(r)} &= p = 1/2 - \epsilon/2\\
        \Prob{W(r+1)=w(r)-1~|~W(r)=w(r)} &= q = 1/2 + \epsilon/2 . \lefttag{\text{and}}
    \end{align*}

    To show the statement we now define $Z'(r) = Z(r)- u^* + \epsilon \cdot n$ and couple $Z'(r)$ with $W(r)$.
    From the definitions of the random walks $Z'(r)$ and $W(r)$ the following two statements follow.
    If $Z'(r) < 0$, then $Z'(r+1) \leq W(r+1)$ since $W(r)$ has a reflecting barrier at $0$.
    Otherwise, $Z'(r+1) \leq W(r+1)$ follows from the coupling between $Z'(r)$ and $W(r)$ since $\Prob{Z'(r+1) = z(r) + 1 ~|~ Z'(r) = z(r)} \leq \Prob{W(r+1) = w(r) + 1 ~|~ W(r) = w(r)}$ for any $z(r)$ and $w(r)$.
    It therefore follows that, deterministically, $Z(r) \leq u^* + \epsilon \cdot n + W(r)$.

    We now proceed to prove that $W(r) \leq \epsilon \cdot n$ \whp.
    It is straightforward to verify that the stationary distribution $W$ of $W(r)$ is given by $\Prob{W=n} = (p/q)^n \cdot (1 - p/q)$.
    Therefore $\Prob{W\geq n} = (p/q)^n$.
    When we start with $W(0) = 0$, it holds 
    from a union bound over $n^3$ steps that $\Prob{\exists t \in [n^3] \colon W(t)\geq m} \leq n^3 \cdot \Prob{W\geq m} \leq n^3 \cdot (p/q)^m$ for some value $m > 0$.
    Setting $\epsilon = 3 \cdot \sqrt{\log n/n}$ in $p$ and $q$ and plugging in $m = \epsilon \cdot n = 3n \cdot \sqrt{\log n/n}$ yields
    \begin{align*}
    \Prob{\exists t \in [n^3] : W(t)\geq 3n \cdot \sqrt{\log n/n}} \numberthis\label{eq:dcm-rw-1}
        &\leq n^{3} \cdot \left( \frac{1/2-\epsilon/2}{ 1/2 + \epsilon/2} \right)^{\epsilon \cdot n}
        \leq n^{3} \cdot (1-\epsilon)^{\epsilon \cdot n}
        < n^{-3}.
    \end{align*}

It remains to show that the lemma statement follows out of this bound.
We remark that $u^* = n\cdot (k-1)/(2k-1)$, which is monotonically increasing in $k$ (for $k \geq 1$). With the upper bound of $k \leq c \cdot \sqrt{n}/\log^2(n)$, we have 
    \begin{align*}
        u^* + 3 \cdot \sqrt{n \log n} 
        & = n \cdot \frac{k-1}{2k-1} + 3 \cdot \sqrt{n \log n}\\
        & \leq n \cdot \frac{c \cdot \sqrt{n}/\log^2(n) - 1}{2 c \cdot  \sqrt{n}/\log^2(n) - 1} + 3 \cdot \sqrt{n \log n}\\
        & = \frac{n}{2} - \sqrt{n} \log(n) \cdot \left( \frac{1}{4c/\log(n)-2\log(n)/\sqrt{n}} - \frac{3}{\sqrt{\log n}}\right).
    \end{align*}
    The last expression in parentheses is larger than $1/(5c)$ for sufficiently large $n$, which proves the second inequality for $u(t)$.
\end{proof}

\subsection{Omitted Proofs of Section 4 (Phase 2)}
\label{apx:omitted-proofs-phase2}
\lemmaPhaseTwoUndecidedLowerBound*
\begin{proof}
    Recall that we defined $Z(t) = n - 2 u(t) - x_{\max}(t)$ and that $Z(T_1) \leq 0$.
    In the following we show that \whp\ $Z(t) \leq c \sqrt{n \cdot \log n}$ for all $T_1\leq t \leq n^3$.
    
    We follow the proof idea of Theorem 6 in \cite{lengler2018drift}.
    We define a new set of random variables with $Y(t) = \exp(\eta \cdot Z(t))$ for $t \geq T_1$ and $\eta=\sqrt{\ln n/n}$
    and let $z_0 = 4\eta \cdot n$. 
    
    Fix an arbitrary $i \geq 0$.
    We first give a bound for $\Ex{Y(i+1)-Y(i) ~|~ Z(i) = z}$.
    Note that $Z(i+1)-Z(i) \in [-2,2]$.
    We get
    \begin{align*}
        \Ex{Y(i+1)-Y(i) ~|~ Z(i) = z}
        & = \Ex{e^{\eta \cdot Z(i+1)} - e^{\eta \cdot Z(t)} ~|~ Z(i) = z}\\
        & = e^{\eta \cdot z} \cdot \Ex{e^{\eta\cdot(Z(i+1)-z)}-1 ~|~ Z(i) = z}\\
        & = e^{\eta \cdot z} \cdot \sum_{j\in [-2,2]} (e^{\eta \cdot j} -1) \cdot \Prob{Z(i+1)-z=j ~|~ Z(i) = z}\\
    \end{align*}

    We derive the following bound for $\exp(\eta \cdot j)-1$.
    Since $\exp(x) \leq 1+x+x^2$ for $x \leq 1$ and $\eta \rightarrow 0$ for large $n$, we have
    $\exp(2\eta) \leq 1+2\eta+(2\eta)^2 = 1+2\eta+\eta \cdot z_0/n$.
    For $j \in [-2,2]$, we thus have $\exp(\eta j)-1 \leq \eta j + \eta \cdot z_0/n$.
    In \cref{lem:phase1} we calculated $\Ex{Z(i+1)-Z(i) ~|~ Z(i) = z} \leq -\frac{z}{n}$.

    Thus, for all $z \geq z_0$ we have    
    \begin{align*}
        \MoveEqLeft \Ex{Y(i+1)-Y(i) ~|~ Z(i) = z}\\
        & \leq e^{\eta \cdot z} \cdot \sum_{j\in [-2,2]} (\eta \cdot j + \eta \cdot z_0/n) \cdot \Prob{Z(i+1)-z=j ~|~ Z(i) = z}\\
        & =  e^{\eta \cdot z} \cdot \eta \cdot (\Ex{Z(i+1)-Z(i) ~|~ Z(i) = z} + z_0/n) \leq 0.
    \end{align*}
    
    In total, we get 
    \begin{align*}
        \Ex{Y(t)}
        = \Ex{Y(0)} + \sum_{i=0}^{t-1} \Ex{Y(i+1)-Y(i)}
        \leq 1.
    \end{align*}

    
    Since $\forall t \geq 0: Y(t) \geq 0$, we can apply Markov's inequality.
    Thus,
    \begin{align*}
        \Prob{Z(t) \geq 2z_0} = \Prob{Y(t) \geq \exp(2\eta z_0)} \leq \frac{\Ex{Y(t)}}{n^{8}} \leq n^{-8}.
    \end{align*}
    Finally, we apply the union bound over $n^3-T_1 \leq n^3$ interactions.
\end{proof}

\lemmaPhaseTwoMaxNoShrinking*
\begin{proof}
    
    We show that $x_{\max}(t) \geq x_{\max}(T_1)/2$ for all $t\in [T_1,T_1+ cn^2 \log n /x_{\max}(T_1)]$.
    First we bound the number of productive interactions  w.r.t.\ $x_{\max}(t)$ within $cn^2 \log n /x_{\max}(T_1)$ interactions and then we bound its affect on the support of the largest opinion. 
    Let 
    \begin{align*}
        \hat{T} = \inf\set{t\geq T_1 ~|~ u(t) \notin [(n-x_{\max}(t'))/2-8\cdot\sqrt{n\ln n}, n/2]}
    \end{align*}
    be a stopping time and let $(\hat{X}(t))_t$ denote the process with $\hat{X}(t) = X(t)$ for all $t\leq \hat{T}$ and $\hat{X}(t) = X(\hat{T})$ for $t>\hat{T}$.
    From  \cref{lem:undecided_general_bounds_v1} and  \cref{lem:undecided>n/2-x1/2-o(n)}  it follows $\hat{T}-t =\Omega(n^2/x_{\max}(t) \cdot \log n)$ \whp.
    Thus, $(\mathbf{X}(t))_t$ and $(\hat{\mathbf{X}}(t))_t$ behave the same between time $t$ and $t + O(n^2/x_{\max}(T_1)\cdot \log n) $.

    As long as $\hat{x}_{\max}(t) \leq 2 \cdot \hat{x}_{\max}(T_1)$  an interaction is productive w.r.t.\ $x_{\max}(t)$ with probability
    \begin{align*}
        \frac{\hat{u} \cdot \hat{x}_{\max} + \hat{x}_{max} \cdot (n-\hat{u}-\hat{x}_{\max})}{n^2}
        = \frac{\hat{x}_{\max} \cdot (n-\hat{x}_{\max})}{n^2}
        \leq 2 \cdot \frac{\hat{x}(T_1)}{n}
    \end{align*}
    It follows from an application of Chernoff bounds that within a sequence of $c\cdot n^2\cdot \log n/x_{\max}(T_1)$ interactions the number of such productive interactions 
    is at most  $4\cdot c \cdot n\log n$ with probability at least $1-n^{-10}$.
    Now consider $\tau = 4\cdot c\cdot n\log n$ such productive interactions and let $Z_t$ denote the change w.r.t.\ $\hat{x}_{\max}(t)$, i.e., the support of the largest opinion increase or decrease by one, respectively.
    That is, assuming the next interaction is such a productive interaction for $\hat{x}(t)$ we have
    \begin{align*}
        &\Prob{Z_t = 1} 
        = \frac{\hat{u} \cdot \hat{x}_{\max}}{ \hat{u}\cdot \hat{x}_{\max} + \hat{x}_{\max} \cdot (n-\hat{u}-\hat{x}_{\max})}
        = \frac{\hat{u}  }{ (n-\hat{x}_{\max})} \\
        &\Prob{Z_t = -1}
        = 1- \Prob{Z_t = 1}
    \end{align*}
    Therefore
    \begin{align*}
        \Ex{Z_t} 
        = \frac{\hat{u}-(n-\hat{u}-\hat{x}_{\max})}{n-\hat{x}_{\max}}
        = \frac{2\cdot \hat{u} + \hat{x}_{\max} - n}{n-\hat{x}_{\max}}
        \geq -48 \cdot \frac{\sqrt{n\ln n}}{n}
    \end{align*}
    Let $Z$ be the sum of  $Z_t$ for all $t\in [1,\tau]$. 
    Then it follows from Hoeffding bound with $\lambda = \hat{x}_{\max}(T_1)/2 - 200 \cdot \sqrt{n}\ln^{3/2} n$
    \begin{align*}
        \Prob{Z < -\frac{1}{2} \cdot \hat{x}_{\max}(T_1)} 
        \leq \Prob{Z < \Ex{Z}-\lambda}
        \leq e^{-\frac{2\lambda^2}{4\tau}}
        \leq n^{-10}
    \end{align*}
    Note that if (ever) $\hat{x}_{\max}(t') > 2\cdot \hat{x}_{\max}(T_1)$ for some $t' \in [T_1,T_1+T]$ the statement hold as well by the union bound and the previous part.
    Thus, starting with $\hat{x}_{\max}(T_1)$  throughout the next $ c\cdot n^2/x_{\max}(T_1)\cdot \log n $ interactions $\hat{x}_{\max}(t) \geq \hat{x}_{\max}(T_1)/2$ with probability at least $1-n^{-5}$. 
\end{proof}

\LemmaSmallOpinionsDoNotGrow*
\begin{proof}
    Let 
    \begin{align*}
        \hat{T} = \inf\set{t\geq t_0 ~|~ u(t) \notin [(n-x_{\max}(t))/2-8\cdot\sqrt{n\ln n}, n/2] }
    \end{align*}
    be a stopping time and let $(\hat{X}(t))_t$ denote the process with $\hat{X}(t) = X(t)$ for all $t\leq \hat{T}$ and $\hat{X}(t) = X(\hat{T})$ for $t>\hat{T}$.
    From  \cref{lem:undecided_general_bounds_v1} and \cref{lem:undecided>n/2-x1/2-o(n)}  it follows $\hat{T} \geq T_1+T$ \whp.
    Thus, $(\mathbf{X}(t))_t$ and $(\hat{\mathbf{X}}(t))_t$ behave the same in the time interval $[t_0,T_1+T]$.

    Now we start with the first statement.
    First we bound the number of $i$-productive interactions in the interval $[t_0,T_1+T]$.
    Recall that only $i$-productive interactions change the support of Opinion $i$.
    As long as $\hat{x}_{i}(t) \leq 40\sqrt{n\log n}$ for $t\in [t_0,T_1+T]$  an interaction is $i$-productive  with probability
    \begin{align*}
        \frac{\hat{u} \cdot \hat{x}_{i} + \hat{x}_{i} \cdot (n-\hat{u}-\hat{x}_{i})}{n^2}
        = \frac{\hat{x}_{i} \cdot (n-\hat{x}_{i})}{n^2}
        \leq   \frac{40\sqrt{n\log n}}{n}
    \end{align*}
    It follows from an application of Chernoff bounds that  the number of such productive interactions 
    is at most  $ n/\log^{1/4} n$ \whp.

    Now consider $\tau = n/\log^{1/4} n$ $i$-productive interactions and let $Z_\ell$ denote the respective change of the $\ell$th $i$-productive interaction.
    That is, 
    \begin{align*}
        &\Prob{Z_\ell = 1} 
        = \frac{\hat{u} \cdot \hat{x}_{i}}{ \hat{u}\cdot \hat{x}_{i} + \hat{x}_{i} \cdot (n-\hat{u}-\hat{x}_{i})}
        = \frac{\hat{u}  }{ (n-\hat{x}_{i})} \\
        &\Prob{Z_\ell = -1}
        = 1- \Prob{Z_\ell = 1}
    \end{align*}
    Therefore
    \begin{align*}
        \Ex{Z_\ell} 
        = \frac{\hat{u}-(n-\hat{u}-\hat{x}_{i})}{n-\hat{x}_{i}}
        = \frac{2\cdot \hat{u} + \hat{x}_{i} - n}{n-\hat{x}_{i}}
        \leq \frac{40\sqrt{n\log n}}{n-40\sqrt{n\log n}}
        \leq \frac{42\sqrt{n\log n}}{n}
    \end{align*}
    Let $Z$ be the sum of  $Z_\ell$ for all $\ell\in [1,\tau]$. 
    Then it follows from Hoeffding bound with $\lambda = \sqrt{n\log n}-\Ex{Z}$
    \begin{align*}
        \Prob{Z > \sqrt{n\log n}} 
        \leq \Prob{Z > \Ex{Z}-\lambda}
        \leq e^{-\frac{2\lambda^2}{4\tau}}
        \leq n^{-6}
    \end{align*}
    Thus, starting with $\hat{x}_{i}(t_0) \leq 20\sqrt{n\log n}$ it holds that $\hat{x}_i(t) \leq 40\sqrt{n\log n}$ for all $t\in [t_0,T_1+\tau]$ \whp. 

\bigskip
We now show the second statement.

    Our proof follows the analysis of the classical Gambler's ruin problem on the quantity $ \hat{x}_{max}(t)-\hat{x}_j(t)$. 
    That is, for $T$ interactions we show that  $x_{\max}(t)-x_j(t) \geq (x_{\max}(t_0)-x_j(t_0))/2$ for all $t \in [t_0,T_1+T]$.
    Let $B(t) = \set{i ~|~ \hat{x}_i(t) \geq \hat{x}_{max}(t)}$ denotes the set of all opinions $i$ with maximum support at time $t$.
    Note that $\abs{B(t)} \geq 1$ for all $t$.
    Consider an arbitrary time $t \in [t_0,T_1+T]$.
    By the definition of USD it follows that
    {\dense \small 
    \begin{align}
        \label{eq:nobias-maintainbias-probs}
        &\Prob{\hat{X}_{max}(t+1)-\hat{X}_j(t+1) = \hat{x}_{max}-\hat{x}_j +1 ~|~ \hat{\mathbf{X}}(t) = \hat{\mathbf{x}}(t)}  
        = \begin{cases}
            \frac{\hat{u}\cdot \sum_{i \in B} \hat{x}_i + \hat{x}_j\cdot (n-\hat{u}-\hat{x}_j)}{n^2} &, \abs{B} > 1 \\
            \frac{\hat{u} \cdot \hat{x}_{max} + \hat{x}_j\cdot (n-\hat{u}-\hat{x}_j)}{n^2} &, \abs{B} = 1
        \end{cases} \\
        &\Prob{\hat{X}_{max}(t+1)-\hat{X}_j(t+1) = \hat{x}_{max}-\hat{x}_j-1 ~|~ \hat{\mathbf{X}}(t) = \hat{\mathbf{x}}(t)}
        = \begin{cases}
            \frac{\hat{u}\cdot \hat{x}_j}{n^2} &, \abs{B} > 1 \\
            \frac{\hat{x}_{max} \cdot (n-\hat{u}-\hat{x}_{max}) + \hat{u}\cdot \hat{x}_j}{n^2} &, \abs{B} = 1
        \end{cases} \nonumber 
    \end{align}
    }
    Let $p_1(t)$ and $p_2(t)$ denote the first and second probability from  \cref{eq:nobias-maintainbias-probs}, respectively.
    Now  assuming for $\hat{\mathbf{x}}(t)$  the next interaction is productive w.r.t.\ $\hat{x}_{max}(t)-\hat{x}_j(t)$ then we have
    \begin{align*}
        &\Prob{\hat{X}_{max}(t+1)-\hat{X}_j(t+1) = \hat{x}_{max}-\hat{x}_j +1 ~|~ \hat{\mathbf{X}}(t) = \hat{\mathbf{x}}(t)}  
        = \frac{p_1(t)}{p_1(t) + p_2(t)} \\
    \end{align*}

    We consider two cases.
    In the first case assume $x_j(t) \geq 20 \cdot \sqrt{n\ln n}$ for all $t \in [t_0,T_1+T]$.
    Then for $p_1'(t) = (\hat{u}(t) \cdot \hat{x}_{max}(t) + \hat{x}_j(t')\cdot (n-\hat{u}(t)-\hat{x}_j(t)))\cdot n^{-2}$ and $p_2'(t) = (\hat{x}_{max}(t) \cdot (n-\hat{u}(t)-\hat{x}_{max}(t)) + \hat{u}(t)\cdot \hat{x}_j(t)) \cdot n^{-2}$ we have
    \begin{align*}
        \frac{p_1}{p_1 + p_2}
        \geq \frac{p_1'}{p_1'+p_2'} 
        &= \frac{1}{2} + \frac{p_1'-p_2'}{2(p_1'+p_2')} \\
        &= \frac{1}{2} + \frac{(2\cdot \hat{u} + \hat{x}_{max}+\hat{x}_j -n) \cdot (\hat{x}_{max}-\hat{x}_j)}{2((\hat{x}_{max}+\hat{x}_j) \cdot n -\hat{x}_{max}^2-\hat{x}_j^2)} \\
        &\geq \frac{1}{2} + \frac{(\hat{x}_j - 16\sqrt{n\ln n}) \cdot (\hat{x}_{\max}-\hat{x}_j)}{2((\hat{x}_{\max}+\hat{x}_j) \cdot n } \\
        &\geq \frac{1}{2} +  \frac{(\hat{x}_{max}-\hat{x}_j)}{25n} \\
    \end{align*}
    Observe that if ever $\hat{x}_{\max}(t')-\hat{x}_{j}(t') \leq (\hat{x}_{\max}(t_0)-\hat{x}_{j}(t_0))/2$ for some $t' \in [t_0,T_1+T]$ the probability to increase $\hat{x}_{\max}(t')-\hat{x}_{j}(t')$ by one is at least $1/2 +  (\hat{x}_{\max}(t_0)-\hat{x}_{j}(t_0))/50n \geq 1/2 + (c' \cdot \sqrt{n}\log (n)) / (50n)$ assuming a productive interaction.
    Finally an application of \cref{lem:random_walk_no_ruin_probability} for $b = (\hat{x}_{\max}(t_0)-\hat{x}_j(t_0))/2 $ and $p = 1/2 + c\sqrt{n}\log n / 50n $ yields that the probability $\hat{x}_{\max}(t')-x_j(t') \geq (\hat{x}_{\max}(t)-\hat{x}_j(t))/2$ for all $t' \in (t,T]$ is ever violated is at most 
    \begin{align*}
        \left(\frac{1-p}{p} \right)^b
        = \left( \frac{25n-2c'\sqrt{n}\log n}{25n+2c'\sqrt{n}\log n} \right)^b
        = \left( 1-\frac{4c'\sqrt{n}\log n}{25n+2c'\sqrt{n}\log n} \right)^b
        \leq n^{-6} .
    \end{align*}
    It remains to show the second case.
    That is,  there exists a time $t\in [t_0,T_1+\tau]$ such that $\hat{x}_j(t) < 20 \sqrt{n\ln n}$.
    From the first statement  it follows that $\hat{x}(t') \leq 40\sqrt{n\log n} $ for all $t' \in [t,T_1+\tau]$.
    Additionally we know that $\hat{x}_{\max}(t') \geq \cdot \sqrt{n}\log n$ and hence, the statement follows by a union bound over both cases \whp.
\end{proof}

\lemmaPhaseTwoDoublingImportant*
 \begin{proof}
    We define 
    \begin{align*}
        \hat{T} = \inf\set{t\geq T_1 ~|~ u(t) \notin \left[\frac{n-x_{\max}(t)}{2}-8\cdot \sqrt{n\ln n}, \frac {n}{2}\right] \mbox{ or } x_{\max}(t)<x_{\max}(T_1)/3}
    \end{align*}
    as a stopping time and $(\hat{\mathbf{X}})_{t\ge T_1}$ as the process with $\hat{\mathbf{X}}(t) = \mathbf{X}(t)$ for all $t\leq \hat{T}$ and $\hat{\mathbf{X}}(t) = \mathbf{X}(\hat{T})$ for $t>\hat{T}$.
    From  \cref{lem:undecided>n/2-x1/2-o(n)} it follows that $u(t) \geq (n-x_{\max}(t))/2-8\cdot \sqrt{n\ln n} $ for all $t \in [T_1,n^3]$, \whp.
    From \cref{lem:undecided_general_bounds_v1} it follows that $ u(t) \leq n/2$  for all $t \in [T_1,n^3]$, \whp.
    Finally, \cref{lem:phase2-max-no-shrinking} gives us that $ x_{\max}(t) \geq x_{\max}(T_1)/3$ for all $t \in [T_1,T_1+ cn^2 \log n / x_{\max}(T_1)]$, \whp.
    Thus, $\hat{T}-T_1 =\Omega(n^2\cdot \log n/x_{\max}(T_1) )$ \whp and we can assume that $(\mathbf{X})_{t\ge T_1}$ and $(\hat{\mathbf{X}})_{t\ge T_1}$ are identical for $t \in [T_1,T_1 + O(n^2\cdot \log n/x_{\max}(T_1))]$.
    We consider a pair of opinions $i$ and $j$ which are important at time $t_0$ and track the evolution  of the difference between the support of $i$ and $j$ within the next $ T$ interactions.
    First we  bound the number of $(i,j)$-productive interactions in the time  interval $[t_0,t_0+T]$ interactions.
    Recall that only $(i,j)$-productive interactions change the support of Opinion $i$ or Opinion $j$.
    An interaction is $(i,j)$-productive with probability
    \begin{align*}
           \MoveEqLeft\frac{\hat{u}(t)\cdot \hat{x}_i(t) + \hat{x}_i(t) \cdot (n-\hat{u}(t) -\hat{x}_i(t)) + \hat{u}(t)\cdot \hat{x}_j(t) + \hat{x}_j(t) \cdot (n-\hat{u}(t)-\hat{x}_j(t))}{n^2} \\
            &= \frac{(\hat{x}_i(t)+\hat{x}_j(t))\cdot n - \hat{x}_i(t)^2-\hat{x}_j(t)^2}{n^2} \\
            &= \frac{(\hat{x}_i(t)+\hat{x}_j(t))\cdot (n+\hat{x}_i(t)-\hat{x}_j(t)) - 2\hat{x}_i(t)^2}{n^2} \\
            &\geq \frac{(2\hat{x}_{\max}-8\alpha\sqrt{n}\log n) \cdot n -2\hat{x}_{\max}^2}{n^2}
            \geq \frac{2\hat{x}_{\max}}{n} \cdot \left(1-\frac{\hat{x}_{\max}}{n} - \frac{8\alpha\sqrt{n}\log n}{2\hat{x}_{\max}}\right) 
            \\ &\geq \frac{\hat{x}_{\max}(T_1)}{2n}
    \end{align*}
    In the first inequality we used that $i$ and $j$ are both important. 
    Additionally we use $\hat{x}_{max}(t)\geq \hat{x}_{max}(T_1)/2$ and  $\hat{x}_{max}(t) \leq 2n/3$.
    Thus, assuming $i$ and $j$ both remain important during the whole time interval an application of Chernoff bounds provides at least $ 16\cdot n$ many $(i,j)$-productive interactions in $[t_0, t_0+T]$, \whp.
    For $1\le i\le  16\cdot n $ we define $t_i$ as the $i$th $(i,j)$-productive interaction in $[t_0,t_0+T]$.

    Recall that only $(i,j)$-productive interactions change the quantity $x_i(t)-x_j(t)$ but other interactions may change the remainder of the configuration, e.g.,\ an additional undecided agent is created.
    If an interaction  is not $(i,j)$-productive then $\Prob{\hat{X}_i(t+1)-\hat{X}_j(t+1) \neq \hat{x}_i-\hat{x}_j ~|~ \hat{\mathbf{X}}(t) = \hat{\mathbf{x}}}$ = 0.
    If an interaction  is $(i,j)$-productive then 
    \begin{align*}
        \MoveEqLeft\Prob{\hat{X}_i(t+1)-\hat{X}_j(t+1) = \hat{x}_i-\hat{x}_j+1 ~|~ \hat{\mathbf{X}}(t) = \hat{\mathbf{x}}} \\ 
        &= \frac{1}{2} + \frac{\hat{u}\cdot \hat{x}_i + \hat{x}_j\cdot(n-\hat{u}-\hat{x}_j) - (\hat{u}\cdot \hat{x}_j + \hat{x}_i\cdot(n-\hat{u}-\hat{x}_i)) }{2(\hat{u}\cdot \hat{x}_i + \hat{x}_j\cdot(n-\hat{u}-\hat{x}_j) + (\hat{u}\cdot \hat{x}_j + \hat{x}_i\cdot(n-\hat{u}-\hat{x}_i)))} \\
        &= \frac{1}{2} + \frac{(\hat{x}_i + \hat{x}_j + 2\hat{u} - n)\cdot (\hat{x}_i - \hat{x}_j)}{2((\hat{x}_i + \hat{x}_j) \cdot n - \hat{x}_i^2+\hat{x}_j^2)}
    \end{align*}
    Now we consider two cases.
    In the first case assume $\hat{x}_i(t_0)-\hat{x}_j(t_0)  < 4\alpha\sqrt{n}$.
    W.l.o.g.\ we assume for the rest of the proof that $\hat{x}_i(t)\geq \hat{x}_j(t)$ (otherwise we simply switch the roles of $i$ and $j$).
    We consider an (arbitrary) $(i,j)$-productive interaction $t_i$ and refine the probability from above in the following way
    \begin{align*}
        \MoveEqLeft\Prob{\hat{X}_i(t_i+1)-\hat{X}_j(t_i+1) = \hat{x}_i-\hat{x}_j+1 ~|~ \hat{\mathbf{X}}(t_i) = \hat{\mathbf{x}}} \\
        &=\frac{1}{2} + \frac{(\hat{x}_i + \hat{x}_j + 2\hat{u} - n)\cdot (\hat{x}_i - \hat{x}_j)}{2((\hat{x}_i + \hat{x}_j) \cdot n - (\hat{x}_i^2+\hat{x}_j^2))} \\
        &\geq \frac{1}{2} + \frac{(2\hat{x}_{max} - 8\alpha\sqrt{n}\log n + 2\hat{u}-n)\cdot (\hat{x}_i-\hat{x}_j)}{2((\hat{x}_i + \hat{x}_j) \cdot n - (\hat{x}_i^2+\hat{x}_j^2))} \\
        &\geq \frac{1}{2} + \frac{(\hat{x}_{max} - 8\alpha\sqrt{n}\log n -16\sqrt{n\ln n})\cdot (\hat{x}_i-\hat{x}_j)}{2((\hat{x}_i + \hat{x}_j) \cdot n - (\hat{x}_i^2+\hat{x}_j^2))} \\
        &\geq \frac{1}{2}
    \end{align*}
    where we use that $\hat{x}_i(t_i),\hat{x}_j(t_i) \geq \hat{x}_{max}(t_i)-4\alpha\sqrt{n}\log n$ and $\hat{x}_{max}(t_i) \geq 8\alpha\sqrt{n}\log n -16\sqrt{n\ln n}$.
    Thus, the evolution of $\hat{x}_i(t)-\hat{x}_j(t)$ over a sequence of $16\cdot n$ many $(i,j)$-productive interactions can be viewed as tossing biased coins with success probability larger than $1/2$ via standard coupling argument between biased and fair coins.
    Applying \cref{lemma:reverse_chernoff_v2} with $\delta = \alpha/(2\sqrt{n})$  yields
    \begin{align*}
        \Prob{\BinDistr(16\cdot n,1/2) \geq \frac{n}{8} + 4\alpha\sqrt{n}}
        \geq e^{-9\delta^2 \cdot  8\cdot n} 
        = e^{-\frac{9 \cdot \alpha^2}{32}}
    \end{align*}
    and hence, the first statement follows by the union bound with the high probability events from above.

    In the second case we assume $\hat{x}_i(t)-\hat{x}_j(t)  \geq 4\alpha\sqrt{n}$.
    Similar to the first case we refine the probability from above assuming a $(i,j)$-productive interaction occur
    \begin{align*}
        \MoveEqLeft\Prob{\hat{X}_i(t_i+1)-\hat{X}_j(t_i+1) = \hat{x}_i-\hat{x}_j+1 ~|~ \hat{\mathbf{X}}(t_i) = \hat{\mathbf{x}}} \\
        &=\frac{1}{2} + \frac{(\hat{x}_i + \hat{x}_j + 2\hat{u} - n)\cdot (\hat{x}_i - \hat{x}_j)}{2((\hat{x}_i + \hat{x}_j) \cdot n - (\hat{x}_i^2+\hat{x}_j^2))} \\
        &\geq \frac{1}{2} + \frac{(2\hat{x}_{max} - 8\alpha\sqrt{n}\log n + 2\hat{u}-n)\cdot (\hat{x}_i-\hat{x}_j)}{2((\hat{x}_i + \hat{x}_j) \cdot n - (\hat{x}_i^2+\hat{x}_j^2))} \\
        &\geq \frac{1}{2} + \frac{(\hat{x}_{max} - 8\alpha\sqrt{n}\log n -8\cdot \sqrt{n\ln n})\cdot (\hat{x}_i-\hat{x}_j)}{4 \cdot \hat{x}_{max} \cdot n )} \\
        &\geq\frac{1}{2} + \left(1-\frac{4\alpha\cdot \sqrt{n}\log n + 16\cdot \sqrt{n\ln n}}{x_{\max}(t_0)}\right)\cdot \frac{\hat{x}_i-\hat{x}_j}{4n} \\
        &\geq \frac{1}{2} +\frac{\hat{x}_i-\hat{x}_j}{12n} 
    \end{align*}
    Thus,  the quantity $\hat{x}_i(t_i)-\hat{x}(t_i)_j$ increases by $1$ with probability at least $1/2 + (\hat{x}_i(t)-\hat{x}(t)_j)/(12n) $ and decreases by $1$, otherwise.
    Observe that starting at time $t_0$ with $\Delta = \hat{x}_i(t_0)-\hat{x}_j(t_0)$ as long as $\hat{x}_i(t_i)-\hat{x}_j(t_i) \geq (3/4) \cdot \Delta$ for the first $i\leq 16\cdot n$ $(i,j)$-productive interactions in $[t_0,t_0+T]$
    the evolution of $\hat{x}_i(t)-\hat{x}_j(t)$ can be viewed as a biased random walk on the line starting at $\Delta$ with success probability (i.e., ''right step'') $p =\frac{1}{2} +\frac{\Delta}{16n}  $.

    Let $T_{min} = \inf\{ t' \in [t_1,t_{c_3\cdot n}] ~|~ \hat{x}_i(t')-\hat{x}(t')_j = (3/4)\cdot(x_i(t_0)-x_j(t_0))\}$ and $T_{max} = \inf\{ t'\geq [t_1,t_{16\cdot n}] ~|~ \hat{x}_i(t')-\hat{x}(t')_j = 2(\hat{x}_i(t_0)-\hat{x}_j(t_0))\}$.

    First we bound $\Prob{T_{max} > T_{min}}$.
    It follows from \cref{lem:random_walk_no_ruin_probability} the probability of ever having an excess of $\Delta/8$ ''left steps'' to ''right steps'' is at most 
    \begin{align*}
        \left(\frac{1-p}{p}\right)^{\Delta/8}
        = \left( \frac{8n-\Delta}{8n+\Delta} \right)^{\Delta/8}
        = \left(1-\frac{2\Delta}{8n+\Delta}\right)^{\Delta/8}
        \leq e^{-\frac{\Delta^2}{4\cdot (8n+\Delta)}}
    \end{align*}
    Next we bound $\Prob{T_{max} > 16 \cdot n}$.
    Again we use the assumption $\hat{x}_i(t_i)-\hat{x}_j(t_i) \geq (3/4) \cdot \Delta$. 
    Now consider  $\tau = 16\cdot n$ independent Poisson trials ($S_i \in \{-1,1\}$ for all $i \leq 16 \cdot n$) each with success probability $p= 1/2+\Delta/16n$.
    Let $S = \sum_{i=1}^{16\cdot n} S_i$.
    Using the Hoeffding bound (\cref{thm:chernoff-hoeffding-bound}) for $\lambda = \Delta$ we get 
    \begin{align*}
        \Prob{T_{max} > 16 \cdot n}
        &\leq \Prob{S < \Delta } \\
        &= \Prob{S-\Ex{S} < \Delta - \Ex{S} }
     \\   &\leq\Prob{\abs{S - \Ex{S}} > \Ex{S} - \Delta}
    \\  &\leq 2\cdot e^{-\frac{2\Delta^2}{4\cdot 16 \cdot n  }}
    \\  &\leq 2\cdot e^{-\frac{\Delta^2}{32 \cdot n  }}
    \end{align*}

    At last we compute the probability of the event $\mathcal{E}$ that if there exists a time $t\in[T_{max},t_0+T]$,i.e.,\ $\hat{x}_i(t) -\hat{x}(t)_j = 2\Delta$, then $ \hat{x}_i(t') -\hat{x'}_j \geq 3/2\cdot \Delta$ for all $t' \in [t,t_0+T]$.
    We can compute this probability in a similar way with \cref{lem:random_walk_no_ruin_probability} as we have shown $\Prob{T_{max} > T_{min}}$.
    In fact we can simply use $\Prob{T_{max} > T_{min}}$ as an upper bound for $\Prob{\bar{\mathcal{E}}}$. 
    
    In order to conclude the second statement we have to show that 
    \begin{align}
    \label{eq:nobias_double_bias_success_prob}
        \Prob{T_{max} \leq c_3 \cdot n \land T_{max} \leq T_{min} \land \mathcal{E}}
        \geq 1-e^{-(\hat{x}_i(t)-\hat{x}_j(t))/\sqrt{n}}
    \end{align}

     it remains to show
    \begin{align*}
     \Prob{T_{max} > 16 \cdot n} + \Prob{T_{max} > T_{min}} + \Prob{\bar{\mathcal{E}}}
     \leq   e^{-(\hat{x}_i(t)-\hat{x}_j(t))/\sqrt{n}}
    \end{align*}
    To do so, recall $\Delta = \hat{x}_i(t)-\hat{x}_j(t) \geq 4\alpha\sqrt{n}$.
    Then starting from the left hand side we have
    {\dense \begin{align*}
        &2\cdot e^{-\frac{\Delta^2}{32n}} + 2\cdot e^{-\frac{\Delta^2}{4\cdot (8n+\Delta)}}
        = 2\cdot \left( e^{-\frac{\Delta^2}{32n}} +  e^{-\frac{\Delta^2}{4\cdot (8n+\Delta)}}\right)
        \leq 2 \cdot \left( e^{-\frac{\Delta^2}{32n}} +  e^{-\frac{\Delta^2}{36n}}\right) \cdot \left( e^{-\frac{\Delta}{\sqrt{n}}} \cdot e^{\frac{\Delta}{\sqrt{n}}}\right) \\
        &\phantom{={}}=2 \cdot \left( e^{\frac{\Delta}{\sqrt{n}} \cdot \left(1-\frac{\Delta}{32\sqrt{n}}\right)} +  e^{\frac{\Delta}{\sqrt{n}} \cdot \left(1-\frac{\Delta}{36\sqrt{n}}\right)}\right) \cdot e^{-\frac{\Delta}{\sqrt{n}}}
        \leq 2 \cdot \left( \frac{1}{10} + \frac{1}{10} \right) \cdot e^{-\frac{\Delta}{\sqrt{n}}}
        \leq \frac{1}{5} \cdot e^{-\frac{\Delta}{\sqrt{n}}}
    \end{align*} }
    where we use that the constant $\alpha$ (from the definition of the additive bias) is sufficiently large.
    Hence, the second statement follows by the union bound with the high probability events from above.
\end{proof}

\subsection{Omitted Proofs of Section 5 (Phase 3)}
\label{apx:omitted-proofs-phase3}

\LemmaPhaseThreeMaxNoShrinking*

\begin{proof}
    Let 
    \begin{align*}
        \hat{T} = \inf\set{t\geq T_2 ~|~ u(t) \notin [(n-x_{\max}(t'))/2-8\cdot\sqrt{n\ln n}, n/2]}
    \end{align*}
    be a stopping time and let $(\hat{X}(t))_t$ denote the process with $\hat{X}(t) = X(t)$ for all $t\leq \hat{T}$ and $\hat{X}(t) = X(\hat{T})$ for $t>\hat{T}$.
    From  \cref{lem:undecided_general_bounds_v1} and  \cref{lem:undecided>n/2-x1/2-o(n)}  it follows $\hat{T}-t =\Omega(n^2/x_{\max}(t) \cdot \log n)$ \whp.
    Thus, $(\mathbf{X}(t))_t$ and $(\hat{\mathbf{X}}(t))_t$ behave the same between time $t$ and $t + O(n^2/x_{\max}(T_2)\cdot \log n) $.
    As long as $\hat{x}_{1}(t') \leq 2 \cdot \hat{x}_{1}(T_2)$  an interaction is $1$-productive  with probability
    \begin{align*}
        \frac{\hat{u} \cdot \hat{x}_{1} + \hat{x}_{1} \cdot (n-\hat{u}-\hat{x}_{1})}{n^2}
        = \frac{\hat{x}_{1} \cdot (n-\hat{x}_{1})}{n^2}
        \leq 2 \cdot \frac{\hat{x}(T_2)}{n}
    \end{align*}
    It follows from an application of Chernoff bounds that within a sequence of $c\cdot n^2\cdot \log n/x_{1}(T_2)$ interactions the number of $1$-productive interactions 
    is at most  $4\cdot c \cdot n\log n$ with probability at least $1-n^{-10}$.
    Now consider $\tau = 4\cdot c \cdot n\log n$ such productive interactions and let $Z_t$ denote the change w.r.t.\ $\hat{x}_{1}(t)$, i.e., the support of the largest opinion increase or decrease by one, respectively.
    That is, assuming the next interaction is a $1$-productive interaction for $\hat{x}(t)$ we have
    \begin{align*}
        &\Prob{Z_t = 1} 
        = \frac{\hat{u} \cdot \hat{x}_{1}}{ \hat{u}\cdot \hat{x}_{1} + \hat{x}_{1} \cdot (n-\hat{u}-\hat{x}_{1})}
        = \frac{\hat{u}  }{ (n-\hat{x}_{1})} \\
        &\Prob{Z_t = -1}
        = 1- \Prob{Z_t = 1}
    \end{align*}
    Therefore
    \begin{align*}
        \Ex{Z_t} 
        = \frac{\hat{u}-(n-\hat{u}-\hat{x}_{1})}{n-\hat{x}_{1}}
        = \frac{2\cdot \hat{u} + \hat{x}_{1} - n}{n-\hat{x}_{1}}
        \geq -48 \cdot \frac{\sqrt{n\ln n}}{n}
    \end{align*}
    Let $Z$ be the sum of  $Z_t$ for all $t\in [1,\tau]$. 
    Then it follows from Hoeffding bound with $\lambda = \hat{x}_{1}(T_2)/2 - 200 \cdot \sqrt{n}\ln^{3/2} n$
    \begin{align*}
        \Prob{Z < -\frac{1}{2} \cdot \hat{x}_{1}(T_1)} 
        \leq \Prob{Z < \Ex{Z}-\lambda}
        \leq e^{-\frac{2\lambda^2}{4\tau}}
        \leq n^{-10}
    \end{align*}

    Note that if (ever) $\hat{x}_1(t') > 2\cdot \hat{x}_1(T_2)$ for some $t'\in [T_2,T_2+T]$ the statement hold  by the union bound and the previous part. 
    Thus, starting with $\hat{x}_{1}(T_2)$  throughout the next $ c\cdot n^2/x_{\max}(T_2)\cdot \log n $ interactions $\hat{x}_{\max}(t) \geq \hat{x}_{\max}(T_2)/2$ with probability at least $1-n^{-5}$. 
\end{proof}

\lemmaPhaseThreeDoubleBiasSingleSubphase*
\begin{proof}
    Our proof follows the analysis of the classical Gambler's ruin problem that within $O(n^2/x_1(t_0))$  interactions 
    we track the evolution of $x_1(t)-x_i(t)$ and show it reaches $2(x_1(t_0)-x_i(t_0))$ before $(x_1(t_0)-x_i(t_0))/2$  as long as $x_i(t)$ remains larger than $20\sqrt{n\log n}$.
    Let
    \begin{align*}
        \hat{T} = \inf\set{t\geq t_0 ~|~ u(t) \notin \left[\frac{n-x_{\max}(t)}{2}-8\cdot \sqrt{n\ln n}, \frac {n}{2}\right] \mbox{ or } x_{1}(t)<x_{1}(T_2)/2}
    \end{align*}
    be a stopping time. We define $(\hat{\mathbf{X}})_{t\ge t_0}$ as the process with $\hat{\mathbf{X}}(t) = \mathbf{X}(t)$ for all $t\leq \hat{T}$ and $\hat{\mathbf{X}}(t) = \mathbf{X}(\hat{T})$ for $t>\hat{T}$.
    From  \cref{lem:undecided>n/2-x1/2-o(n)} it follows that $u(t) \geq (n-x_{\max}(t))/2-8\cdot \sqrt{n\ln n} $ for all $t \in [T_2,n^3]$, \whp.
    From \cref{lem:undecided_general_bounds_v1} it follows that $ u(t) \leq n/2$  for all $t \in [T_2,n^3]$, \whp.
    Finally, \cref{lem:phase3-max-no-shrinking} gives us that $ x_{\max}(t) \geq x_{\max}(T_2)/2$ for all $t \in [T_2,T_2+ cn^2 \log n / x_{\max}(T_2)]$, \whp.
    Thus, $\hat{T} -T_2=\Omega(n^2\cdot \log n/x_{1}(t_0) )$ \whp and we can assume that $(\mathbf{X})_{t\ge t_0}$ and $(\hat{\mathbf{X}})_{t\ge t_0}$ are identical for $t \in [t_0,T_2 + O(n^2\cdot \log n/x_{\max}(T_2))]$.

    First we bound the number of $(1,i)$-productive interactions in the interval $[t_0,t_0+T]$.
    Assume for the remainder of the proof that $\hat{x}_i(t) \geq 20\sqrt{n\log n}$ for all $t \in [t_0,t_0+T]$ (otherwise the statement follows immediately).
    Recall that only $(1,i)$-productive interactions change the quantity $\hat{x}_1(t)-\hat{x}_i(t)$ but other interactions may change the remainder of the configuration, e.g.,\ an additional undecided agent is created.

    An interaction is $(1,i)$-productive with probability
        \begin{align*}
            \MoveEqLeft\frac{\hat{u}(t) \cdot \hat{x}_1(t) + \hat{x}_1(t)\cdot(n-\hat{u}(t) -\hat{x}_1(t)) + \hat{u}(t) \cdot \hat{x}_i(t) + \hat{x}_i(t)(n-\hat{u}(t)-\hat{x}_i(t))}{n^2} \\
                &= \frac{(\hat{x}_1+\hat{x}_i) \cdot n - \hat{x}_1^2-\hat{x}_i^2}{n^2} 
                = \frac{\hat{x}_1 \cdot (n-\hat{x}_1) + \hat{x}_i \cdot (n-\hat{x}_i)}{n^2} 
                \geq \frac{\hat{x}_1 \cdot (n-\hat{x}_1)}{n^2}
                \geq \frac{\hat{x}_1(T_2)}{6n}
        \end{align*}
        where we use $\hat{x}_1(T_2)/2 \leq \hat{x}_1(t) \leq 2n/3$. 
    Thus,  an application of Chernoff bounds provides for $c_1 = c/7$ at least $c_1\cdot n$ many $(1,i)$-productive interactions in $[t_0,t_0+T]$  \whp.
    
    For $1\leq i \leq c_1\cdot n$ we define $t_i$ as the $i$th $(1,i)$-productive interaction in $[t_0,t_0+\tau]$.
    Then for an arbitrary $i\in [1,c_1\cdot n]$ we have
    \begin{align*}
        \MoveEqLeft\Prob{\hat{X}_1(t_i+1)-\hat{X}_i(t_i+1) = \hat{x}_1-\hat{x}_i+1 ~|~ \hat{\mathbf{X}}(t_i) = \hat{\mathbf{x}}} \\ 
        &= \frac{1}{2} + \frac{\hat{u}\cdot \hat{x}_1 + \hat{x}_i\cdot(n-\hat{u}-\hat{x}_i) - (\hat{u}\cdot \hat{x}_i + \hat{x}_1\cdot(n-\hat{u}-\hat{x}_1)) }{2(\hat{u}\cdot \hat{x}_1 + \hat{x}_i\cdot(n-\hat{u}-\hat{x}_i) + (\hat{u}\cdot \hat{x}_i + \hat{x}_1\cdot(n-\hat{u}-\hat{x}_1)))} \\
        &= \frac{1}{2} + \frac{(\hat{x}_1 + \hat{x}_i + 2\hat{u} - n)\cdot (\hat{x}_1 - \hat{x}_i)}{2((\hat{x}_1 + \hat{x}_i) \cdot n - (\hat{x}_1^2+\hat{x}_i^2))} \\
        &\geq \frac{1}{2} +\frac{(\hat{x}_i-16\sqrt{n\ln n})(\hat{x}_1-\hat{x}_i)}{2(n(\hat{x}_1+\hat{x}_i)-(\hat{x}_1^2+\hat{x}_i^2))} \\
        &\geq \frac{1}{2} +\frac{(\hat{x}_i-16\sqrt{n\ln n})(\hat{x}_1-\hat{x}_i)}{2n(4\hat{x}_i+\hat{x}_i)}
        = \frac{1}{2} + \left(1-\frac{16\sqrt{n\ln n}}{\hat{x}_i}\right)\cdot \frac{(\hat{x}_1-\hat{x}_i)}{10n}\\
        &\geq \frac{1}{2}+\frac{\hat{x}_1(t_0)-\hat{x}_i(t_0)}{60n} 
    \end{align*}
    where we use $\hat{x}_1 < 4\cdot \hat{x}_i$ and $\hat{x}_i > 20\sqrt{n\log n}$ (otherwise the statement follows immediately).
    Additionally note that the last inequality holds as long as $\hat{x}_1(t)-\hat{x}_i(t) \geq (5/6) \cdot (\hat{x}_1(t_0)-\hat{x}_i(t_0)) $.

 Thus, the quantity $\hat{x}_1(t_i)-\hat{x}_i(t_i)$ increases by $1$ with probability at least $p = 1/2 + (\hat{x}_1(t_0)-\hat{x}_i(t_0))/(60n) $ and decreases by $1$, otherwise.
    Observe that starting at time $t_0$ with $\Delta = \hat{x}_1(t_0)-\hat{x}_i(t_0)$ as long as $\hat{x}_1(t_i)-\hat{x}_i(t_i) \geq \Delta/2$ for the first $i\leq c_1\cdot n$ many $(1,i)$-productive interactions in $[t_0,t_0+T]$
    the evolution of $\hat{x}_1(t_i)-\hat{x}_i(t_i)$ can be viewed as a biased random walk on the line starting at $\Delta$  where   a ''right step'' happens with probability $p$ and ''left step'' with probability $1-p$, otherwise. 
    The correctness follows from a standard coupling argument between two biased coins.
    Formally let $T_{min} = \inf\{ t' \geq t_0 ~|~ \hat{x}_1(t')-\hat{x}_i(t') = (5/6)\cdot (\hat{x}_1(t_0)-\hat{x}_i(t_0))\}$ and $T_{max} = \inf\{ t'\geq t_0 ~|~ \hat{x}_1(t')-\hat{x}_i(t') = 2(\hat{x}_1(t_0)-\hat{x}_i(t_0))\}$.
    First we bound $\Prob{T_{max} > T_{min}}$.
    It follows from \cref{lem:random_walk_no_ruin_probability} the probability of ever having an excess of $(5/12) \cdot \Delta$ ''left steps'' to ''right steps'' is at most 
    \begin{align*}
        \left(\frac{1-p}{p}\right)^{(5/12)\cdot \Delta}
        = \left( \frac{30n-\Delta}{30n+\Delta} \right)^{(5/12)\cdot \Delta}
        =\left(1-\frac{2\Delta}{30n+\Delta}\right)^{(5/12)\cdot \Delta}
        \leq e^{-\frac{(5/6)\cdot \Delta^2}{30n+\Delta}} 
        \leq n^{-5}
    \end{align*}
    where we use $\Delta \geq \alpha\sqrt{n}\log n$.

    Next we bound $\Prob{T_{max} > c_1 \cdot n}$.
    Again we use the assumption $\hat{x}_i(t_i)-\hat{x}_j(t_i) \geq (1/2) \cdot \Delta$. 
    Now consider  $\tau = c_1\cdot n$ independent Poisson trials ($S_i \in \{-1,1\}$ for all $i \leq c_1 \cdot n$) each with success probability $p= 1/2+\Delta/60n$.
    Let $S = \sum_{i=1}^{c_1\cdot n} S_i$.
    Using the Hoeffding bound (\cref{thm:chernoff-hoeffding-bound}) for $\lambda = \Delta$ we get 
    \begin{align*}
        \Prob{T_{max} > c_1 \cdot n}
        &\leq \Prob{S < \Delta } \\
        &= \Prob{S-\Ex{S} < \Delta - \Ex{S} }
     \\   &\leq\Prob{\abs{S - \Ex{S}} > \Ex{S} - \Delta}
    \\  &\leq 2\cdot e^{-\frac{2\Delta^2}{4\cdot c_1 \cdot n  }}
    \\  &\leq 2\cdot e^{-\frac{\Delta^2}{2\cdot c_1 \cdot n  }}
    \\ &\leq n^{-5}
    \end{align*}
    Hence, the statement follows by the union bound over the high probability events from above.
\end{proof}

\claimPhaseThreeLossOfMultiBias*
\begin{proof}
    Recall that we showed for ``small'' opinions with $\hat{x}_i < 20 \cdot \sqrt{n\log n}$ that the multiplicative bias is always larger than a constant.
    Furthermore recall that $\tau = 420 \cdot n^2 \cdot \log n/\hat{x}_1(T_2)$ and $T_2$ is the end of Phase 2.
    Assume w.l.o.g.\ that we start with the analysis at time $t_0=0$.
    Let 
    \begin{align*}
        \hat{T} = \inf\set{t\geq 0 ~|~ u(t) \notin [(n-x_{\max}(t))/2-8\cdot\sqrt{n\ln n}, n/2]}
    \end{align*}
    be a stopping time and let $(\hat{X}(t))_t$ denote the process with $\hat{X}(t) = X(t)$ for all $t\leq \hat{T}$ and $\hat{X}(t) = X(\hat{T})$ for $t>\hat{T}$.
   From \cref{lem:undecided_general_bounds_v1} and \cref{lem:undecided>n/2-x1/2-o(n)} it follows that $(\mathbf{X}(t))_t$ and $(\hat{\mathbf{X}}(t))_t$ behave the same for the duration of at least two subphases.
    
    An interaction is productive w.r.t.\ to $x_1$ and $x_i$ (meaning that either $x_1$ or $x_i$ change) with probability
    {\dense
    \begin{align*}
        p = \frac{\hat{u} \cdot \hat{x}_{1} + \hat{x}_{1} \cdot (n-\hat{u}-\hat{x}_{1}) + \hat{u} \cdot \hat{x}_i + \hat{x}_i \cdot (n - \hat{x}_i) }{n^2}
        = \frac{\hat{x}_1 \cdot (n-\hat{x}_1) + \hat{x}_i \cdot (n-\hat{x}_i)}{n^2}
        \leq 3 \cdot \frac{\hat{x}_1(T_2)}{n}
    \end{align*} }
    for $\hat{x}_1 \geq 2 \cdot \hat{x}_i$.
    It follows from an application of Chernoff bounds that within a sequence of $T$ interactions the number of $1$-productive interactions 
    is at most  $2T \cdot p=6 T\cdot (\hat{x}(T_2)) \cdot \log n/n \leq 2520 n \cdot \log n$ with probability at least $1-n^{-10}$. We define $\tau' = 2520 n \cdot \log n$
    and consider $\tau'$ productive interactions.  Let $Z(t) = \hat{x}_{1}(t)-2\hat{x}_i(t)$.
   Our goal is to use  the Hoeffding bound (\cref{lem:hoeffding-dependent-tailbound_v1}) to show that this quantity does not decrease significantly throughout $\tau'$ productive interactions. Hence, we have to calculate the probability that $Z(t)$ increases or decreases. Note that the maximum one step change in $\in [-2,2]$.
    Assuming the next interaction is a $1$-productive interaction for $\hat{x}(t)$ we have
    {\dense
    \begin{align*}
        \Prob{Z(t+1)-Z(t) = 1 ~|~ \mathbf{X}(t) = \mathbf{x}} 
        &= \frac{1}{p}\cdot \frac{\hat{x}_1 \cdot \hat{u} }{n^2}
        &= \frac{\hat{x}_1 \cdot \hat{u}}{ \hat{x}_1 \cdot (n-\hat{x}_1) + \hat{x}_i \cdot (n-\hat{x}_i)} \\
        \Prob{Z(t+1)-Z(t) = -1 ~|~ \mathbf{X}(t) = \mathbf{x}}
        &=  \frac{1}{p} \cdot \frac{\hat{x}_1 \cdot (n-u-\hat{x}_1) }{n^2}
        &= \frac{\hat{x}_1 \cdot (n-\hat{x}_1) }{ \hat{x}_1 \cdot (n-\hat{x}_{1}) + \hat{x}_i \cdot (n-\hat{x}_i)} \\
        \Prob{Z(t+1)-Z(t) = -2 ~|~ \mathbf{X}(t) = \mathbf{x}}
        &=  \frac{1}{p} \cdot \frac{\hat{x}_i \cdot \hat{u} }{n^2} 
        &= \frac{\hat{x}_i \cdot (n-\hat{x}_i) }{ \hat{x}_1 \cdot (n-\hat{x}_{1}) + \hat{x}_i \cdot (n-\hat{x}_i)} \\
        \Prob{Z(t+1)-Z(t) = 2 ~|~ \mathbf{X}(t) = \mathbf{x}}
        &= \frac{1}{p} \cdot \frac{\hat{x}_i \cdot (n-\hat{u}-\hat{x}_i) }{n^2}
        &= \frac{\hat{x}_1 \cdot \hat{u}}{ \hat{x}_1 \cdot (n-\hat{x}_{1}) + \hat{x}_i \cdot (n-\hat{x}_i)}
        \end{align*} }
        Therefore
        \begin{align*}
        \Ex{Z(t+1)-Z(t) ~|~ \mathbf{X}(t) = \mathbf{x}} 
        & = \frac{\hat{x}_1 \cdot \hat{u} - 2 \hat{x}_i \cdot \hat{u} - \hat{x}_1 (n-\hat{u}-\hat{x}_{1}) + 2 \hat{x}_i (n-\hat{u}-\hat{x}_{i})}{\hat{x}_1 \cdot (n-\hat{x}_{1})+\hat{x}_i\cdot(n-\hat{x}_i)}\\
        & \geq \frac{(\hat{x}_1-2\hat{x}_i) \cdot (2\hat{u}-n+\hat{x}_1) + \hat{x}_1 \cdot \hat{x}_i}{n-\hat{x}_{1}+\hat{x}_i\cdot(n-\hat{x}_i)}
    \intertext{where we used that $\hat{x}_1 \geq 2 \cdot \hat{x}_i$.
    Since $\hat{x}_i \geq 20 \cdot \sqrt{n\log n}$ and $\hat{u} \geq n/2-\hat{x}_1/2-8\cdot\sqrt{n\log n}$ (\cref{lem:undecided>n/2-x1/2-o(n)}), we get}
        \Ex{Z(t+1)-Z(t) ~|~ \mathbf{X}(t) = \mathbf{x}}
        &\geq \frac{(\hat{x}_1-2\hat{x}_i) \cdot (2\hat{u}-n+\hat{x}_1) + \hat{x}_1 \cdot \hat{x}_i}{n-\hat{x}_{1}+\hat{x}_i\cdot(n-\hat{x}_i)}\\
        & \geq \frac{(\hat{x}_1-2\hat{x}_i) \cdot (-16 \sqrt{n\log n}) + 4\hat{x}_1 \cdot \hat{x}_i/5 + \hat{x}_1\cdot \hat{x}_i/5}{n-\hat{x}_{1}+\hat{x}_i\cdot(n-\hat{x}_i)}\\
        & \geq \frac{2 \hat{x}_i \cdot 16 \sqrt{n\log n} + \hat{x}_1\cdot \hat{x}_i/5}{n-\hat{x}_{1}+\hat{x}_i\cdot(n-\hat{x}_i)}
        > 0        
    \end{align*}
    Thus, we have $\Ex{Z(t+1)-Z(t) ~|~ \mathbf{X}(t) = \mathbf{x}} \geq 0$ if $\hat{x}_1 \geq 2 \cdot \hat{x}_i$ and $\hat{x}_i \geq 20 \cdot \sqrt{n\log n}$.

    Now we are ready to apply the Hoeffding bound from \cref{lem:hoeffding-dependent-tailbound_v1}. Observe that $\abs{Z(t+1)-Z(t)} \leq 2$ for all $t\in[0,\tau'-1]$ and
    \begin{align*}
        S = \sum_{t=0}^{\tau'-1} Z(t+1)-Z(t) = Z(\tau')-Z(0)
    \end{align*}
    Then it follows from Hoeffding bound (\cref{lem:hoeffding-dependent-tailbound_v1}) with $\lambda = Z(0) \geq \hat{x}_1(0)/2$ that
    \begin{align*}
        \Prob{S < Z(0) - c_1 \cdot Z(0)} 
        \leq \Prob{S - \Ex{S} < -\lambda}
        \leq \exp\left(-\frac{2\lambda^2}{16\tau'}\right)
        \leq n^{-c \cdot \log^2(n)}
    \end{align*}
    for some constant $c$.
    Thus, we have that \whp $Z(\tau') \geq  Z(0)$.
    Then,
    \begin{align*}
        \frac{\hat{x}_1(\tau')}{\hat{x}_i(\tau')} 
        = \frac{\hat{x}_1(\tau')-2\hat{x}_i(\tau')}{\hat{x}_i(\tau')} + \frac{2\hat{x}_i(\tau')}{\hat{x}_i(\tau')}
        = \frac{\hat{x}_1(\tau')-2\hat{x}_i(\tau')}{\hat{x}_i(\tau')} + 2
        = \frac{Z(\tau')}{\hat{x}_i(\tau')} + 2
        \geq 2 .
    \end{align*}
    Thus, \whp $x_1(\tau') \geq x_i(\tau')$.
    The claim follows from the union bound over all $\tau' < n^3$ interactions.
\end{proof}

\subsection{Omitted Proofs of Section 6 (Phase 4)}
\label{apx:omitted-proofs-phase4}

\lemmaPhaseFourXmaxNotShrinking* 
\begin{proof}
    Let 
    \begin{align*}
        \hat{T} = \inf\set{t\geq T_3 ~|~ u(t) \notin [(n-x_{\max}(t'))/2-8\cdot\sqrt{n\ln n}, n/2]}
    \end{align*}
    be a stopping time and let $(\hat{X}(t))_t$ denote the process with $\hat{X}(t) = X(t)$ for all $t\leq \hat{T}$ and $\hat{X}(t) = X(\hat{T})$ for $t>\hat{T}$.
    From  \cref{lem:undecided_general_bounds_v1} and  \cref{lem:undecided>n/2-x1/2-o(n)}  it follows $\hat{T}-t =\Omega(n^2/x_{\max}(t) \cdot \log n)$ \whp.
    Thus, $(\mathbf{X}(t))_t$ and $(\hat{\mathbf{X}}(t))_t$ behave the same between time $t$ and $t + O(n^2/x_{\max}(T_3)\cdot \log n) $.
    As long as $\hat{x}_{1}(t') \leq 2 \cdot \hat{x}_{1}(T_3)$  an interaction is $1$-productive  with probability
    \begin{align*}
        \frac{\hat{u} \cdot \hat{x}_{1} + \hat{x}_{1} \cdot (n-\hat{u}-\hat{x}_{1})}{n^2}
        = \frac{\hat{x}_{1} \cdot (n-\hat{x}_{1})}{n^2}
        \leq 2 \cdot \frac{\hat{x}(T_3)}{n}
    \end{align*}
    It follows from an application of Chernoff bounds that within a sequence of $c\cdot n^2\cdot \log n/x_{1}(T_3)$ interactions the number of $1$-productive interactions 
    is at most  $4\cdot c \cdot n\log n$ with probability at least $1-n^{-10}$.
    Now consider $\tau = 4\cdot c \cdot n\log n$ such productive interactions and let $Z_t$ denote the change w.r.t.\ $\hat{x}_{1}(t)$, i.e., the support of the largest opinion increase or decrease by one, respectively.
    That is, assuming the next interaction is a $1$-productive interaction for $\hat{x}(t)$ we have
    \begin{align*}
        &\Prob{Z_t = 1} 
        = \frac{\hat{u} \cdot \hat{x}_{1}}{ \hat{u}\cdot \hat{x}_{1} + \hat{x}_{1} \cdot (n-\hat{u}-\hat{x}_{1})}
        = \frac{\hat{u}  }{ (n-\hat{x}_{1})} \\
        &\Prob{Z_t = -1}
        = 1- \Prob{Z_t = 1}
    \end{align*}
    Therefore
    \begin{align*}
        \Ex{Z_t} 
        = \frac{\hat{u}-(n-\hat{u}-\hat{x}_{1})}{n-\hat{x}_{1}}
        = \frac{2\cdot \hat{u} + \hat{x}_{1} - n}{n-\hat{x}_{1}}
        \geq -48 \cdot \frac{\sqrt{n\ln n}}{n}
    \end{align*}
    Let $Z$ be the sum of  $Z_t$ for all $t\in [1,\tau]$. 
    Then it follows from Hoeffding bound with $\lambda = \hat{x}_{1}(T_2)/2 - 200 \cdot \sqrt{n}\ln^{3/2} n$
    \begin{align*}
        \Prob{Z < -\frac{1}{2} \cdot \hat{x}_{1}(T_1)} 
        \leq \Prob{Z < \Ex{Z}-\lambda}
        \leq e^{-\frac{2\lambda^2}{4\tau}}
        \leq n^{-10}
    \end{align*}

    Note that if (ever) $\hat{x}_1(t') > 2\cdot \hat{x}_1(T_3)$ for some $t'\in [T_3,T_3+T]$ the statement hold  by the union bound and the previous part. 
    Thus, starting with $\hat{x}_{1}(T_3)$  throughout the next $ c\cdot n^2/x_{\max}(T_2)\cdot \log n $ interactions $\hat{x}_{\max}(t) \geq \hat{x}_{\max}(T_3)/2$ with probability at least $1-n^{-5}$. 
\end{proof}

\lemmaPhaseFourUndecidedImprovedBoundMultiplicativeBias* 
\begin{proof}
    The proof is similar to the proof of \cref{claim:phase3-loss-of-bias} using $Z(t) = x_1(t) - 7x_i(t)/4$ instead of $Z(t) = (t) - 2x_i(t)$.
    We have $Z(0) = x_1(0)-7x_i(0)/4 \geq x_1(0)/8$ and $\Ex{Z(t+1)-Z(t)} \geq 0$.
\end{proof}

\lemmaPhaseUndecidedGrowImprovedBound* 
\begin{proof}
    To bound $T_u-T_3$ we follow the proof of \cref{lem:phase1}. 
    Let $\alpha = 7/8$ and let $Z(t) = n - 2 u(t) - \alpha \cdot x_1(t)$ and let $r^2 = \sum_{i \in [k]} x_i^2$.
    Then 
    {\small \dense
    \begin{align*}
        \MoveEqLeft \Ex{ Z(t)-Z(t+1) ~|~ \mathbf{X}(t)=\mathbf{x} }\\
        & = - \frac{x_1 \cdot u}{n^2} \cdot (2-\alpha) - \sum_{i = 2}^k \frac{x_i \cdot u}{n^2} \cdot 2 - \frac{x_1(n-u-x_1)}{n^2}\cdot (-2+\alpha) - \sum_{i=2}^k \frac{x_i(n-u-x_i)}{n^2} \cdot (-2)\\
        & = (2-\alpha) \cdot \frac{-(x_1 \cdot u) + x_1 \cdot (n-u-x_1)}{n^2} + 2 \cdot \sum_{i=2}^k \frac{-(x_i \cdot u) + x_i \cdot (n-u-x_i)}{n^2}\\
        & = (2-\alpha) \cdot \frac{x_1 \cdot (n-2u-x_1)}{n^2} + 2 \cdot \sum_{i=2}^k \frac{x_i \cdot (n-2u) -x_i^2}{n^2}\\
        & = 2 \cdot \frac{x_1 \cdot (n-2u) - x_1^2}{n^2} - \alpha \cdot \frac{x_1 \cdot (n-2u-x_1)}{n^2} + 2 \cdot \frac{(n-u-x_1)(n-2u)}{n^2} - 2 \cdot \frac{r^2-x_1^2}{n^2}\\
        & = 2 \cdot \frac{x_1 \cdot (n-2u) + (n-u-x_1)(n-2u) - r^2}{n^2} - \frac{\alpha \cdot x_1\cdot (n-2u-x_1)}{n^2}\\
        & = 2 \cdot \frac{(n-2u)\cdot (n-u)-r^2}{n^2} -  \frac{\alpha \cdot x_1 \cdot (n-2u-x_1)}{n^2}\\
        & = 2 \cdot \frac{(n-2u-\alpha \cdot x_1) \cdot (n-u) + \alpha \cdot x_1 \cdot (n-u)-r^2}{n^2} - \frac{2\alpha \cdot x_1 \cdot (n-u) - \alpha \cdot x_1 \cdot (n+x_1)}{n^2}\\
        & = 2 \cdot \frac{(n-2u-\alpha \cdot x_1) \cdot (n-u)}{n^2} + \frac{2 \alpha \cdot x_1 \cdot (n-u)-2r^2}{n^2} - \frac{2 \alpha \cdot x_1 \cdot (n-u) - \alpha \cdot x_1 \cdot (n+x_1)}{n^2}\\
        & = 2 \cdot \frac{Z(t)\cdot (n-u)}{n^2} + \frac{\alpha \cdot x_1 \cdot (n+x_1)-2r^2}{n^2}\\
        & = \frac{Z(t)}{2n} + \frac{1}{n^2} \cdot \left( 3n^2/2-2n\cdot u + \alpha \cdot x_1 \cdot n + \alpha \cdot x_1^2 - 2r^2 \right)
    \end{align*}
    }
    Note that $r^2 = \sum_{i=1}^k x_i^2 \leq x_1^2 + (4/7) \cdot x_1 \cdot \sum_{i=2}^k x_i = x_1^2 + (4/7) \cdot x_1 \cdot (n-u-x_1)$. 
    Furthermore, by \cref{lem:undecided>n/2-x1/2-o(n)} and \cref{lem:undecided_general_bounds_v1} and using $x_1 \leq 2n/3$, we have \whp $u < n/2$ and $u \geq n/2-x_1/2-o(x_1) \geq n/8$ for sufficiently large $n$.
    For the last expression in parentheses we calculate 
    \begin{align*}
         \MoveEqLeft 3n^2/2-2n\cdot u + \alpha \cdot x_1 \cdot n + \alpha \cdot x_1^2 - 2r^2\\
         & \geq 3n^2/2-2n\cdot u + \alpha \cdot x_1 \cdot n + \alpha \cdot x_1^2 - 2 (x_1^2 + (4/7) \cdot x_1 \cdot (n-u-x_1)) \\
         & \geq 3n^2/2 -n^2 + \alpha \cdot x_1 \cdot n + \alpha \cdot x_1^2 - 2 (x_1^2 + (4/7) \cdot x_1 \cdot ((7/8) \cdot n-x_1))\\
        \geq 0
    \end{align*}
    for $\alpha = 7/8$.

    The remainder of the proof is identical to that of \cref{lem:phase1} except that we note that either $\Ex{Z(t)-Z(t+1) ~|~ \mathbf{X}(t)=\mathbf{x} } \geq Z/(n)$ or at some time $t \in [T_3,n^3]: x_1(t) < 7/4 \cdot x_i(t)$ for some $i > 1$.
    The latter event is ruled out \whp by \cref{lem:phase4_multiplicative_bias}.
    
    We now apply \cref{thm:mult_drift_tail_lengler_18} with $r = 3 \ln n$, $s_0 = n-2u(0)-7/8 \cdot x_1(0) \leq n$, $s_{min} = 1$, $\delta = 1/(2n)$ and get with $T = \inf\set{t \geq T_3 ~|~ Z(t) \leq 0}$
    %
    \begin{align*}
    \Prob{T-T_3 > \lceil 7 n\ln n \rceil} 
    & \leq \Prob{T-T_3 > \left \lceil{ \frac{6 \cdot \ln n + \ln(n-2u(T_3)- 7/8 \cdot x_1(T_3))}{1/(2n)} }\right \rceil }\\
    &\leq e^{-3 \cdot \ln(n)}
    = n^{-3} \; .
    \end{align*}
    Note that if ever $x_1(t) \geq 2n/3$ for $t < \lceil 7 n\ln n \rceil$, we have $T_4 \leq \lceil 7 n\ln n \rceil$.
    Otherwise, we have shown that $T_u \leq \lceil 7 n\ln n \rceil$.
    Hence, overall we get $\min\{T_u,T_4\}-T_3\le \lceil 7 n\ln n \rceil$.
    \end{proof}

\claimPhaseUndecidedImprovedBoundHoldsUntilNextPhase*
    \begin{proof}
    We follow the proof idea of Theorem 6 in \cite{lengler2018drift}.
    We define a new set of random variables with $Y(t) = \exp(\eta \cdot Z(t))$ for $t \geq T$ and $\eta=\sqrt{\ln n/n}$
    and let $z_0 = 4\eta \cdot n$. 
    
    Fix an arbitrary $i \geq 0$.
    We first give a bound for $\Ex{Y(i+1)-Y(i) ~|~ Z(i) = z}$.
    Note that $Z(i+1)-Z(i) \in [-2,2]$.
    We get
    \begin{align*}
        \MoveEqLeft \Ex{Y(i+1)-Y(i) ~|~ Z(i) = z}\\
        & = \Ex{e^{\eta \cdot Z(i+1)} - e^{\eta \cdot Z(t)} ~|~ Z(i) = z}\\
        & = e^{\eta \cdot z} \cdot \Ex{e^{\eta\cdot(Z(i+1)-z)}-1 ~|~ Z(i) = z}\\
        & = e^{\eta \cdot z} \cdot \sum_{\mathclap{j\in \set{-2,-9/8,0,9/8,2}}} (e^{\eta \cdot j} -1) \cdot \Prob{Z(i+1)-z=j ~|~ Z(i) = z}\\
    \end{align*}

    We derive the following bound for $\exp(\eta \cdot j)-1$.
    Since $\exp(x) \leq 1+x+x^2$ for $x \leq 1$ and $\eta \rightarrow 0$ for large $n$, we have
    $\exp(2\eta) \leq 1+2\eta+(2\eta)^2 = 1+2\eta+\eta \cdot z_0/n$.
    For $j \in [-2,2]$, we thus have $\exp(\eta j)-1 \leq \eta j + \eta \cdot z_0/n$.
    We know that $\Ex{Z(i+1)-Z(i) ~|~ Z(i) = z} \leq -\frac{z}{n}$ \whp from Part 1.

    Thus, for all $z \geq z_0$ we have    
    \begin{align*}
        \MoveEqLeft \Ex{Y(i+1)-Y(i) ~|~ Z(i) = z}\\
        & \leq e^{\eta \cdot z} \cdot \sum_{\mathclap{j\in \set{-2,-9/8,0,9/8,2}}} (\eta \cdot j + \eta \cdot z_0/n) \cdot \Prob{Z(i+1)-z=j ~|~ Z(i) = z}\\
        & =  e^{\eta \cdot z} \cdot \eta \cdot (\Ex{Z(i+1)-Z(i) ~|~ Z(i) = z} + z_0/n) \leq 0.
    \end{align*}
    
    In total, we get 
    \begin{align*}
        \Ex{Y(t)}
        = \Ex{Y(0)} + \sum_{i=0}^{t-1} \Ex{Y(i+1)-Y(i)}
        \leq 1.
    \end{align*}

    
    Since $\forall t \geq 0: Y(t) \geq 0$, we can apply Markov's inequality.
    Thus,
    \begin{align*}
        \Prob{Z(t) \geq 2z_0} = \Prob{Y(t) \geq \exp(2\eta z_0)} \leq \frac{\Ex{Y(t)}}{n^{8}} \leq n^{-8}.
    \end{align*}
    Finally, we apply the union bound over $n^3-T \leq n^3$ interactions.
\end{proof}

\claimPhaseFourDoubleSupportMaximumSingleSubphase* 
\begin{proof}
     The proof is similar to the proof of \cref{lem:additive_bias_growth_v2} but instead of analyzing the quantity $x_1(t)-x_i(t)$ we only analyzing the growth of $x_1(t)$ directly.
    Let 
    {\dense
    \begin{align*}
        \hat{T} = \inf\set{t\geq T_3 +7n\ln n ~|~ u(t) \notin \left[\frac{\left(n-7/8 \cdot x_{1}(t)\right)}{2}-8\cdot \sqrt{n\ln n}, \frac{n}{2}\right] \mbox{ or } x_1(t) < \frac{x_1(T_3)}{2}}
    \end{align*} }
    be a stopping time and let $(\hat{\mathbf{X}}(t))_{t\geq T_3+7n\ln n}$ denote the process with $\hat{\mathbf{X}}(t) = \mathbf{X}(t)$ for all $t\leq \hat{T}$ and $\hat{\mathbf{X}}(t) = \mathbf{X}(\hat{T})$ for $t>\hat{T}$.
    From  \cref{lem:undecided_grow_improved_bound} it follows that $u(t) \geq (n-x_{1}(T_3+7n\ln n))/2-8\cdot \sqrt{n\ln n} $ for all $t \in [T_3+7n\ln n,n^3]$, \whp.
    From \cref{lem:undecided_general_bounds_v1} it follows that $ u(t) \leq n/2$  for all $t \in [T_1,n^3]$, \whp.
    Finally, \cref{lem:phase2-max-no-shrinking} gives us that $ x_{1}(t) \geq x_{1}(T_3)/2$ for all $t \in [T_3,T_3+ cn^2 \log n / x_{1}(T_3)]$, \whp.
    Thus, $\hat{T}-(T_3+7n\ln n) =\Omega(n^2/x_{1}(T_3) )$ \whp and we can assume that $(\mathbf{X})_{t\ge T_3+t_0}$ and $(\hat{\mathbf{X}})_{t\ge T_3+t_0}$ are identical for $t \in [t_0,T_3+7n\ln n + O(n^2\cdot \log n/x_{1}(T_3))]$.


    First we bound the number of $1$-productive interactions in the interval $[t_0,t_0+\tau]$ for $\tau = 111 \cdot n^2/\hat{x}_1(t_0)$.
    Assume for the remainder of the proof that $\hat{x}_1(t) < 2n/3$ for all $t \in [t_0,t_0+\tau]$ (otherwise the statement follows immediately).
    Recall that only $1$-productive interactions change the quantity $x_1(t)$ but other interactions may change the remainder of the configuration, e.g.,\ an additional undecided agent is created.

    An interaction is $1$-productive with probability
        \begin{align*}
            \MoveEqLeft\frac{\hat{u}(t) \cdot \hat{x}_1(t) + \hat{x}_1(t)\cdot(n-\hat{u}(t) -\hat{x}_1(t)) }{n^2} 
                \geq \frac{\hat{x}_1(t) \cdot (n-\hat{x}_1(t))}{n^2}
                \geq \frac{\hat{x}_1(t)}{3n} 
                \geq \frac{\hat{x}_1(t_0)}{10n}
        \end{align*}    
    where we use $\hat{x}_1(t_0)/2 \leq \hat{x}_1(t)<2n/3$.
    
    Thus, an application of Chernoff bounds provides for $c_1 = 110$ at least $c_1\cdot n$ many $1$-productive interactions in $[t_0,t_0+\tau]$  \whp.
    For $1\leq i \leq c_1\cdot n$ we define $t_i$ as the $i$th $1$-productive interaction in $[t_0,t_0+\tau]$.
    Then for an arbitrary $i\in [1,c_1\cdot n]$ we have

    \begin{align*}
        \MoveEqLeft \Prob{\hat{X}_1(t_i+1) = \hat{x}_1+1 ~|~ \hat{\mathbf{X}}(t_i) = \hat{\mathbf{x}} }\\
        &= \frac{1}{2} + \frac{\hat{x}_1 \cdot \hat{u} - \hat{x}_1(n-\hat{u}-\hat{x}_1)}{2 \cdot (\hat{x}_1 \cdot \hat{u} +\hat{x}_1(n-\hat{u}-\hat{x}_1)) }
        = \frac{1}{2} + \frac{2\hat{u}-n+\hat{x}_1}{2(n-\hat{x}_1)}
       \geq \frac{1}{2} + \frac{\hat{x}_1(t_0)}{110n} 
    \end{align*}  
    Note that the last inequality holds as long as $\hat{x}_1(t) \geq \hat{x}_1(t_0)/2$.
    Thus, the quantity $\hat{x}_1(t_i)$ increases by $1$ with probability at least $p = 1/2 + \hat{x}_1(t_0)/(110n) $ and decreases by $1$, otherwise.
    Observe that starting at time $t_0$ with $\Delta = \hat{x}_1(t_0)$ as long as $\hat{x}_1(t_i) \geq \Delta/2$ for the first $i\leq c_1\cdot n$ many $1$-productive interactions in $[t_0,t_0+\tau]$
    the evolution of $\hat{x}_1(t_i)$ can be viewed as a biased random walk on the line starting at $\Delta$  where   a ''right step'' happens with probability $p$ and ''left step'' with probability $1-p$, otherwise. 
    The correctness follows from a standard coupling argument between two biased coins.
    Formally let $T_{min} = \inf\{ t' \geq t_0 ~|~ \hat{x}_1(t') = \hat{x}_1(t_0)/2\}$ and $T_{max} = \inf\{ t'\geq t_0 ~|~ \hat{x}_1(t') = 2\hat{x}_1(t_0)\}$.

    First we bound $\Prob{T_{max} > T_{min}}$.
    It follows from \cref{lem:random_walk_no_ruin_probability} the probability of ever having an excess of $\Delta/4$ ''left steps'' to ''right steps'' is at most 
    \begin{align*}
        \left(\frac{1-p}{p}\right)^{\Delta/4}
        = \left( \frac{55n-\Delta}{55n+\Delta} \right)^{\Delta/4}
        =\left(1-\frac{2\Delta}{55n+\Delta}\right)^{\Delta/4}
        \leq e^{-\frac{\Delta^2}{2(55n+\Delta)}}
        \leq n^{-5}
    \end{align*}
    where we use $\Delta \geq \hat{x}_1(t_0)/2$.

    Next we bound $\Prob{T_1 > c_1 \cdot n}$.
    Now consider  $\tau = c_1\cdot n$ independent Poisson trials ($S_i \in \{-1,1\}$ for all $i \leq c_1 \cdot n$) each with success probability $p=  1/2 + \hat{x}_1(t_0)/(36n)$.
    Let $S = \sum_{i=1}^{c_1\cdot n} S_i$.
    Using the Hoeffding bound (\cref{thm:chernoff-hoeffding-bound}) for $\lambda = \Delta$ we get 
    \begin{align*}
        \Prob{T_1 > c_1 \cdot n}
        &\leq \Prob{S < \Delta } \\
        &= \Prob{S-\Ex{S} < \Delta - \Ex{S} }
     \\   &\leq\Prob{\abs{S - \Ex{S}} > \Ex{S} - \Delta}
    \\  &\leq 2\cdot e^{-\frac{2\Delta^2}{4\cdot c_1 \cdot n  }}
    \\  &\leq 2\cdot e^{-\frac{\Delta^2}{2\cdot c_1 \cdot n  }}
    \\ &\leq n^{-5}
    \end{align*}
    Hence, the statement follows by the union bound over the high probability events from above.
\end{proof}

\newpage

\subsection{Omitted Proofs of Section 7 (Phase 5)}
\label{apx:omitted-proofs-phase5}

In this appendix we present the remaining cases from the proof of \cref{lem:k-undecided-part5}.

\paragraph{ Case 2: $i\le a$ and $j > a$}
\mbox{}

\begin{center}
\begin{tabular}{llll|ll}
\toprule
$v_i(t)$  & $\tilde{v}_i(t)$  & $v_j(t)$ & $\tilde{v}_j(t)$ & $v_i(t+1)$ & $\tilde{v}_i(t+1)$\\
\midrule
1 & 1 & $\bot$ & 2 &               1  & $\bot$  \\
1 & 1 & 1 & $\bot$ &               1  & 1 \\
1 & 1 & 1 & 2 &                    1  & $\bot$ \\
$\bot$ & $\bot$ & $\bot$ & 2 & $\bot$ & 2 \\
$\bot$ & $\bot$ & 1 & $\bot$ &     1  & $\bot$ \\
$\bot$ & $\bot$ & 1 & 2 &          1  & 2 \\
2 & 2 & $\bot$ & 2 &               2  & 2 \\
2 & 2 & 1 & $\bot$ &           $\bot$ & 2 \\
2 & 2 & 1 & 2 &                $\bot$ & 2 \\
$>2$ & 2 & $\bot$ & 2 &              $>2$ & 2 \\
$>2$ & 2 & 1 & $\bot$ &          $\bot$ & 2 \\
$>2$ & 2 & 1 & 2 &               $\bot$ & 2 \\
\bottomrule
\end{tabular}
\end{center}

\paragraph{Case 3: $i > a$ and $j\le a$}
\mbox{}

\begin{center}
\begin{tabular}{llll|ll}
\toprule
$v_i(t)$  & $\tilde{v}_i(t)$  & $v_j(t)$ & $\tilde{v}_j(t)$ & $v_i(t+1)$ & $\tilde{v}_i(t+1)$\\
\midrule
$\bot$ & 2 & 1 & 1 &                        1      & $\bot$ \\
1 & $\bot$ & 1 & 1 &                        1      & 1      \\
1 & 2 &  1 & 1 &                            1      & $\bot$ \\
$\bot$ & 2 & $\bot$ & $\bot$ &              $\bot$ & 2      \\
1 & $\bot$ & $\bot$ & $\bot$ &              1      & $\bot$ \\
1 & 2 & $\bot$ & $\bot$ &                   1      & 2      \\
$\bot$ & 2 & 2 & 2 &                        2      & 2      \\
1 & $\bot$ & 2 & 2 &                        $\bot$ & 2      \\
1 & 2 & 2 & 2 &                             $\bot$ & 2      \\
$\bot$ & 2 & $>2$ & 2 &                       $>2$     & 2      \\
1 & $\bot$ & $>2$ & 2 &                       $\bot$ & 2      \\
1 & 2 &  $>2$ & 2 &                           $\bot$ & 2      \\
\bottomrule
\end{tabular}
\end{center}
    
\paragraph{ Case 3: $i,j > a$. }

\mbox{}
\begin{center}
\begin{tabular}{llll|ll}
\toprule
$v_i(t)$  & $\tilde{v}_i(t)$  & $v_j(t)$ & $\tilde{v}_j(t)$ & $v_i(t+1)$ & $\tilde{v}_i(t+1)$\\
\midrule
1 & $\bot$ & 1 & $\bot$ &   1      &  $\bot$ \\
1 & $\bot$ & 1 & 2      &    1      &  2 \\
1 & $\bot$ & $\bot$ & 2  &   1      &  2 \\
1 & 2      & 1 & $\bot$ &   1      &  2 \\
1 & 2      & 1 & 2      &    1      &  2 \\
1 & 2      & $\bot$ & 2  &   1      &  2 \\
$\bot$ & 2 & 1 & $\bot$ &   1      &  2 \\
$\bot$ & 2 & 1 & 2      &    1      &  2 \\
$\bot$ & 2 & $\bot$ & 2 &    $\bot$ &  2 \\
\bottomrule
\end{tabular}
\end{center}

It is easy to see that in all three cases we have $x_1(t)\ge \tilde{x}(t)$ and 
$x_1(t)+u(t)\ge \tilde{x}(t)\tilde{u}(t)$.
This holds since our coupling maintains majorization: whenever $\tilde{u}(t)$ is increased, $x_1(t)$ is increased.

\section[Comparison of Convergence Rates]{Comparison of Convergence Rates With Becchetti et al.\ \cite{DBLP:conf/soda/BecchettiCNPS15}}
\label{apx:becchetti-compare}

We show that given an initial configuration
with a multiplicative bias, our convergence rate
from \cref{thm:main_theorem} improves over the analogous rate from Becchetti et al.\ \cite{DBLP:conf/soda/BecchettiCNPS15} whenever the initial support of the largest opinion $x_1$ is close to the average opinion size, that is, $x_1 \leq n/k \cdot \log n$.

In the regime of 
an initial multiplicative bias, the analysis of Becchetti et al.\ 
of the USD in the gossip model shows the process achieves
plurality consensus in  $O(\text{md}(\x(0)) \cdot \log n)$ rounds, where
(assuming $x_1$ has largest initial support)
\begin{equation*}
\text{md}(\x(0)) = \sum_{i\in [k]} \Big( \frac{x_i(0)}{x_1(0)} \Big)^2 \;.
\end{equation*}
On the other hand, recall our result from \cref{thm:main_theorem}, which shows convergence
towards plurality consensus in the population protocol model in 
$O(n \log n + n^2/x_1(0))$ 
interactions, which is equivalent to 
$O(\log n + n/x_1(0))$ \emph{parallel} time.

Considering the range of $k$ for which their result holds, our
convergence rate  improves  the one
of Becchetti et al. To see this, consider an initial
configuration $\x$ and assume that w.l.o.g.\ $x_1 \geq x_i $ for all $2 \leq i \leq n$. We calculate
\[
\text{md}(\x) \log n = \sum_{i=1}^{k}x_i^2/x_1^2 \log n \geq  \frac{k\cdot (n/k)^2}{x_1^2} \log n = 
\frac{n^2}{k \cdot x_1^2} \log n = \frac{n\cdot \log n}{k\cdot x_1} \cdot \frac{n}{x_1}.
\]
Hence, $ \text{md}(\x) \log n $ gives the better running time if
\[
 x_1> \frac{n\cdot \log n}{k}.
\]

\end{document}